\documentclass[aps,pre,reprint,superscriptaddress,floatfix]{revtex4-2}
\usepackage{amsmath,amssymb,amsthm,amsfonts}
\usepackage{bm,braket,graphicx,booktabs,tikz}
\usetikzlibrary{arrows.meta,calc,positioning}
\usepackage{hyperref}
\AtBeginDocument{\setcounter{secnumdepth}{3}}

\newcommand{\E}{\mathbb{E}}
\newcommand{\Cov}{\mathrm{Cov}}
\newcommand{\Var}{\mathrm{Var}}
\newcommand{\Tr}{\mathrm{Tr}}
\newcommand{\adj}{\mathrm{adj}}
\newcommand{\dd}{\mathrm{d}}
\newcommand{\1}{\mathbf 1}
\newcommand{\F}{\textsc{f}}
\newcommand{\Wg}{\mathrm{Wg}}
\newcommand{\cG}{\mathcal{G}}

\begin{document}

\title{Exact Replica-Sector Hierarchy of Multi-Resolvent Spectral-Weight Fluctuations in Random Free Fermions}

\author{Zhiqiang Huang}
\email{zqhuang@hubu.edu.cn}
\affiliation{School of Physics, Hubei University, Wuhan 430062, China}

\date{\today}

\begin{abstract}
The one-point ETH envelope fixes the smooth part of a spectral density but not how the spectral weight of an eigenstate is distributed across channels; its higher-order correlations are joint fluctuations of those weights and are carried by connected products of channel resolvents.  In random free fermions every channel overlap is a squared Slater minor of a Haar-orthogonal one-body eigenvector matrix, and the multi-resolvent hierarchy has an exact microscopic realization there.  The $r$-resolvent moment admits a decomposition into finitely many overlap sectors whose values are fixed by the channel and eigenstate overlap patterns and whose weights are exact finite-$N$ orthogonal Weingarten functions: the sector theorem holds at arbitrary finite multiplicity, it is realized in full at $r=3$ ($k=2,3,4$) and through exact sector slices at $r=4,5$, and the sector moments lift to connected resolvent cumulants and to the irreducible Feshbach self-energy vertices, whose deep plane begins at the order $z^{-(3r-2)}$.  At $r=2$ this structure reproduces and completes an exact benchmark in which a flat envelope coexists with a non-Porter--Thomas intensity hierarchy and closes the two-resolvent covariance, and under interaction it separates into a deep plane fixed by Hamiltonian channel moments and a bulk that requires eigenstate-resolved information.
\end{abstract}

\maketitle

\section{Introduction}
\label{sec:intro}

Eigenstate thermalization fixes the smooth envelope of a few-body
observable: its diagonal matrix elements, smoothed over a narrow energy
window, equal the microcanonical average, and its off-diagonal elements
are controlled by the same envelope
\cite{Deutsch1991,Srednicki1994,Dalessio2016}, with deviations beyond the envelope and matrix-model constructions of ETH treated in a separate line \cite{Wang2022,Jafferis2023}.  The statement is
one-point in structure.  What it does not fix is how the spectral
weight of a given many-body eigenstate is distributed among the
channels of an operator, nor how those weights fluctuate jointly
across channels and across neighbouring eigenstates.  These
higher-order correlations are the fluctuation sector of the hierarchy
\cite{HOETH}, and they are genuinely independent of the envelope: in
the exact random free-fermion problem a perfectly flat envelope
coexists with weights that are strongly non-Porter--Thomas.

A formulation that can reach them has to keep several eigenstate
labels at once.  Connected products of channel resolvents do: the
$r$-resolvent moment with distinct spectral parameters probes the
$r$-th order fluctuation off shell, and the energy resolution carried
by the parameters resolves eigenstate structure that any
single-argument object averages out.  The resolvent therefore organizes
these statistics in a single analytic object rather than offering a
shorter route to the same numbers.  The channel resolvent $R_a(z)=\sum_n p_n^a/(z-\lambda_n)$
encodes the overlaps, the energies and their resolution in one
analytic function, and a direct moment-by-moment description is
recovered from it as a boundary projection (Sec.~\ref{sec:whyres}).

The gap this addresses is specific.  The one-point envelope and its
beyond-envelope matrix-model refinements are well fixed
\cite{Deutsch1991,Srednicki1994,Dalessio2016,Wang2022,Jafferis2023};
higher-order ETH correlations have been organized either by
free-cumulant identities for a single operator \cite{Pappalardi2022} or
by closure schemes that presuppose the eigenstate statistics
\cite{Mirlin}.  Neither route fixes the joint distribution of the
channel weights across channels and across neighbouring eigenstates,
which is what a multi-resolvent average measures, and neither delivers
it in closed form at finite $N$.  The hierarchy introduced in
Ref.~\cite{HOETH} was a framework; the question it left open is
whether its objects can be evaluated exactly in a solvable model.
Random free fermions make that question decidable, and this article
settles it there.

The central claim of this article is that in the random free-fermion
problem this hierarchy is not only formal.  Every channel overlap is a
squared Slater minor of a Haar-orthogonal one-body eigenvector matrix,
and the $r$-resolvent moment then decomposes exactly into finitely many
overlap sectors: classes of channel and eigenstate overlap patterns,
each carrying a combinatorial coefficient fixed by the pattern and an
$O(N)$ Weingarten weight that is exact at every finite $N$
(Sec.~\ref{sec:replica}).  The decomposition is a theorem at arbitrary
finite multiplicity.  It is realized in full at $r=3$ ($k=2,3,4$) and
through exact sector slices at $r=4,5$
(Table~\ref{tab:exactmap}).  The sector moments lift to connected
resolvent cumulants and to the irreducible Feshbach vertices, and the
deep plane of the connected $r$-resolvent object begins at the order
$z^{-(3r-2)}$.  The Weingarten contraction is a finite-dimensional
average, so the replica index counts resolvents rather than copies of
the system, and the ensemble average is evaluated exactly at every
finite $N$.

Four results carry the argument.  The sector theorem organizes the $r$-resolvent moment for all $r$
(Sec.~\ref{sec:replica}).  At $r=2$ the structure reproduces and completes the benchmark: a flat envelope coexisting with a non-Porter--Thomas intensity hierarchy, an eigenstate-to-eigenstate overlap kernel closing the two-resolvent covariance, and a projected architecture reconstructing the irreducible vertex (Sec.~\ref{sec:benchmark}).  At $r=3$ the general assembly is
evaluated in full $k=2,3,4$ sector arrays, and at $r=4,5$ in sector slices of the higher-order kernels (Sec.~\ref{sec:kernel}).  The sector moments are then mapped into connected resolvent cumulants and irreducible vertices
(Sec.~\ref{sec:vertices}), whose deep-plane coefficients are
Hamiltonian-moment covariances computable without eigenstates, while
the bulk sector requires eigenstate-resolved information
(Sec.~\ref{sec:physical}).  The last point is the structural boundary
of the construction: it separates what the exact sector algebra
determines from what interaction leaves to the eigenstate basis.

The article is organized around the evidence chain.  Section~\ref{sec:replica} develops the
generating functional, the resolvent multiplicity, the sector theorem,
and the orthogonal Weingarten calculus behind the exact weights.
Section~\ref{sec:benchmark} treats the exact two-resolvent benchmark.
Section~\ref{sec:kernel} carries the three-resolvent implementation and
the general-$r$ assembly.  Section~\ref{sec:vertices} lifts sector
moments to connected resolvent cumulants and to the irreducible
Feshbach vertices.  Section~\ref{sec:kernel4} presents the four-point
content of the same object.  Section~\ref{sec:physical} states the
interaction deformation, the moment rigidity and the bulk/deep-plane
boundary.  Section~\ref{sec:disc} concludes.  Exact algebra and proofs
are collected in Sec.~\ref{app:algebra}, numerical protocols in
Sec.~\ref{app:numerics}, derivations in
Sec.~\ref{app:derivations}, and the data and code entry points in
Sec.~\ref{app:models}.

\section{Multi-resolvent hierarchy and the exact sector theorem}
\label{sec:replica}

This section builds the object the rest of the article transforms: the multi-resolvent generating functional, the replica index carried by the resolvent multiplicity, the sector theorem that organizes the $r$-resolvent moment, and the orthogonal Weingarten calculus for the sector weights.  It closes with the object dictionary (Sec.~\ref{sec:dictionary}) and the resolvent organization of the higher-order fluctuations (Sec.~\ref{sec:whyres}).

\subsection{The multi-resolvent generating functional}

We work in the channel basis of the framework
\cite{HOETH} and of Sec.~\ref{sec:benchmark}.  $\{\ket{\varphi_a}\}$, $a=1,\dots,D$, an
orthonormal basis of the $k$-particle sector (for $N$ modes,
$D=\binom Nk$), with channel resolvents
\begin{equation}
\mathcal R_a(z)=\sum_n\frac{p_n^a}{z-\lambda_n}, \qquad p_n^a=|\braket{\varphi_a|\psi_n}|^2,
\label{eq:Rdef}
\end{equation}
$\lambda_n$ the many-body levels.  Physically, $\mathcal R_a(z)$ is the
Green function of the channel state $\ket{\varphi_a}$, the Stieltjes transform of its spectral measure \cite{ErdosYau,Tao2012,PottersBouchaud,Pastur}.  Its spectral
density $\rho_a(E)=\sum_np_n^a\delta(E-\lambda_n)=-\tfrac1\pi
\Im\mathcal R_a(E+\mathrm i0^+)$ is the channel-resolved local density
of states (strength function), and $p_n^a$ is the spectral weight with
which the many-body eigenstate $\ket{\psi_n}$ contributes to the
channel.  We write
$\hat{\mathcal R}_0(z)=D^{-1}\Tr(z-H)^{-1}$ for the mean-sector
(trace) resolvent and define the resolvent fluctuation about it,
\begin{equation}
\delta\mathcal R_a(z)\equiv\mathcal R_a(z)-\hat{\mathcal R}_0(z)=\sum_n\frac{\delta p_n^a}{z-\lambda_n}, \quad \delta p_n^a=p_n^a-\frac1D,
\label{eq:dRdef}
\end{equation}
the centering that the overlap kernel below makes natural: $\E_U[p_n^a]
=1/D$ for every eigenstate, so at fixed spectrum
$\E_U[\mathcal R_a\,|\,\Lambda]=\hat{\mathcal R}_0$ identically and
$\delta\mathcal R_a$ has zero conditional mean.  The centering is
conditional on the spectrum, not a deterministic shift: no term of the
M\"obius inversion that defines a cumulant of the $\delta\mathcal R$
fields carries a single factor, so $\E_\Lambda$ of a fixed-$\Lambda$
cumulant is a cumulant of the fluctuations under $\E_H$, and for
$r\le3$ it is the plain moment of Eq.~\eqref{eq:cumdef}.  The hierarchy
is therefore built from conditional objects, which is exactly what the
fixed-$\Lambda$ Haar sector theorem supplies.  The cumulant of the
uncentered resolvent is larger: at $r=2$ it exceeds $\mathcal C^{(2)}$
by the spectral-trace covariance
$\Cov_\Lambda[\hat{\mathcal R}_0(z),\hat{\mathcal R}_0(z')]$, the
second term of Eq.~\eqref{eq:bridge}.  The fluctuation sector of the
hierarchy is the sequence of connected multi-resolvent cumulants
\begin{equation}
\begin{aligned}
\mathcal C^{(2)}_{ab}&=\E_H[\delta\mathcal R_a\delta\mathcal R_b],\qquad \mathcal C^{(3)}_{abc}\\
&\quad =\E_H[\delta\mathcal R_a\delta\mathcal R_b \delta\mathcal R_c],\qquad\ldots,
\end{aligned}
\label{eq:cumdef}
\end{equation}
organized by the generating functional at fixed spectrum
\begin{equation}
\begin{aligned}
\mathcal Z[J\,|\,\Lambda]&=\E_U\Bigl[\exp\Bigl(\sum_a J_a\,\delta\mathcal R_a(z_a)\Bigr)\Big|\Lambda\Bigr],\\
W[J\,|\,\Lambda]&=\ln\mathcal Z[J\,|\,\Lambda],
\end{aligned}
\label{eq:ZJ}
\end{equation}
whose derivatives at $J=0$ are the $U$-conditional cumulants:
$W^{(1)}=0$ by Eq.~\eqref{eq:dRdef},
$W^{(2)}=\Cov_U(\mathcal R_a,\mathcal R_b\,|\,\Lambda)$,
$W^{(3)}=\kappa^{(U)}_3(\delta\mathcal R_a,\delta\mathcal R_b,\delta\mathcal R_c)$
at fixed $\Lambda$, with $\mathcal C^{(r)}=\E_\Lambda W^{(r)}$ as in
Eq.~\eqref{eq:cumdef} \cite{HOETH}.  Equivalently, the cumulants are the M\"obius coefficients of the partition-lattice
moment--cumulant inversion over the $r$ resolvent arguments.  The
hierarchy is therefore graded by the number of simultaneously
correlated resolvent channels, not by a perturbative order.  Three
levels have to be distinguished: the raw moments $\mathcal M^{(r)}$,
the connected content $\mathcal C^{(r)}$, and the
Feshbach-irreducible vertices $\Gamma^{(r)}$ of
Sec.~\ref{sec:gamma3}.  They are related by two different operations,
the M\"obius inversion above and the Schur dressing below, so that
$\mathcal M^{(r)}\neq\mathcal C^{(r)}\neq\Gamma^{(r)}$ in general; with
centered arguments $\mathcal M^{(r)}=\mathcal C^{(r)}$ for $r\le3$, and
the first reducible content appears at $r=4$.  The two-point level is
the covariance governed by the ladder equation, and $\Gamma^{(r)}$
denotes the cumulant of the self-energy fluctuation $\delta\cG$ in the
Feshbach representation: it is irreducible with respect to the geometric resolvent dressing, not one-particle-irreducible (Appendix~\ref{app:algebra} separates the three notions).  The partition-lattice
expansion, the non-derivative structure of the channel-resolved
objects, and the explicit $r=4$ reducible term are worked out in
Sec.~\ref{app:algebra}.

\subsection{Resolvent multiplicity as a replica index}

The spectral representation of Eq.~\eqref{eq:Rdef} makes the
correspondence explicit.  Expanding the $r$-th cumulant,
\begin{equation}
\mathcal C^{(r)}_{a_1\cdots a_r}(z_1,\dots,z_r)=\sum_{n_1,\dots,n_r} \frac{\E[\delta p_{n_1}^{a_1}\cdots\delta p_{n_r}^{a_r}]} {\prod_i(z_i-\lambda_{n_i})},\\
\label{eq:specr}
\end{equation}
each resolvent factor carries one eigenstate index and one channel
index.  In random free fermions the one-body eigenvector matrix
$U\in O(N)$ is Haar distributed and independent of the spectrum
\cite{Mehta,Magan2016}, and the overlaps are squared Slater minors
\begin{equation}
\begin{aligned}
p_n^a&=|\det U[A_n,C_a]|^2, \qquad |A_n|=|C_a|=k,
\end{aligned}
\label{eq:minor}
\end{equation}
with $A_n$ the set of one-body levels occupied in eigenstate $n$ and
$C_a$ the set of modes occupied in channel $a$, $U[A,C]$ being the
submatrix of rows $A$ and columns $C$ (the transposed reading of
Eq.~\eqref{eq:pminor} below, whose rows are modes, has the same Haar
averages because $U^{\mathsf T}$ is Haar whenever $U$ is, but the two
conventions are not interchangeable factor by factor).  Equivalently,
$p_n^a=|\langle A_n|\wedge^kU|C_a\rangle|^2$ is the modulus-squared
matrix coefficient of the exterior-power representation $\wedge^k$ of
$O(N)$ between the Slater basis vectors built on $A_n$ and $C_a$.  The
sector labels
therefore carry direct microscopic meaning.  $A_n$ is the one-body
configuration of the many-body eigenstate, $C_a$ the mode
configuration of the channel, and the overlap pattern measures how
much one-body structure the states and the probes share, the
microscopic geometry behind the spectral-weight correlations.  The
$r$-resolvent
moment is therefore a Haar integral of $2r$ determinants,
$\int_{O(N)}\dd U\prod_{i=1}^r\det U[A_i,C_i]^2$.  For real $U$ each
determinant is real, so $|\det U[A,C]|^2=\det U[A,C]^2$ and each factor
is a second copy of the same minor, not the transposed minor
$\det U[C_i,A_i]$ (the two factors are not interchangeable:
$\E\det U[A,C]\det U[C,A]\neq\E\det U[A,C]^2$ for $A\neq C$).  Each
squared minor is a product over two
permutation expansions,
$|\det U[A,C]|^2=\sum_{\sigma,\tau\in S_k}\mathrm{sgn}(\sigma\tau)
\prod_{j=1}^k U_{A_{\sigma(j)},C_j}U_{A_j,C_{\tau(j)}}$.  The
orthogonal Weingarten formula \cite{Weingarten,CM2009} evaluates
this Haar integral as a double sum over the pair partitions
$\pi,\pi'$ of the $2kr$ slots,
\begin{equation}
\begin{aligned}
\E_U\prod_{i=1}^{2kr} U_{x_i,y_i}&=\sum_{\pi,\pi'\in M_{2kr}} \Delta_\pi(x)\Delta_{\pi'}(y)\Wg^{O(N)}(\pi^{-1}\pi'),
\end{aligned}
\label{eq:weingarten}
\end{equation}
with $\Delta_\pi(x)=\prod_{\{i,j\}\in\pi}\delta_{x_i,x_j}$.  The
$r$ resolvents of Eq.~\eqref{eq:specr} are thus $r$ replicas.
Each carries its own $2k$ slots, the disorder (Haar) average generates
the coupling between replicas through the Weingarten weight
$\Wg^{O(N)}(\pi^{-1}\pi')$, and the pairing sectors $\pi,\pi'$ are the
contraction patterns of the replica fields.  Here ``replica'' denotes repeated copies of the resolvent observable; the replica index counts resolvents.    The Weingarten sectors are thus the
exact bookkeeping of shared microscopic randomness.  All many-body
overlaps are built from the same one-body eigenvector matrix, and the
pairing sectors record which overlaps share which piece of that
randomness, the microscopic origin of the replica organization.

\subsection{The sector theorem}

The weight $\Wg^{O(N)}(\pi^{-1}\pi')$ depends on the pair
$(\pi,\pi')$ only through the coset type $\mu\vdash kr$ of
$\pi^{-1}\pi'$ in the hyperoctahedral double cosets
\cite{CM2009}, the partition of $kr$ whose parts are the half-lengths
of the connected components of the union graph $\pi\cup\pi'$.  The
remaining sums are the $\delta$-constraints of
Eq.~\eqref{eq:weingarten}.  $\Delta_\pi$ forces the row labels of each
$\pi$-pair to coincide, and the row labels are the level labels
$A_i(\cdot)$.  Hence the row sum depends on the sets $A_1,\dots,A_r$
only through their overlap pattern, and the column sum on
$C_1,\dots,C_r$ only through theirs.  This is the finite-$N$
replica-sector invariance of the replica organization:
\begin{quote}
\noindent\emph{Proposition (finite-$N$ replica-sector invariance).}
For fixed $N$, $r$ and $k$, the $r$-replica Haar sector depends on the
$r$ level sets and the $r$ mode sets only through their overlap
patterns.  The sector value is a function of the two Venn occupancies
alone, and the permutation (coset) classes
of $M_{2kr}$ are the replica contraction sectors, the $r$-point kernel
closing as $\mathcal K_r=\sum_c T^{(r)}_c(\mathcal P_A,\mathcal P_C)
\,w^{(r)}_c(N)$ [Eq.~\eqref{eq:sectgen} below].
\end{quote}
\noindent\emph{Proof.}  A relabeling of the modes within one Venn cell
permutes the valid pairing set of Eq.~\eqref{eq:weingarten} while
preserving both the pairing-compatibility constraints of
$\Delta_\pi,\Delta_{\pi'}$ (labels that must coincide are relabeled
coherently) and the double-coset class of $\pi^{-1}\pi'$ (the union
graph $\pi\cup\pi'$ is unchanged up to relabeling).  The Weingarten
double sum is therefore unchanged, and its value depends on the
patterns alone.
For $r=2$ the two-copy sectors are the classes $(m,\ell)$ with
$m=|A_1\cap A_2|$, $\ell=|C_1\cap C_2|$, the overlap geometry of the
two-point benchmark, whose kernel $L(k;m,\ell)$ is the two-copy
restriction of the sector sum (Sec.~\ref{sec:benchmark}).  For $r=3$ the
sectors are the sixteen classes
$(m_{12},m_{13},m_{23},t)$ and $(\ell_{12},\ell_{13},\ell_{23},t_\ell)$
of Sec.~\ref{sec:kernel}, where $t=|A_1\cap A_2\cap A_3|$ is the
three-copy intersection.  The connected kernel of level $r$ is obtained from the raw $r$-replica sector sum by moment--cumulant subtraction.  Its fully coincident sector (all sets equal) is the
single-channel $r$-th moment, the projection that yields $q_r$
(Sec.~\ref{sec:q3}).  The notation is collected in the object dictionary of Table~\ref{tab:dict}.

\subsection{Orthogonal Weingarten sector calculus}
\label{sec:weingarten}

The pairings $M_{2d}$ of $2d$ slots ($|M_{2d}|=(2d-1)!!$) carry the
Gram matrix $G(\pi,\pi')=N^{\kappa(\pi^{-1}\pi')}$, where $\kappa$ is
the number of connected components of the union graph, and the
Weingarten function is its inverse,
$\Wg^{O(N)}(\pi^{-1}\pi')=[G^{-1}]_{\pi,\pi'}$
\cite{Weingarten,CM2009}.  Three statements have to be kept apart.
The Haar average of Eq.~\eqref{eq:weingarten}, and with it the sector
invariance proved above, is defined for every finite $N$ and $r$.  The
pairing Gram matrix is nonsingular off a finite set of exceptional $N$
(its determinant is a polynomial in $N$ with unit leading term), and
the reduced class system [Eq.~\eqref{eq:reduced}] or the zonal form
[Eq.~\eqref{eq:zonal}] below evaluates the weights as rational functions
of $N$ in lowest terms on that complement; at an exceptional $N$ the
weights remain defined by the Haar integral, but the inverse formula is
unavailable there.  Every $N$ used in this article is nonexceptional,
and each weight quoted as exact at finite $N$ is the value of the
rational function at that $N$.
Since $(S_{2d},H_d)$ is a Gelfand pair, the
double cosets of the hyperoctahedral group $H_d$ form a commutative
association scheme on $M_{2d}$.  The inverse lies in the double-coset
algebra, so $\Wg^{O(N)}$ is constant on coset classes and the
$10395\times10395$ inversion for $d=6$ ($2kr=12$ slots at $k=2$,
$r=3$) collapses to an $11\times11$ system over the classes $\mu\vdash
6$,
\begin{equation}
\sum_{\mu,\mu'} N^{\ell(\mu)}\,c^{\nu}_{\mu,\mu'}\,w_{\mu'}(N)
=\delta_{\nu,1^6},
\qquad
\Wg^{O(N)}(\mu)=w_\mu(N),
\label{eq:reduced}
\end{equation}
with $c^{\nu}_{\mu,\mu'}=\#\{\rho\in M_{12}:\mathrm{type}(\pi_0,\rho)
=\mu,\ \mathrm{type}(\rho,\tau)=\mu'\}$ the intersection numbers for
any $\pi_0,\tau$ with $\mathrm{type}(\pi_0,\tau)=\nu$, $\ell(\mu)$ the
number of parts of $\mu$, and $1^6$ the identity class
(Sec.~\ref{app:wgtab}).  The reduced system was verified against
the full Gram inversion at $d=3,4$ (numerically to machine precision) and against the
classical values at $d=1,2$.  The eleven functions $w_\mu(N)$ are rational in $N$ and
collected in Sec.~\ref{app:wgtab} in lowest terms with factored
denominators.  Their common degree-14 denominator is the exact
finite-$N$ content of the three-copy average, the promotion of the
two-copy denominators $N(N-1)(N+2)$ of the benchmark's kernel.

On a double-coset sector the orthogonal Weingarten function is a
rational function of $N$: the class reduction turns the pairing-space
average into a finite linear system [Eq.~\eqref{eq:reduced}], and the
zonal expansion gives the closed form [Eq.~\eqref{eq:zonal}].  At depth
six the reduced system has eleven classes; at depth nine it has thirty.
The three-copy sector sum then assembles the $r=3$ weights from the
two-copy ones [Eq.~\eqref{eq:sectorsum}].  The derivation of the class
system, the zonal evaluation, and the worked depth-six table are
collected in Sec.~\ref{app:algebra}.

\subsection{The object dictionary}
\label{sec:dictionary}

The objects and operations used throughout the article are collected in Table~\ref{tab:dict}. Its three blocks give, in order, the states and the correlation functions of the hierarchy, the sector symbols with their multiplicities and covariances, and the operations that connect the levels; the right-hand column points to the section in which each object is defined or used. The scope of exactness of each result is mapped separately in Table~\ref{tab:exactmap} of Sec.~\ref{app:algebra}, so that the two tables together let the main text be read as a single argument.

\begin{table*}[t]
\caption{Dictionary of the objects and operations of the multi-resolvent hierarchy. The scope of exactness of each object is mapped separately in Table~\ref{tab:exactmap}.}
\label{tab:dict}
\small
\begingroup
\setlength{\tabcolsep}{3pt}
\begin{ruledtabular}
\begin{tabular}{lp{3.55in}}
object / operation & role\\
\hline
$p_n^a=|\braket{\varphi_a|\psi_n}|^2$ & spectral weight of eigenstate $n$ in channel $a$\\
$\mathcal R_a(z)=\sum_n p_n^a/(z-\lambda_n)$ & channel resolvent; the Level-I mean carries the envelope\\
$\mathcal K_r$, $\Cov_r$ & raw $r$-overlap moment; connected overlap cumulant (Sec.~\ref{sec:replica})\\
$\mathcal M^{(r)}$ & raw $r$-resolvent moment (Sec.~\ref{sec:replica})\\
$\mathcal C^{(r)}$ & connected $r$-resolvent cumulant (Sec.~\ref{sec:vertices})\\
$r$, $k$ & resolvent multiplicity; mode-set size probed by each resolvent\\
$\mathcal P_A,\mathcal P_C$ & overlap patterns of the level sets and mode sets\\
$\rho_s^{(r)}$, $N_s$, $\Cov_r(s)$ & sector spectral density, multiplicity, and sector covariance (Sec.~\ref{sec:bulk})\\
$\Gamma^{(r)}$ & Feshbach irreducible self-energy vertex (Sec.~\ref{sec:gamma3})\\
$K_{rs}$ & deep-plane moment coefficients of $\mathcal C^{(2)}$ (Sec.~\ref{sec:rigidity})\\
sector $c$ & overlap pattern of the $r$ eigenstate labels and channel modes, with multiplicity $N_c$, combinatorial coefficient $T_c$, weight $w_c$\\
$L(k;m,\ell)$ & exact two-point eigenstate-to-eigenstate overlap kernel (Sec.~\ref{sec:benchmark})\\
$S^{(r)}$ & energy-resolved spectral cumulants behind $\mathcal C^{(r)}$ (Sec.~\ref{sec:spectral})\\
\hline
Haar/Slater structure & $p_n^a$ = squared minor of one Haar-orthogonal matrix\\
Weingarten/zonal expansion & exact finite-$N$ sector weights $w_c$ (Sec.~\ref{sec:weingarten})\\
occupancy/transfer assembly & combinatorial sector coefficients $T_c$ (Sec.~\ref{sec:assembly})\\
M\"obius moment--cumulant inversion & $\mathcal K_r\to\Cov_r$ and $\mathcal M^{(r)}\to\mathcal C^{(r)}$\\
Schur/ladder dressing & $\mathcal C^{(r)}\to\Gamma^{(r)}$ (Sec.~\ref{sec:gamma3})\\
spectral dictionary & $\Cov_r\to S^{(r)}$ (spectral transform), $S^{(r)}\to\mathcal C^{(r)}$ (Stieltjes) (Sec.~\ref{sec:spectral})\\
\end{tabular}
\end{ruledtabular}
\endgroup
\end{table*}

\subsection{The resolvent organization of ETH fluctuations}
\label{sec:whyres}

The sector moments are the microscopic data; the resolvent is the
analytic object in which those data are the coefficients.

In the solvable benchmark the higher-order overlap statistics follow
directly from the Haar/Slater structure, which raises a fair question:
what does a resolvent hierarchy add over a direct computation of
higher-order ETH correlations?  
Writing the expectation as $\E=\E_\Lambda\E_U$ over eigenvalues and
eigenvectors, the law of total covariance gives the connected
two-point function of channel resolvents exactly,
\begin{equation}
\begin{aligned}
\Cov\bigl[R_a(z),R_b(z')\bigr]&=\sum_{A,B}\frac{\E_\Lambda[\Cov_U(p_A^a,p_B^b)]}{(z-\lambda_A)(z'-\lambda_B)}\\
&\quad+\Cov_\Lambda\Bigl[\sum_A\frac{\E_U p_A^a}{z-\lambda_A},\sum_B\frac{\E_U p_B^b}{z'-\lambda_B}\Bigr],
\end{aligned}
\label{eq:bridge}
\end{equation}
the first term the overlap-kernel contribution and the second the
spectral-trace one.  The first term is exactly $\mathcal C^{(2)}_{ab}$
of Eq.~\eqref{eq:cumdef}, the $\E_\Lambda$ average of the
fixed-spectrum covariance; the second is the trace-resolvent
fluctuation that the conditional centering of
Eq.~\eqref{eq:dRdef} removes, so the full covariance of the uncentered
resolvents exceeds the cumulant of the hierarchy by that term.  In the
benchmark the eigenvector matrix is
independent of the spectrum, so the right-hand overlap covariance is
the exact two-point kernel while the flat weight is $1/D$; the
$\eta\to0^+$ boundary value of Eq.~\eqref{eq:bridge} is then the
energy-resolved fluctuation kernel $K_{\alpha\beta}(E,E')$ and its
difference-resolved form $F_{\alpha\beta}(\omega)$
(Secs.~\ref{app:bench} and~\ref{app:derivations}).  The relation
is one of containment: the direct ETH statistics are boundary
observables of the resolvent object, which retains the full
$(z,z',\eta)$ analytic content.

Three structural properties follow that a direct moment-by-moment
description does not naturally organize: energy resolution built into
the spectral parameters, a hierarchy indexed by resolvent multiplicity
rather than moment order, and connected content organized from the
start, with the failure of an ansatz defined as an irreducible residual (Sec.~\ref{sec:closure:arch}).  The same organization pays off
under interaction: the resolvent formulation separates the fluctuation
content into a Hamiltonian-moment sector, whose deformation is
determined at the moment level, and an eigenstate-resolved sector
whose perturbative deformation is singular
(Secs.~\ref{sec:rigidity} and~\ref{sec:bulkb}).  The comparison is
collected in Table~\ref{tab:directvsres}.

\begin{table}[t]
\caption{Direct higher-order ETH statistics versus the resolvent hierarchy. }
\label{tab:directvsres}
\centering
\footnotesize
\begingroup
\setlength{\tabcolsep}{3pt}
\begin{ruledtabular}
\begin{tabular}{lll}
\hline\hline
 & \parbox{2.15cm}{\raggedright direct higher-order ETH} & \parbox{2.45cm}{\raggedright resolvent hierarchy}\\
\hline
basic object & \parbox{2.15cm}{\raggedright overlap moments $|O_{AB}|^2$, $q_r$, $C_\ell$, $L$} & \parbox{2.45cm}{\raggedright channel resolvent $R_a(z)$}\\
hierarchy & \parbox{2.15cm}{\raggedright moment / cumulant order} & \parbox{2.45cm}{\raggedright resolvent multiplicity}\\
energy information & \parbox{2.15cm}{\raggedright external binning} & \parbox{2.45cm}{\raggedright built into $z,z'$}\\
spectral resolution & \parbox{2.15cm}{\raggedright post-processing} & \parbox{2.45cm}{\raggedright $\Im z=\eta$}\\
disconnected terms & \parbox{2.15cm}{\raggedright separately organized through cumulants} & \parbox{2.45cm}{\raggedright naturally organized in connected sectors}\\
closure & \parbox{2.15cm}{\raggedright correlator by correlator} & \parbox{2.45cm}{\raggedright projected hierarchy, exact equations}\\
residual & \parbox{2.15cm}{\raggedright not natural} & \parbox{2.45cm}{\raggedright irreducible residual [Eq.~\eqref{eq:residual}]}\\
interaction response & \parbox{2.15cm}{\raggedright eigenstate statistics recomputed} & \parbox{2.45cm}{\raggedright moment vs kernel sectors separated}\\
dynamics & \parbox{2.15cm}{\raggedright requires new equations} & \parbox{2.45cm}{\raggedright operator identity provides the interface}\\
\hline
\end{tabular}
\end{ruledtabular}
\endgroup
\end{table}

\emph{Formal statements.}  The containment, the coefficient identification, the sector-resolved diagnostic and the deep-plane/bulk split are stated as three propositions and a corollary in Sec.~\ref{app:containment}.

\section{The exact two-resolvent benchmark}
\label{sec:benchmark}

The sector theorem is general; the first place its consequences can be read off exactly is the two-resolvent level, where every object is known in closed form.  This section states the $r=2$ benchmark, the ground-truth input for the reconstruction and deformation results of Sec.~\ref{sec:physical}.  It supplies the microscopic coordinates used everywhere below, and all entries below are the two-copy special cases of the sector algebra of Secs.~\ref{sec:weingarten} and \ref{sec:kernel}, computed here along the direct route.  Four results follow below.

\subsection{Ensemble and channel dictionary}
\label{sec:bench:dict}

We consider $N$ fermionic modes with a real symmetric one-body Hamiltonian $h=(h_{ab})$ drawn from the GOE in the Mehta normalization ($\E h_{aa}^2=2$, $\E h_{ab}^2=1$, spectral radius $2\sqrt N$) \cite{Mehta}. Diagonalizing $h=U\Lambda U^\dagger$ gives one-body energies $\varepsilon_0,\dots,\varepsilon_{N-1}$ and a Haar-distributed orthogonal eigenvector matrix $U$.  The many-body eigenstates are the Slater determinants
\begin{equation}
\ket{\psi_A}=\prod_{a\in A}d_a^\dagger\ket0,\qquad
d_a^\dagger=\sum_{b=0}^{N-1}U_{ba}c_b^\dagger,
\label{eq:slater}
\end{equation}
labeled by occupation sets $A\subset\{0,\dots,N-1\}$ with $|A|=k$ and energies $\lambda_A=\sum_{a\in A}\varepsilon_a$.  The Hilbert-space dimension is $D=\binom Nk$ and the filling $\nu=k/N$. The system is mode $0$ and the bath the modes $1,\dots,N-1$. A channel $a=(\mu,i)$ is the Fock state $\ket{\varphi_{\mu i}}=(c_0^\dagger)^i\prod_{b\in C_\mu}c_b^\dagger\ket0$ with $i\in\{0,1\}$ and $C_\mu\subset\{1,\dots,N-1\}$ of size $|C_\mu|=k-i$, so that the channel overlap is a minor of $U$:
\begin{equation}
p_A^a=|\braket{\varphi_a|\psi_A}|^2=|\det U[C_a,A]|^2,
\label{eq:pminor}
\end{equation}
with $C_a=C_\mu$ for $i=0$ and $C_a=\{0\}\cup C_\mu$ for $i=1$.  Each sector contains $d_B=\binom{N-1}{k-i}$ channels.

In the dictionary of Ref.~\cite{HOETH} the sector populations $D_A^i$, the off-diagonal matrix elements $\sigma_{AB}^{ji}$, the diagonal baseline $D_{ji}$, the smooth function $f_{ji}^2$ with its correlation correction $g_{ji}$, and the fluctuation parameters ($q$, $V_{\rm dg}$, $V_{\rm off}$, $\bar C_i$) are all given in closed form, together with the two exact structural identities, the partition $D_A^0+D_A^1=1$ and the complementarity $\sum_{B\ne A}|\sigma_{AB}^{ii}|^2=D_A^i(1-D_A^i)$, in Sec.~\ref{app:bench}. Because the GOE eigenvectors are independent of the eigenvalues ($U\perp\Lambda$), every dictionary entry is energy independent.  Random free fermions are the energy-independent fixed point of the framework, whose diagonal and off-diagonal ETH structure, including its energy dependence and deformations, is analyzed in the companion random-matrix study \cite{THF}. The two objects built in this section (the one-point hierarchy and the two-point kernel) are the ground-truth inputs supplied to the projected closure of Sec.~\ref{sec:closure}, and their moment-weighted part will be recognized in Sec.~\ref{sec:rigidity} as exactly moment-determined.

\subsection{One-point law: flat envelope and the intensity hierarchy}
\label{sec:bench:onepoint}

The channel envelope is exactly flat, $f^{\mu i}=1$ for every channel and sector.  By Muirhead's theorem the Gram matrix behind $p_A^a$ has the matrix-variate beta law, whose Jacobi eigenvalue ensemble and Selberg moments telescope at $t=1$ to $M_k(1)=1/D$ (Sec.~\ref{app:bench}, which also records the particle--hole complementation $k\mapsto N-k$). The normalized second-moment ratio, to be read against the Porter--Thomas value $q=3$ of a Gaussian intensity \cite{Alt1995}, is
\begin{equation}
\boxed{\;q_{\rm FF}
=\frac{(k+1)(k+2)(N-k+1)(N-k+2)}{2(N+1)(N+2)}\;}
\label{eq:qff}
\end{equation}
for all $0\le k\le N$.  At $k=1$ it reduces to $3N/(N+2)$, the Porter--Thomas value up to the finite-$N$ normalization correction, while for fixed $k$ and $N\to\infty$ it tends to $(k+1)(k+2)/2$, the value for the squared determinant of a $k\times k$ matrix of independent standard Gaussians.  The departure from Porter--Thomas statistics is the intrinsic product structure of a determinant, not a finite-size correction. The full intensity hierarchy is
\begin{equation}
q_r:=\frac{\E[(p_A^a)^r]}{\E[p_A^a]^r}
=\prod_{j=1}^{k}\prod_{i=1}^{r-1}
\frac{(j+2i)(N-k+j)}{(N-k+j+2i)\,j},
\label{eq:qr}
\end{equation}
with normalized cumulants departing from the Porter--Thomas values by factors growing polynomially in $k$ [Eq.~\eqref{eq:Kr}]. The single-channel density carries the algebraic small-intensity tail and the power-law cutoff recorded in Sec.~\ref{app:bench}.  ETH smoothness coexists with non-Porter--Thomas channel intensities. The fluctuation strength is set by the determinant depth $k$ rather than by the exponential channel count $D$.  For fixed $k$ the ratio $q_{\rm FF}$ is a function of $k$ alone as $N\to\infty$, and in the extensive-filling limit
\begin{equation}
q_{\rm FF}=\frac12\Bigl(\frac zN\Bigr)^2\Bigl[1+O\Bigl(\frac1k\Bigr)\Bigr],
\label{eq:qz}
\end{equation}
where $z=k(N-k)$ is the particle--hole connectivity of the one-body hopping [Eq.~\eqref{eq:complementarity} of Sec.~\ref{app:bench}], a structural correspondence, not a control relation. This fixes the one-point targets ($q_{\rm FF}$ and the hierarchy $q_r$) that the covariance of Sec.~\ref{sec:closure} must reproduce in its diagonal bin.

\subsection{Channel geometry and the central kernel}
\label{sec:bench:twopoint}

Classifying ordered channel pairs by the shared-mode count $\ell=|C_a\cap C_b|$ and eigenstate pairs by the shared-level count $m=|A\cap B|$, the channel distance acting as a principal-overlap measure between mode sets \cite{AbsilEdelmanKoev}, the two-point covariance becomes a function of the channel distance,
\begin{equation}
C_\ell:=\E\bigl[p_A^ap_A^b\,\big|\,\ell\bigr]-\frac1{D^2},
\label{eq:Cell}
\end{equation}
which is not flat.  It increases monotonically with $\ell$ and changes sign near $\ell\simeq k/2$, so channels sharing more bath modes are positively correlated while distant channels are anticorrelated. For $k\le5$ the profile is exact in closed form, $D^2C_\ell=P_\ell(N)/[(k!/2)\binom{N+2}{2}]$ with the polynomials $P_\ell$ of Eq.~\eqref{eq:Cellkk}.  For $k=2$,
\begin{equation}
D^2C_0=-\frac{2(2N+1)}{(N+1)(N+2)},\quad
D^2C_1=\frac{N^2-5N-2}{(N+1)(N+2)}.
\label{eq:Cellk2}
\end{equation}

\emph{Channel-pair selection rule.} The channel-pair dependence reduces to the shared-mode count $\ell$. For every fixed eigenstate pair $(A,B)$, the two-point kernel $K^{AB}_{ab}:=\E[p_A^a p_B^b]-1/D^2$ is a function of the ordered pair of row sets $(C_a,C_b)$ alone [Eq.~\eqref{eq:pminor}], and the Haar measure of $U$ is invariant under left multiplication by permutation matrices, so $K^{AB}_{\pi a,\pi b}=K^{AB}_{ab}$ for every mode permutation $\pi\in S_N$. Ordered pairs of $k$-subsets of the $N$ modes are classified up to $S_N$ by $(\delta_{ab},\ell_{ab})$, so every $m$-bin, the energy-resolved kernels [Eqs.~\eqref{eq:Fomega}--\eqref{eq:rescov}], and the full resolvent covariance depend on the channel pair only through $(\delta_{ab},\ell_{ab})$.  Symmetry fixes the labeling, and the sign-structured $C_\ell$ profile [Eq.~\eqref{eq:Cellk2}] is the dynamical content within it.

The rule transfers to the interacting moment sectors of
Sec.~\ref{sec:rigidity} whenever the interaction law is mode-permutation
invariant (Sec.~\ref{app:bench}).

The mean sector is likewise closed in explicit form [Sec.~\ref{app:bench}].  The correlation correction $g_{ii}<0$ is a negative, order-one term of purely normalization (idempotency) origin (a quadratic Hamiltonian has no many-body interaction vertices that could generate it) and a flat ETH envelope does not imply the absence of multi-channel correlation [Eq.~\eqref{eq:counterexample} of Sec.~\ref{app:bench}].

The central object is the eigenstate-to-eigenstate overlap kernel in the fixed-depth (Gaussian-minor) limit.  Its finite-$N$ corrections are computed separately [Eqs.~\eqref{eq:closedm}]. For two $k$-row sets $A,B$ sharing $m$ rows and two $k$-column sets $C_1,C_2$ sharing $\ell$ columns of a matrix of independent standard Gaussians, write
$L(k;m,\ell):=\E[|\det G[A,C_1]|^2|\det G[B,C_2]|^2]/(k!)^2-1$. Expanding both determinants along the shared rows and closing the shared-row block moments by the Bartlett decomposition (derivation in Sec.~\ref{app:bench}) gives
\begin{equation}
\boxed{\begin{aligned}
L(k;m,\ell)&=\binom{k}{m}^{\!-2}\sum_{j=0}^{m}
N(k,m,\ell,j)\,L(m;m,j),\\
L(m;m,j)&=\frac{j(2m-j+3)}{(m-j+1)(m-j+2)},
\end{aligned}}
\label{eq:Lgeneral}
\end{equation}
with the combinatorial count $N(k,m,\ell,j)$ of Eq.~\eqref{eq:Ncount}. The kernel is symmetric in $m$ and $\ell$, vanishes when either overlap is empty, $L(k;0,\ell)=0$, the independence of the disjoint rows of the eigenvector matrix, which is what makes $m$ a genuine microscopic eigenstate-connectivity variable, and reduces at $m=k$ to the same-eigenstate channel profile, $L(k;k,\ell)=\ell(2k-\ell+3)/[(k-\ell+1)(k-\ell+2)]$. At the Gaussian fixed point it closes the two-state overlap covariance, $\E[p_A^ap_B^b]=(1/D^2)[1+L(k;m,\ell)]$ in the fixed-depth limit.

The energy resolution of the kernel factorizes into this geometry and the classical spectral-difference convolution of the one-body semicircle.  With $\mathcal N_m=\binom{N}{m}\binom{N-m}{k-m}\binom{N-k}{k-m}$ the number of eigenstate pairs sharing $m$ levels,
\begin{equation}
\begin{aligned}
F_{\alpha\beta}(\omega):=\sum_{A,B}D^2\Bigl(\E[p_A^\alpha p_B^\beta]-\frac1{D^2}\Bigr)\times\delta(\lambda_B-\lambda_A-\omega)\\
=D\,L(k;k,\ell)\,\delta(\omega)+\sum_{m=1}^{k-1}\mathcal N_m\,L(k;m,\ell)\,\rho_{\rm sc}^{*2(k-m)}(\omega),
\end{aligned}
\label{eq:Fomega}
\end{equation}
with $\rho_{\rm sc}^{*n}$ the $n$-fold convolution of the unit semicircle [Eq.~\eqref{eq:rhoconv}], supported on $|\omega|\le 2n$.  The zero-energy $\delta$ carries the same-eigenstate channel-distance correlations, and each off-diagonal class contributes a ridge of total width $4(k-m)$ whose weight is fixed by Eq.~\eqref{eq:Lgeneral}. The prediction is confirmed by direct sampling [Sec.~\ref{app:bench}]. The bridge to the hierarchy is the two-variable kernel $K_{\alpha\beta}(E,E'):=\sum_{A,B}D^2\E[\delta p_A^\alpha\,\delta p_B^\beta\,\delta(E-\lambda_A)\,\delta(E'-\lambda_B)]$, whose fixed-depth form
\begin{equation}
\begin{aligned}
K_{\alpha\beta}(E,E')&=D\,L(k;k,\ell)\,\rho_{\rm mb}(E)\,\delta(E-E')\\
&\quad+\sum_{m=1}^{k-1}\mathcal N_m\,L(k;m,\ell)\,\rho_{AB}^{(m)}(E,E'),
\end{aligned}
\label{eq:Kab}
\end{equation}
with $\rho_{\rm mb}=\rho_{\rm sc}^{*k}$ the many-body density and $\rho_{AB}^{(m)}$ the joint density of the two many-body energies for pairs sharing $m$ levels [Eq.~\eqref{eq:rhoAB}], makes the two-resolvent covariance its Cauchy transform.  With $R_\alpha(z)=\sum_Ap_A^\alpha/(z-\lambda_A)$,
\begin{equation}
\begin{aligned}
D^2\Cov\bigl(R_\alpha(z),R_\beta(z')\bigr)
&=\iint\frac{K_{\alpha\beta}(E,E')}{(z-E)(z'-E')}\,\dd E\,\dd E'\\
&\quad+C_{\alpha\beta}^{\rm flat}(z,z'),
\end{aligned}
\label{eq:rescov}
\end{equation}
where $C_{\alpha\beta}^{\rm flat}$ is the flat-weight many-body resolvent covariance carried by the spectral correlations of the one-body GOE spectrum [Eq.~\eqref{eq:Cflat}]. The chain
\begin{equation*}
\begin{aligned}
&\text{Haar/Slater minors}\;\longrightarrow\;
M_k(t),\,q_r,\,C_\ell\;\longrightarrow\; L(k;m,\ell)\\
&\quad\longrightarrow\; F_{\alpha\beta}(\omega)\;\longrightarrow\;
C_{\alpha\beta}(z,z')
\end{aligned}
\end{equation*}
therefore closes the loop of the benchmark without any intermediate modeling. The kernel $L(k;m,\ell)$ is thus the microscopic object whose reconstruction is the task of Sec.~\ref{sec:closure} and whose deformation under interaction is the subject of Secs.~\ref{sec:rigidity} and \ref{sec:bulkb}.

\subsection{What the benchmark fixes}
\label{sec:bench:status}

The benchmark separates exact statements from dictionary-level correspondences [Table~\ref{tab:status}]. What is exact: the flat envelope, the hierarchy $q_r$, the channel-distance covariance $C_\ell$, the kernel $L(k;m,\ell)$, the energy-resolved and resolvent-level embeddings [Eqs.~\eqref{eq:Fomega}--\eqref{eq:rescov}], and the structural identities. The remaining entries record a correspondence (Table~\ref{tab:status}): the identification of the remaining projections with the vertex-resolved level $g^{(2)}$, which for a quadratic Hamiltonian is not organized by interaction vertices and is left open (the Mag\'an relation, Sec.~\ref{app:bench}). The benchmark supplies the two ingredients a projected vertex closure needs, the $(m,\ell)$ sector geometry and the ground-truth amplitudes against which every step of that closure is tested (Sec.~\ref{sec:closure}). Anticipating Secs.~\ref{sec:rigidity} and \ref{sec:bulkb}, the chain already prefigures the central split of the paper.  Its leading moment coefficients are determined exactly by Hamiltonian moments and rigid under interaction, while the full kernel is eigenstate-resolved and beyond ordinary perturbation theory.

\section{Exact three-resolvent correlations and sector assembly}
\label{sec:kernel}

The first multiplicity beyond the benchmark is $r=3$, and it is where
the general assembly rule becomes visible: the three-point overlap
kernel is derived here and checked against the exact two-resolvent
limit, and the assembly rule is then applied at general $r$.  The full
sector arrays and the item-by-item checks are collected in
Sec.~\ref{app:algebra} and in the machine-readable data of
Sec.~\ref{app:models}.

\subsection{Triple geometry and the kernel}

Label eigenstate triples by the overlap pattern
$(m_{12},m_{13},m_{23},t)$ with $m_{ij}=|A_i\cap A_j|$ and
$t=|A_1\cap A_2\cap A_3|$, and channel triples by
$(\ell_{12},\ell_{13},\ell_{23},t_\ell)$ analogously.  For $k=2$ the
sixteen patterns of each kind are listed in Sec.~\ref{app:kernel}.
The three-point overlap kernel is the connected three-copy moment,
with the shorthands $\mathbf m=(m_{12},m_{13},m_{23})$ and
$\bm\ell=(\ell_{12},\ell_{13},\ell_{23})$,
\begin{equation}
\begin{aligned}
&\Cov_3(\mathbf m,t;\bm\ell,t_\ell)
=\E[\delta p_n^a\delta p_m^b\delta p_l^c\,|\,\text{sector}]\\
&=\E[p_n^ap_m^bp_l^c]-\frac1D\Bigl(\E[p_n^ap_m^b]+\E[p_n^ap_l^c]+\E[p_m^bp_l^c]\Bigr)+\frac{2}{D^3},
\end{aligned}
\label{eq:cov3}
\end{equation}
with $\delta p=p-1/D$ and the two-point terms the benchmark kernel
$L(k;m,\ell)$ in the same notation (Sec.~\ref{sec:benchmark}).  By the sector
theorem the value depends only on the sector, and the $16\times16$
array $D^3\Cov_3$ is a table of rational functions of $N$ whose
closed forms are collected in Sec.~\ref{app:kernel}.  The
representative diagonal entries read, with
$D=\binom N2=N(N-1)/2$,
\begin{equation}
\begin{aligned}
&D^3\Cov_3^{2222}=\frac{(N-2)(74N^3-120N^2-170N-24)}
{(N+1)(N+2)(N+3)(N+4)},\\[4pt]
&D^3\Cov_3^{1111}=\frac{P_1(N)}{\Delta_1},\\[2pt]
&P_1(N)=N^6-7N^5+31N^4+N^3-248N^2+414N+288,\\[4pt]
&D^3\Cov_3^{0000}
=\frac{8(44N^3+84N^2-188N-180)}{\Delta_0},
\end{aligned}
\label{eq:diag3}
\end{equation}
with $\Cov_3^{abcd}\equiv\Cov_3(a,b,c,d;a,b,c,d)$,
$\Delta_1=(N+1)(N-3)(N+4)(N^2-4)(N^2-9)$ and
$\Delta_0=(N+1)(N-5)(N+4)(N^2-4)(N^2-9)$.
Each has its $N\to\infty$ limit the corresponding three-minor moment
of a Gaussian matrix (Sec.~\ref{app:kernel}).

\subsection{Exact consistency tests}

Four exact identities anchor the table.  \emph{(i) Selberg.}  The
fully coincident sector is the single-channel third moment, and the
Selberg hierarchy of the benchmark gives
$q_3=\prod_{j=1}^k\prod_{i=1}^2\frac{(j+2i)(N-k+j)}{(N-k+j+2i)j}$
[Eq.~\eqref{eq:qr}].  The computed sector satisfies
\begin{equation}
\E[p^3]=\frac{q_3}{D^3}
=\frac{90N^2(N-1)^2}{(N+1)(N+2)(N+3)(N+4)D^3},
\label{eq:selbergcheck}
\end{equation}
exactly for all $N$ tested ($N=8,14,30,\dots,52$), and the connected
coincident sector reproduces the cumulant
$K_3=q_3-3q_2+2$.  \emph{(ii) Marginals.}  Summing the kernel over the
third eigenstate with its sector multiplicities reproduces the
two-point kernel exactly,
\begin{equation}
\sum_{A_3}\E[p_1p_2p_3]=\E[p_1p_2],
\label{eq:marginal3}
\end{equation}
verified for every pair of row and column patterns at $N=14$ with zero
deviation, closing the benchmark's two-point table as the marginal of
the three-point one.  \emph{(iii) Conservation.}  The channels form a
basis, so $\sum_l\delta p_l^c=0$.  The channel-side marginal
$\sum_{C_3}\Cov_3=0$ holds exactly with the sector multiplicities
at $N=14$.  \emph{(iv) Replica symmetry.}  The table is invariant
under the symmetric group of the three replicas and under the
transpose (level $\leftrightarrow$ mode) exchange, as required by
$\mathrm{law}(U)=\mathrm{law}(U^T)$.

\subsection{The replica-sector assembly at general $r$}
\label{sec:assembly}

The levels $r=2,3,4$ and the five-point rung above are not isolated
calculations.  They are explicit realizations of one algebraic
construction, which we now state for general $(r,k)$.  Let the $r$
resolvents probe eigenstate sets $A_1,\dots,A_r$ and channel sets
$C_1,\dots,C_r$, each of size $k$, and let $\mathcal P_A$ ($\mathcal
P_C$) denote the overlap pattern of the $A$-sets (of the $C$-sets),
the occupancies of their Venn cells.  The exact $r$-point kernel then
admits the sector assembly
\begin{equation}
\mathcal K_r(A_1,\dots,A_r;C_1,\dots,C_r)
=\sum_{c} T^{(r)}_c(\mathcal P_A,\mathcal P_C)\,w^{(r)}_c(N),
\label{eq:sectgen}
\end{equation}
the sum running over the permutation sectors $c$ of the pairings of
the $2kr$ slots, with $w^{(r)}_c(N)$ the depth-$kr$ orthogonal
Weingarten class weight of the zonal route (Sec.~\ref{sec:weingarten})
and $T^{(r)}_c$ a purely combinatorial coefficient.
Eq.~\eqref{eq:sectgen} is the coefficient form of the finite-$N$
replica-sector invariance of Sec.~\ref{sec:replica}.  For $r=2$ this
is the $(m,\ell)$ sector statement of the benchmark
(Sec.~\ref{sec:benchmark}) and of Sec.~\ref{sec:kernel}, and each of the
tables of this section realizes it at one level.

The sector coefficients close through the occupancy-matrix algebra.
The $2kr$ slots are organized into $r$ rows (the replicas) of $2k$
columns, and the slot-incidence matrix splits as
$m(\sigma,\tau)=M_1(\sigma)+M_2(\tau)$, where $M_1$ collects the
$\sigma$-slots, whose row labels are permuted by
$\sigma\in S_k^{\,r}$, and $M_2$ the $\tau$-slots.  The signed census
\begin{equation}
A_m=\sum_{\sigma,\tau}\mathrm{sgn}(\sigma\tau)\,
\delta_{m,\,M_1(\sigma)+M_2(\tau)}
\label{eq:censusgen}
\end{equation}
is thus the convolution of two $k!^{\,r}$-term histograms (packed as
fixed-width integer lanes, mirror symmetry removing half of the
pairs), and
\begin{equation}
T^{(r)}_c(\mathcal P_A,\mathcal P_C)=\sum_m A_m\,N_c^{(r,k)}(m),
\label{eq:assemblygen}
\end{equation}
with $N_c^{(r,k)}(m)=\#\{(\pi,\pi'):\mathrm{type}(\pi\pi')=c\}$ over the
pairing pairs compatible with the row and column groups of $m$.  Every
matrix splits over the connected components of its support graph, and
disconnected blocks factor over the parts of $c$.  On a connected block
the counts follow from a finite-state chain transfer matrix.  Processing
the block row by row, the state is the multiset of open alternating
chains $\{(c_1,c_2,a)\}$, the open $\pi'$-tails at columns $c_1,c_2$ of
a chain holding $a$ pairs, and
\begin{equation}
\mathcal F_{r+1}(s')=\sum_{s\to s'}w(s\to s')\,\mathcal F_r(s),
\label{eq:chaindp}
\end{equation}
with the transitions opening, extending, merging and closing chains, a
closure of age $a$ emitting a cycle of part $a+1$.  The cycle-type
generating function travels with the transfer.  The local transition
types of the state machine are independent of $(r,k)$, while the
number of simultaneously open chains and the width of the packed
state grow with the slot number $2kr$, so
Eqs.~\eqref{eq:sectgen}--\eqref{eq:chaindp} define a constructive
finite-$r$ assembly scheme whose local rules are
multiplicity-independent.  Its realizations in hand
, $r=2$ (all nine two-point sectors), $r=3$ at $k=2,3,4$ ($256$
sectors, the $39\times39$ array, and the $81\times81$ array), $r=4$ at
$k=2$, and $r=5$ at $k=2$ (the diagonal five-point table above), all
obey the same assembly.  The DP reproduces brute-force pairings exactly
in integer arithmetic on the largest brute-force blocks.  The explicit coverage is the scope map of Table~\ref{tab:exactmap}.

The assembly feeds the bulk hierarchy
level by level through the sector decomposition of
Sec.~\ref{sec:bulk}, so the organizing basis of the multi-resolvent
cumulants is this one algebra at every multiplicity.  The general result is the finite-$r$ sector algebra and its assembly rule, realized and stress-tested by the explicit tables through $r=5$; the rapidly growing pattern spaces are computational rather than conceptual limitations.

\paragraph{The $r\to r-1$ consistency hierarchy}
\label{sec:rcons}

Marginalizing one resolvent label reduces the $r$-resolvent moment to
the $(r-1)$-resolvent one, and the reduction is exact at every finite
$N$.  The resulting hierarchy of relations fixes the normalization of
each sector coefficient against its lower-order counterpart; it is the
sector-level version of the marginalization of
Sec.~\ref{sec:marginal} and it is the statement used in
Sec.~\ref{sec:kernel4} to check the four-point content against the
three-point tables.  The complete family of relations, together with
the derivation of the channel-versus-pattern marginalization
equivalence, is given in Sec.~\ref{app:algebra}.

\section{From sector moments to irreducible vertices}
\label{sec:vertices}

The sector moments are the microscopic input; the physics lives in the
objects they generate.  This section follows one chain: sector moments
are summed into the connected three-resolvent cumulant, its deep-plane
content is identified, and the cumulant is dressed by the ladder
equation into the irreducible Feshbach vertex.  The four-point sector
content closes the chain as the last higher-order check.

\subsection{Bulk three-resolvent cumulant}
\label{sec:bulk}

Combining Eq.~\eqref{eq:specr} with the sector theorem gives the
exact bulk representation
\begin{equation}
\begin{aligned}
\mathcal C^{(3)}_{abc}(z_1,z_2,z_3)&=\sum_{s} N_s\,\Cov_3(s)\, \hat{\cG}_3(z_1,z_2,z_3;s),
\end{aligned}
\label{eq:bulk3}
\end{equation}
where the sum runs over the eigenstate-triple sectors $s$ for the
fixed channel triple $(a,b,c)$, $N_s$ is the sector multiplicity, and
\begin{equation}
\hat{\cG}_3(z_1,z_2,z_3;s)=\frac1{N_s}\sum_{(n,m,l)\in s}\frac{1}{(z_1-\lambda_n)(z_2-\lambda_m)(z_3-\lambda_l)}
\label{eq:G3}
\end{equation}
is the three-eigenvalue resolvent average of the sector, computed per
realization without any expansion.  In Eq.~\eqref{eq:bulk3}
$\Cov_3(s)=\E_U[\delta p_n^a\delta p_m^b\delta p_l^c]$ is the
Haar-eigenvector average of the sector and $\hat{\cG}_3$ is a spectral
sum over the one-body levels.  Because the eigenvector matrix $U$ and
the spectrum are independent, the joint average factors, and
Eq.~\eqref{eq:bulk3} is the exact sector decomposition of the
three-resolvent cumulant, the three-point promotion of the two-point
spectral kernel of Sec.~\ref{sec:bulk:kernel}, exact at every
broadening for the random free-fermion ensemble (the representation is
exact, the numerical verification below is Monte Carlo).  The structure
parallels the two-point case.
In the deep plane $\hat{\cG}_3$ collapses to its leading moments and
the coefficient-level equation reduces to the channel-energy
three-point cumulants of Sec.~\ref{sec:rigidity:identity}, while in the bulk the
sector sum is a finite combination of exact spectral functions with
the kernel as its exact vertex content.

\subsection{Deep plane and the $q_3$ projection}

\label{sec:q3}

In the deep plane $z_1,z_2,z_3\to i\infty$ the sector decomposition
collapses to the moment coefficients.  The leading three-resolvent
coefficient is the channel-energy three-point cumulant
$K_{111}(a,b,c)=\E[s_1(a)s_1(b)s_1(c)]$ with
$s_r(a)=\delta(H^r)_{aa}$ [Eq.~\eqref{eq:momentcov}], which vanishes exactly in
free fermions because $s_1$ is linear in the Gaussian one-body matrix.
The first nonzero coefficient $K_{211}$ closes in the piecewise Wick
form of Sec.~\ref{sec:rigidity:identity}.  Both are verified numerically with
bin-averaging over the channel-triple patterns at the percent level
(Sec.~\ref{app:num}).
The deep-plane limit of
Eq.~\eqref{eq:bulk3} is therefore consistent with the moment-field
identity, and the three-resolvent deep plane is carried by the same
Hamiltonian-moment structure that closed the two-resolvent deep
plane, the bulk/deep-plane decoupling of Sec.~\ref{sec:bulkb} extends
to the third level.

The fully coincident sector answers the benchmark's projection
question.  Since $q_r=\E[p^r]/\E[p]^r$ and $\E[p]=1/D$, the diagonal
$r$-replica sector is exactly
\begin{equation}
\begin{aligned}
\E[(\delta p)^r]\big|_{\text{coincident}}&=\sum_{s=0}^{r}\binom rs (-1)^{r-s}\,\frac{q_s}{D^r},
\end{aligned}
\label{eq:qrproj}
\end{equation}
with $q_0=1$.  The moment hierarchy $q_r$ is the sequence of fully
coincident diagonal sectors of the overlap kernel, i.e.\ the
all-equal sector projection of the replica-sector hierarchy.
Equivalently, $q_r=D^{r-1}\E[\mathrm{IPR}_r]$ with
$\mathrm{IPR}_r=\sum_n(p_n^a)^r$ the $r$-th generalized inverse
participation ratio of the channel state in the many-body
eigenbasis, the $q_r$ are the intensity (participation) moments of
the spectral-weight distribution.
At $r=3$ this reads $\E[(\delta p)^3]=K_3/D^3=(q_3-3q_2+2)/D^3$,
verified both by the closed-form table (Selberg check of
Sec.~\ref{sec:kernel}) and by direct Monte Carlo
(Sec.~\ref{app:num}), closing
the
identification that the benchmark posed but did not take.  The same
identification holds one level higher.  The diagonal four-replica
sector, computed with the $d=8$ sector algebra ($2{,}027{,}025$
pairings, $22$ coset classes, Sec.~\ref{app:wgtab}), reproduces
the Selberg hierarchy $\E[p^4]=q_4/D^4$ exactly at $N=8,14,30$
($1/3003$, $1/125970$, $1/30260340$), so the diagonal replica sector
yields $q_r$ at $r=3$ and $r=4$ directly from the sector sum.

The same diagonal sector provides a combinatorial interpretation of
the question that the benchmark
poses for the fluctuation strength.  How the $q_{\rm FF}\sim k^2/8$
growth at half filling organizes into an effective connectivity of the
microscopic Hamiltonian.  The Selberg product telescopes in both
directions.  With $\prod_{j=1}^{k}(j+2i)/j=\binom{k+2i}{2i}$ and
$\prod_{j=1}^{k}\frac{N-k+j}{N-k+j+2i}
=\binom{N-k+2i}{2i}/\binom{N+2i}{2i}$, the full moment hierarchy takes
the binomial form
\begin{equation}
\begin{aligned}
q_r(k,N)&=\prod_{i=1}^{r-1} \frac{\binom{k+2i}{2i}\binom{N-k+2i}{2i}}{\binom{N+2i}{2i}},
\end{aligned}
\label{eq:qrbin}
\end{equation}
verified in exact rational arithmetic against the Selberg product form
of the benchmark for $r\le8$ at every size checked (and against the
sector tables where they exist, $r\le5$).  Each factor of the product is the
fraction of $2i$-subsets of the extended $N+2i$ slot set that split
into a $2i$-subset of the extended minor ($k+2i$) and one of the
extended complement ($N-k+2i$), the $i$-th layer of the diagonal
replica sector contributes the determinant-depth pair count
$\binom{k+2i}{2i}$ dressed by the Fock-space fraction
$\binom{N-k+2i}{2i}/\binom{N+2i}{2i}$.  For $r=2$ this reads
$q_2=\binom{k+2}{2}\binom{N-k+2}{2}/\binom{N+2}{2}$.  At fixed $k$ and
$N\to\infty$ it tends to $\binom{k+2}{2}=(k+1)(k+2)/2$ (the
Porter--Thomas value $3$ at $k=1$), while at half filling
$q_2=(k+1)(k+2)^2/(4(2k+1))$, and the half-filling growth
$k^2/8$ factors exactly into the minor-depth pair count
$\binom{k+2}{2}\simeq k^2/2$ and the Fock-space pair fraction
$\binom{N-k+2}{2}/\binom{N+2}{2}\simeq(N-k)^2/N^2=1/4$.  The effective
connectivity of the moment hierarchy is the pair-count ratio of
Eq.~\eqref{eq:qrbin}, with the determinant depth and the
particle--hole complement playing exactly the roles the benchmark
asked to identify, the Hilbert-space dimension $D$ entering only
through the normalization $\E[p]=1/D$.  The hierarchy is therefore
far from that of structureless Haar-random vectors in Fock space
($q_2=3$), the excess intensity fluctuations are generated by the
Slater determinantal structure itself (not by a failure of one-body
randomness) and this shared microscopic structure is the source of
the higher-order correlations throughout this article.  At the
sector level the same
ratio is assembled class by class.  At $(14,2)$ the five $d=4$ coset
classes contribute to $q_2$ with a single dominant positive class (the
$24$-pairing class, which alone carries the full depth factor
$\binom{k+2}{2}$ as $N\to\infty$) and four subleading corrections.

\subsection{The rung to $\Gamma^{(3)}$ and the four-point content}
\label{sec:gamma3}

\subsubsection{Deep plane of the self-energy}

The Feshbach representation introduces the self-energy
$\cG_a=z-a_a-1/\mathcal R_a$, whose fluctuation about its ensemble
mean obeys the exact geometric identity
$\mathcal R_a=\bar{\mathcal R}_a^0/(1-\bar{\mathcal
R}_a^0\,\delta\cG_a)$ with $\bar{\mathcal R}_a^0=1/(z-a_a-\bar\cG_a)$
(Sec.~\ref{sec:closure:vertex}).  The self-energy is the exact one-channel Schur complement of the Feshbach elimination.  With
$P_a=\ket{\varphi_a}\bra{\varphi_a}$ and $Q_a=\mathbf 1-P_a$, the
exact operator identity
$P_a(z-H)^{-1}P_a=[z-P_aHP_a-P_aHQ_a(z-Q_aHQ_a)^{-1}Q_aHP_a]^{-1}$
\cite{Feshbach1958,Feshbach1962} gives, on the channel matrix element,
\begin{multline}
\mathcal R_a(z)=\bigl[z-a_a-\Sigma_a(z)\bigr]^{-1},\\[2pt]
\Sigma_a(z)=\braket{\varphi_a}{HQ_a\,(z-Q_aHQ_a)^{-1}Q_aH}{\varphi_a},
\label{eq:schur}
\end{multline}
so that $\cG_a=\Sigma_a$ holds at every realization.  The geometric
identity above is the scalar avatar of this operator Schur identity,
and the Feshbach-irreducible vertices are, by construction, the
connected content of an eliminated (Schur-complemented) dynamics
rather than of a Legendre effective action.  This exactness does not by
itself close the ensemble problem.  $\bar{\mathcal R}_a=\E[(z-a_a-
\Sigma_a)^{-1}]$ does not reduce to $[z-a_a-\bar\Sigma_a]^{-1}$, the
difference being generated by the connected fluctuations of
$\Sigma_a$, the very objects $\Gamma^{(r)}$, so the Schur complement
solves the realization-level problem while the hierarchy of this
article solves the closure problem generated by ensemble averaging.
In the deep plane the moment-field identity closes the
expansion of the self-energy itself.  Expanding $1/\mathcal R_a$ in
$1/z$ gives
\begin{equation}
\cG_a(z)=\frac{v_a}{z}+\frac{w_a}{z^2}+\cdots,
\qquad
v_a=\sum_{b\ne a}V_{ab}^2,
\label{eq:Gdeep}
\end{equation}
with $w_a=\sum_{b\ne a}V_{ab}^2a_b+\sum_{b,c\ne a,\,b\ne
c}V_{ab}V_{bc}V_{ca}$,
because the $O(z^0)$ term is $m_1(a)-a_a$, and
$m_1(a)=\sum_np_n^a\lambda_n=(H)_{aa}=a_a$ holds at every
realization, the $r=1$ case of the moment-field identity, verified to
machine precision in the numerics below.  In random free fermions the
channel graph restricts $v_a$ to the edge set
$E_a=\{(c,q):c\in A,\ q\notin A\}$ of channel $a$.
$v_a=\sum_{(c,q)\in E_a}h_{cq}^2$, the total squared coupling through
which channel $a$ hybridizes, so the vertex cumulants below correlate
the hybridization fluctuations of the participating channels.  The
subleading $w_a$ is odd under
$h\to-h$, and the Gaussian measure is even, so any cumulant containing
an odd number of $w$-fields vanishes by symmetry (in particular
$\E[\delta v\,\delta v\,\delta w]=0$ with $\delta v_a=v_a-\E[v_a]$,
$\delta w_a=w_a-\E[w_a]$) while the mixed cumulants entering the
orders retained below vanish under the corresponding Wick contractions
(each $w$-monomial is of odd degree in the independent
diagonal and off-diagonal Gaussian fields).  The
Feshbach irreducible self-energy vertices close in the deep plane as
\begin{align}
&\Gamma^{(3)}_{abc}=\frac{8\,e_{abc}}{z_1z_2z_3}+O(z^{-5}),
\quad
e_{abc}=|E_a\cap E_b\cap E_c|,\label{eq:Gamma3}\\
&\Gamma^{(4)}_{abcd}=\frac{48\,e_{abcd}}{z_1z_2z_3z_4}+O(z^{-6}),
\quad
e_{abcd}=|E_a\cap E_b\cap E_c\cap E_d|,\label{eq:Gamma4}
\end{align}
with the chi-square cumulants $\kappa_3(\chi^2_1)=8$ and
$\kappa_4(\chi^2_1)=48$ as the Wick factors.  The shared-edge counts
have a direct dynamical meaning.  $e_{abc}$ counts the one-body
particle--hole transitions available simultaneously to the three
channels (their shared virtual pathways) so the leading Feshbach
vertex measures, in the deep-plane leading order, the number of shared
hybridization pathways, dressed
by the universal chi-square cumulant of the squared coupling
amplitude.  $e_{abcd}$ is the corresponding four-channel overlap.  The
shared-edge counts
are closed combinatorial objects.  $e_{abc}=\sum_S n_S\,n_{S^c}$ runs
over the four complementary pairs of the eight cells of the Venn
partition of $(A,B,C)$ ($n_S$ the occupancy of cell $S$).  At $k=2$,
$e_{aaa}=k(N-k)=24$, $e=12$ for $A=B$ with $|A\cap C|=1$, $e=10$ for
three channels sharing one mode, and $e=4$ for $A=B$ with
$C\cap A=\varnothing$.  Both closed forms were verified by the exact
monomial Wick engine for all $216$ channel-triple patterns and for
representative quadruples.

\subsubsection{Ladder bookkeeping and the extraction of $\Gamma^{(3)}$}

The fluctuation field of the hierarchy,
$\delta\mathcal R_a=\sum_n\delta p_n^a/(z-\lambda_n)=\mathcal
R_a-\hat{\mathcal R}_0$ with $\hat{\mathcal
R}_0=D^{-1}\Tr(z-H)^{-1}$ the mean-sector (trace) resolvent
[Eq.~\eqref{eq:dRdef}], splits
exactly as $\delta\mathcal R_a=X_a+\Delta_a$ with
$X_a=\bar{\mathcal R}_a^0-\hat{\mathcal R}_0$, where
$\Delta_a=\sum_{r\ge1}(\bar{\mathcal R}_a^0)^{r+1}(\delta\cG_a)^r$ is
the geometric ladder dressing of the self-energy fluctuation.  The
three-resolvent cumulant is therefore exactly
\begin{equation}
\begin{aligned}
\mathcal C^{(3)}_{abc}
=\kappa_3(X_a,X_b,X_c)+\bigl[\kappa_3(X_a,X_b,\Delta_c)\\
+\kappa_3(X_a,\Delta_b,X_c)+\kappa_3(\Delta_a,X_b,X_c)\bigr]+\bigl[\kappa_3(X_a,\Delta_b,\Delta_c)\\
+\kappa_3(\Delta_a,X_b,\Delta_c)+\kappa_3(\Delta_a,\Delta_b,X_c)\bigr]+\kappa_3(\Delta_a,\Delta_b,\Delta_c),
\end{aligned}
\label{eq:ladder3}
\end{equation}
The
deep-plane ordering of Eq.~\eqref{eq:ladder3} is rigid.  Since
$\bar{\mathcal R}_a^0-\hat{\mathcal R}_0=s_1(a)/z^2+[s_2(a)-v_a-\bar
v_a]/z^3+O(z^{-4})$ with $\bar v_a=\E[v_a]$, and
$\Delta_a=\delta v_a/z^3+O(z^{-4})$, with $\delta v_a=v_a-\bar v_a$.  The
mean-field triple is $O(z^{-7})$ and carries the deep-plane content
$K_{211}(a,b,c)=\E[s_2(a)s_1(b)s_1(c)]$ of Sec.~\ref{sec:q3}.  The
mixed terms start at $O(z^{-9})$ (the leading cross cumulants vanish
by the diagonal--off-diagonal independence of $h$).  And the vertex
term is
\begin{equation}
\kappa_3(\Delta_a,\Delta_b,\Delta_c)
=\big(\bar{\mathcal R}_a^0\bar{\mathcal R}_b^0\bar{\mathcal
R}_c^0\big)^2\,\Gamma^{(3)}_{abc}\,\big[1+O(z^{-2})\big],
\label{eq:Delta3}
\end{equation}
which equals $8\,e_{abc}/(z_1^3z_2^3z_3^3)+O(z^{-11})$ by
Eq.~\eqref{eq:Gamma3}: the ladder contribution is the vertex multiplied
by the geometric dressing, so the $\Delta$ sector itself carries no
independent structure.

With $\Gamma^{(r)}=\kappa_r(\delta\cG_{a_1},\dots,\delta\cG_{a_r})$
defined as the cumulant of the self-energy fluctuation, the geometric
identity $\mathcal R_a=\bar{\mathcal R}_a^0/(1-\bar{\mathcal
R}_a^0\,\delta\cG_a)$ exhibits the ladder series as the dressing, so
the vertex is extracted exactly by
$\Gamma^{(3)}=\kappa_3(\delta\cG_a,\delta\cG_b,\delta\cG_c)$.  The two
levels answer different questions: $\mathcal C^{(r)}$ is the
correlation language, which statistical correlations between channel
spectral weights are present, while $\Gamma^{(r)}$ is the
self-consistency language, what a single-channel description must carry
after the complementary-space dynamics is eliminated.  The general-$r$
form of the bookkeeping [Eq.~\eqref{eq:laddergen}], its rigid
deep-plane ordering, and the numerical checks at $(14,2)$ are given in
Sec.~\ref{app:derivations}.

\subsubsection{Four-point content and the FK dictionary}

The four-resolvent connected cumulant admits the same exact deep-plane
moment expansion, with its complete finite-frequency sector
representation carried at the level of the representative
channel-pattern slices of Sec.~\ref{sec:kernel4}:
\begin{equation}
\begin{aligned}
\mathcal C^{(4)}&=\sum_{rstu}K_{rstu}\,z_1^{-(r+1)}z_2^{-(s+1)}\times z_3^{-(t+1)}z_4^{-(u+1)},\\
K_{rstu}&=\E[s_r(a)s_s(b)s_t(c)s_u(d)]_c.
\end{aligned}
\label{eq:C4def}
\end{equation}
The four coefficients at the bottom vanish.
$K_{1111}=0$ (the fourth cumulant of Gaussian linear fields) and
$K_{2111}=0$ identically (five linear factors, odd Wick).  The leading
content appears at total moment order six,
\begin{equation}
K_{2211}=\E_c[s_2s_2s_1s_1],\qquad
K_{3111}=\E_c[s_3s_1s_1s_1],
\label{eq:K2211}
\end{equation}
whose exact Wick values (the fixed-size evaluation of the monomial
Wick engine of Sec.~\ref{app:num}) are nonzero on each of the
four canonical channel quadruples (coincident, pair-sharing
$([12],[13],[12],[13])$, generic $([12],[13],[24],[56])$, and
disjoint), in agreement with Monte Carlo within the heavy-tailed
sampling error of the cubic fields (Sec.~\ref{app:num}).  The exact
rational values are listed in Sec.~\ref{app:num}, and the
pair-sharing value is the chain-saturating coefficient of the
Proposition below.
The connected $r$-resolvent deep plane starts at total
inverse-energy order $3r-2$, and this is a general bound:

\begin{quote}
\noindent\emph{Proposition (connected deep-plane degree).}
For the Gaussian one-body ensemble considered here, scale all
resolvent variables jointly, $z_i=t\zeta_i$, with
$\boldsymbol\zeta=(\zeta_1,\dots,\zeta_r)$ fixed and $\zeta_i\neq0$.
Then the connected $r$-resolvent cumulant $\mathcal C^{(r)}$ obeys
\begin{equation}
\mathcal C^{(r)}(t\boldsymbol\zeta)
=O\bigl(t^{-(3r-2)}\bigr),
\qquad
|t|\to\infty,\,
\deg_t\mathcal C^{(r)}\ge 3r-2,
\label{eq:3r2}
\end{equation}
and the bound is attained by the tree-chain pattern
$s_1-s_2-\cdots-s_2-s_1$.  Whenever its coefficient is nonzero the bound
is saturated, and the Gaussian evaluation below shows the coefficient
to be strictly positive for every $r$:
\begin{equation}
\underbrace{r-1}_{\text{connectivity}}
\Rightarrow
\underbrace{2(r-1)}_{\text{moment degree}}
\xrightarrow{+r\ \text{denominators}}
\underbrace{3r-2}_{\text{resolvent degree}}.
\label{eq:degree}
\end{equation}
\end{quote}

The proof uses only the connected-diagram expansion of cumulants
\cite{Lehner2004} together with the moment-field identity, and it is
given in Sec.~\ref{app:derivations}; for a non-Gaussian one-body
matrix the Wick edges are replaced by higher-order cumulant hyperedges
and the leading scaling may change.

\subsection{The four-point sector kernel and four-resolvent correlations}
\label{sec:kernel4}

This section answers three questions about the $r=4$ level.  What
generalizes (the $d=8$ sector algebra carries the four-point overlap
kernel, and the four-resolvent correlations follow from it), what is
verified (the exact consistency reductions to the three-point kernel
and the deep-plane limits), and how complete the result is, the
explicit tables here are representative channel-pattern slices, not
the full sector array, whose exactness status is recorded in
Table~\ref{tab:exactmap}.

\subsubsection{The kernel and its exact consistency}

The four-point overlap kernel $\E[p_1p_2p_3p_4]$ expands over
$S_2^{\times4}$ sign products of sixteen $U$-factors, so the same
$d=8$ sector algebra that closed the $\lambda=0$ anchor of
Sec.~\ref{sec:interact} applies verbatim.  Grouping the four eigenstate
(row) pairs and the four channel (column) pairs into their canonical
patterns (the restricted-growth classes of four 2-sets, enumerated in
Sec.~\ref{app:num}) the kernel is
\begin{equation}
\E[p_1p_2p_3p_4](P_0,Q_0;N)=\sum_{\gamma\in\mathcal C_8}
T_\gamma(P_0,Q_0)\,\Wg_\gamma^{O(N)},
\label{eq:kernel4}
\end{equation}
with exact integer sector tables $T_\gamma$ and the exact rational
Weingarten functions of the 22-class $d=8$ algebra.  $N$ enters only
through the weights, so every sector is an exact rational function of
$N$.  For $r=4,k=2$ the eigenstate-pattern space holds $1657$ labeled
restricted-growth classes (Sec.~\ref{app:num}).  The present
finite-$N$ calculation resolves every row class against four
representative channel-pattern classes (the coincident, pair-sharing,
generic and disjoint quadruples, the columns of the tables
below) and does not require the full $1657\times1657$ array.  By the
sector theorem the general pattern-pair dependence is carried by the
same $T_\gamma$ assembly, and the selected slices provide the explicit
finite-$N$ realizations and consistency tests.  The table passes four
exact consistency identities, all without
Monte Carlo.  (i) the fully coincident sector reproduces the Selberg
corner $\E[p^4]=q_4/D^4$ at every $N=10,\dots,30$ checked, e.g.\
$1/125970$ at $(14,2)$.  (ii) the three-point marginals
$\sum_{A_4}\E[p_1p_2p_3p_4]=\E[p_1p_2p_3]$ with the fourth eigenstate
channel summed over the full pattern space reproduce the $k=2$
three-point table of Sec.~\ref{sec:kernel} exactly at every checked
$N$ (the sum must run over all classes, the orientation of a 2-set
inside a class is a relabeling degree of freedom, so no
``sorted-pattern'' restriction is imposed).  (iii) replica symmetry
(channel relabeling $1\leftrightarrow2$) sector by sector over all
classes.  (iv) covariance conservation $\sum_{A_4}\kappa_4(\delta
p_1,\delta p_2,\delta p_3,\delta p_4)=0$ for every fixed triple of
channels, with $\Cov_4=\kappa_4(\delta p,\delta p,\delta p,\delta p)$
built by the exact $\delta$-expansion from the $k=2$ three-point and
two-point kernels.

The bulk closure is the four-resolvent analogue of
Eq.~\eqref{eq:bulk3}, completed by the pair-coupling terms that the
fourth joint cumulant forces.  Expanding the centered channel
resolvent fields $\tilde R_s(a)=\sum_n\delta p_n^a/(z_s-\lambda_n)$
and using the multilinearity of $\kappa_4$,
\begin{widetext}
\begin{equation}
\mathcal C^{(4)}(z_1,\dots,z_4)=\sum_{P_0}\Cov_4(P_0;Q_0)\,\hat\cG_4(P_0)
+\sum_{(ij)(kl)}\sum_{n,m,l,r}\kappa_2(\delta p_i,\delta p_j)\kappa_2(\delta p_k,\delta p_l)
\Bigl[\langle w_iw_jw_kw_l\rangle-\langle w_iw_j\rangle\langle w_kw_l\rangle\Bigr],
\label{eq:c4closure}
\end{equation}
where the outer sum over $(ij)(kl)$ runs over the three pairings of
the four channels, $w_s(n)=1/(z_s-\lambda_n)$,
$\kappa_2(\delta p_i,\delta p_j)$ is the exact
two-point kernel of Sec.~\ref{sec:kernel} evaluated on the eigenstate
pair $(n_i,n_j)$ (the channel pair fixed by the quadruple), and
$\langle\cdot\rangle$
denotes the GOE eigenvalue average.  The pair terms are absent at the
two-point level ($\E[\delta p]=0$ makes the $\kappa_2$ factorization
exact there) but the fourth joint cumulant of products does not
factorize, so they are forced at order four.  In the deep plane the
coincident-quadruple content closes on the exact Wick coefficients.
With $\langle\lambda^r\lambda^s\lambda^t\lambda^u\rangle_{P_0}$ the
pattern-binned eigenvalue averages and $D_{rstu}$ the pair-coupling
sum of Eq.~\eqref{eq:c4closure} at
$(W_1,W_2,W_3,W_4)=(\lambda_n^r,\lambda_m^s,\lambda_l^t,\lambda_r^u)$,
\begin{equation}
\sum_{P_0}\Cov_4(P_0;\mathrm{coinc})\langle\lambda^2\lambda^2\lambda\lambda\rangle_{P_0}
+D_{2211}=K_{2211},\qquad
\sum_{P_0}\Cov_4(P_0;\mathrm{coinc})\langle\lambda^3\lambda\lambda\lambda\rangle_{P_0}
+D_{3111}=K_{3111},
\label{eq:deep4}
\end{equation}
\end{widetext}
at $(14,2)$.  The pair terms contribute a substantial fraction of the
two totals (Sec.~\ref{app:num}), the bulk-side realization
of Eq.~\eqref{eq:Gamma4}'s $48\,e_{abcd}$ structure.  Since the sector
tables are $N$-independent and only the Weingarten weights are
re-evaluated, Eq.~\eqref{eq:deep4} is closed at every checked size.
At $N=8,10,12,16$ the combination reproduces the exact Wick values
(Sec.~\ref{app:num}) to within the
eigenvalue--Monte~Carlo error of the binned averages.  The layer structure of the table realizes
the replica-sector organization in closed form.  Each of the 22
Weingarten classes carries the connectedness of the union graph of the
four 2-sets, and the Gram weight of a class is
$N^{\#\,\text{components}}$, the pattern-resolved sectors of
Eq.~\eqref{eq:kernel4} are the four-resolvent entries of the FK
dictionary, labeled by layer exactly as the three-point level of
Sec.~\ref{sec:kernel}.

\subsubsection{Connected content beyond independent factorization}
\label{sec:annealed}

The independent-factorization baseline of every $r$-point sector is
the product of the single-sector averages,
$\E_{\rm ind}=\prod_i\E[p_i]=D^{-r}$.  The deviation of the exact
sector tables from this constant measures the connected content
discarded by factorization.  In
units $\E\cdot D^{r}$ the baseline is $1$
(SBA denotes sectors below the baseline, i.e.\ sectors with
$\E\cdot D^r<1$, 
entries in the last column are SBA/total), 
\begin{center}
\small
\setlength{\tabcolsep}{4pt}
\begin{tabular}{llll}
$r$ & model & $q_r$ corner & SBA/total \\
\hline
2 & $(8,4)$ & $5$ & $16/25$ \\
3 & $(14,2)$ & $8281/204=40.593$ & $84/256$ \\
4 & $(14,2)$ & $544.375$ & $3659/6628$ \\
\end{tabular}
\end{center}
The corner values in the table are the Selberg moments $q_r$
themselves at the two anchors, $q_2,q_3,q_4=5,\,49.495,\,726.85$ at
$(8,4)$ and $4.55,\,40.593,\,544.375$ at $(14,2)$.  The columns of the
table are the four canonical channel quadruples of
Eq.~\eqref{eq:kernel4}, and the SBA counts taken over all pattern
classes (the full sector ranges, column-resolved, are recorded in
Sec.~\ref{app:num}).  The
spread grows with $r$ while the baseline stays pinned at $1$.  The
multi-resolvent hierarchy carries connected content end to end, and
the sector tables
quantify it exactly at every finite $N$.  The mechanism is transparent
at the group level.  Haar averaging does not make different minors
independent, because all minors are matrix coefficients of the same
exterior-power operator $\wedge^kU$, their joint moments are governed
by the full pairing algebra of Eq.~\eqref{eq:weingarten}, not by
products of one-minor averages.

\section{Moment-determined and eigenstate-resolved sectors}
\label{sec:physical}

Everything so far was exact for free fermions.  The question that
decides the physical reach of the construction is what survives when
the interaction is switched on, and the answer is a boundary between two sectors.  This section states the projected closure of
the irreducible vertex, the exact moment rigidity of the deep-plane
vertex, the nonperturbative character of the bulk, and the
sector-resolved response that makes the boundary testable.  The three
questions asked throughout are: which parts of the fluctuation sector
the exact sector structure determines, which parts no description in
terms of Hamiltonian moments alone can determine, and whether the
resulting boundary has a checkable formula or numerical signature.

Throughout, three kinds of evidence are kept distinct: operator and finite-dimensional algebraic identities, exact for any Hamiltonian; exact finite-$N$ free-fermion results; and finite-size numerical verification, whose conventions and error control are fixed in Sec.~\ref{app:num}.

\subsection{Projected closure: reconstructing the irreducible vertex}
\label{sec:closure}

\subsubsection{The four-step architecture}
\label{sec:closure:arch}

We state the closure architecture in the notation of Ref.~\cite{HOETH}. Level~I determines the Feshbach mean-field channel resolvent through the self-consistency
\begin{equation}
\bar{\mathcal R}_a^0=\bigl[z-a_a-\bar{\mathcal G}_a(\bar{\mathcal R}^0)\bigr]^{-1},
\label{eq:level1}
\end{equation}
in general distinct from the exact ensemble mean.  Level~II governs the resolvent covariance through the exact ladder hierarchy
\begin{equation}
\begin{aligned}
\mathcal C^{(2)}&=\mathcal C_0^{(2)}[\bar{\mathcal R}^0]+\mathcal L[\bar{\mathcal R}^0]\,\mathcal C^{(2)}+\mathcal H[\mathcal C^{(2)}],\\
\mathcal C_0^{(2)}&=(\bar{\mathcal R}^0)^2(\bar{\mathcal R}^0)'^{\,2}\Gamma^{(2)},
\end{aligned}
\label{eq:level2}
\end{equation}
with ladder kernel $\mathcal L_{ab,cd}=(\bar{\mathcal R}_a^0)^2(\bar{\mathcal R}_b^0)'^{\,2}|V_{ac}|^2|V_{bd}|^2$, irreducible vertex $\Gamma^{(2)}_{ab}=\E[\delta\mathcal G_a\delta\mathcal G_b]_c$ the connected self-energy covariance, and $\mathcal H$ collecting the higher irreducible contributions. Eq.~\eqref{eq:level2} is exact.  The difficulty is to characterize $\Gamma^{(2)}$ (an object on channel-pair space with two frequency arguments) and $\mathcal H$. A closure produces a usable $\Gamma^{(2)}$ from controlled inputs in four steps.

\emph{Step 1.  Sector basis.} Choose a finite family of invariant sectors $\{\Phi_\alpha\}$ on channel pairs (selected by symmetry, conservation, channel geometry, and the known ETH scaling) and expand
\begin{equation}
\Gamma^{(2)}=\sum_\alpha\gamma_\alpha\,\Phi_\alpha+\Gamma^{(2)}_{\rm rest},
\label{eq:ansatzgen}
\end{equation}
with amplitudes $\gamma_\alpha(z,z')$ carrying the residual frequency dependence. The sectors are the closure coordinates. For free fermions this choice is not a guess.  The exact two-point geometry $L(k;m,\ell)$ of Sec.~\ref{sec:benchmark} establishes the shared-level and shared-mode counts $(m,\ell)$ as the covariance coordinates, and supplies the ground-truth amplitudes against which the closure is tested.

\emph{Step 2.  Projections.} Select observables $Q_\beta$ computable independently of the vertex (conservation identities, normalization constraints, Porter--Thomas-type ratios, fixed-depth limits) and demand that the covariance solving the ladder equation reproduce them,
\begin{equation}
P_\beta[\mathcal C^{(2)}(\gamma)]=Q_\beta,\qquad
\mathcal C^{(2)}(\gamma)=\mathcal C_0^{(2)}(\gamma)+\mathcal L\,\mathcal C^{(2)}(\gamma).
\label{eq:projeq}
\end{equation}

\emph{Step 3.  The parameter system.} Since $\mathcal C^{(2)}$ is linear in $\Gamma^{(2)}$, Eq.~\eqref{eq:projeq} is a finite system, linear in $\gamma$ whenever the projections are linear functionals.  $M_{\beta\alpha}\gamma_\alpha=Q_\beta$, with $M$ built from the ladder dressing and the sector overlaps.

\emph{Step 4.  Residual diagnostic.} The residual is the footprint of irreducible vertex content that is invisible to the chosen constrained sector basis and cannot be generated by ladder dressing. The system may solve and yet be incomplete, the ansatz reproduces its own defining observables by construction, and only its unconstrained entries can be tested against independent data. Define the residual
\begin{equation}
\Delta_\alpha=P_\alpha[\mathcal C^{(2)}_{\rm exact}]-P_\alpha[\mathcal C^{(2)}_{\rm ansatz}]
=\Delta_\alpha^{\rm ladder}+\Delta_\alpha^{\rm irr},
\label{eq:residual}
\end{equation}
split by dressing scaling.  Relative to the leading coefficient of the covariance, one ladder insertion is suppressed by $(\bar{\mathcal R}^0)^2(\bar{\mathcal R}^0)'^{\,2}=O(\eta^{-4})$ ($\eta=\Im z$), so content with an $\eta$-independent ($\eta^0$) limit is irreducible vertex content while content collapsing as $\eta^{-4}$ is ladder dressing (Sec.~\ref{sec:closure:solved}). A nonzero $\Delta^{\rm irr}$ locates a sector missing from the basis, and the architecture iterates by adding that sector. Steps~1--3 close the known sectors.  Step~4 is the discovery mechanism for the unknown ones. The availability of Steps~1--3 is model dependent, and a generic vertex may require infinitely many sectors; the solvable ensemble below shows that the architecture, including the diagnostic step, is executable end to end.

\subsubsection{Channel resolvents, Level I, and the exact vertex decomposition}
\label{sec:closure:vertex}

The channel resolvent is the spectral object of the framework
[Eq.~\eqref{eq:Rdef}]; for a quadratic Hamiltonian it admits the exact
one-body representation [Eq.~\eqref{eq:detres}], and the exact mean is
channel independent at every $N$ [Eq.~\eqref{eq:meanS}].  The
determinantal representation, its fixed-depth limit, and the comparison
with the scalar self-consistent Born approximation are given in
Sec.~\ref{app:derivations}.

The fluctuation expansion of the exact Feshbach identity $\mathcal R_a=[z-a_a-\mathcal G_a(z)]^{-1}$ is performed about the mean-field resolvent
\begin{equation}
\bar{\mathcal R}_a^0(z):=\frac{1}{z-a_a-\bar{\mathcal G}_a(z)},\qquad
\bar{\mathcal G}_a:=\E\,\mathcal G_a,
\label{eq:rbar0}
\end{equation}
about which the exact resolvent identity holds, $\mathcal R_a=\bar{\mathcal R}_a^0/(1-\bar{\mathcal R}_a^0\,\delta\mathcal G_a)$. The self-energy fluctuation splits identically into the one-loop (off-diagonal) part and the cavity remainder, so that the connected self-energy covariance decomposes identically as
\begin{equation}
\E[\delta\mathcal G_a\delta\mathcal G_b]_c=\mathrm{LAD}_{ab}+\Gamma^{(2)}_{ab},\quad
\Gamma^{(2)}_{ab}=\mathrm{ODCC}_{ab}+\mathrm{CCCC}_{ab},
\label{eq:vertexsplit}
\end{equation}
algebraic identities holding for any Hamiltonian [Eq.~\eqref{eq:odccfull}].  The physics resides in the relative weights of the three sectors. Three structural facts stand out [Table~S21]. \emph{(a)} The decomposition is exact in both fluctuation branches, to $\sim10^{-15}$ at every size. \emph{(b)} The cavity--cavity sector has the largest Frobenius-norm magnitude in both branches, $|\mathrm{CCCC}|/|\Gamma^{(2)}|\approx1.0$ (dense GOE) and $1.8$--$3.0$ (free fermions), the ratio exceeding unity because the ODCC sector partially cancels it.  The connected cavity fluctuations of the exact self-energy dominate the irreducible vertex in both branches. (c) The branches separate in the ladder share.  The one-loop ladder carries $9$--$28\%$ of the connected self-energy covariance in free fermions but only $0.2\%$ in the ETH-scaled dense reference, where the coupling itself is $O(1/\sqrt D)$. The factorized ladder is moreover quantitatively modified by the Weingarten correlations between $|V|^2$ and $\delta\mathcal R$ ($\Delta$FACT $=0.38$--$0.73$), which any closure using the factorized ladder must retain.

The one-loop ladder factorizes, and for free fermions the hopping graph
of the channel basis is regular, so its channel-space spectral radius
grows with the connectivity [Eq.~\eqref{eq:rhow}].  The ladder is
therefore supercritical at every accessible size and the geometric
resummation $(1-\mathcal L)^{-1}$ is unavailable: the Level-II equation
is a genuine fixed-point problem in which the vertex carries the
compensating weight.  The spectral radii, the deep-plane crossing
$\eta_c$, and the truncation diagnostics are tabulated in
Sec.~\ref{app:derivations}.

\subsubsection{The closure protocol in action: solve, diagnose, complete}
\label{sec:closure:solved}

In the deep plane the covariance admits the moment expansion $\mathcal C_{ab}(z,z')=\sum_{r,s}K_{rs}^{ab}/(z^{r+1}z'^{s+1})$ with
\begin{equation}
K_{rs}^{ab}:=\sum_{A,B}\E[\delta p_A^a\delta p_B^b]\,\lambda_A^r\lambda_B^s .
\label{eq:Krs}
\end{equation}
Covariance conservation ($\sum_A\delta p_A^a=0$ identically, since the channels form a basis) kills every coefficient with $r=0$ or $s=0$, so the leading nonzero coefficient is the moment-weighted covariance $K_{11}(\ell)$, a function of the channel distance alone. In the fixed-depth large-$N$ approximation the benchmark kernel of Eq.~\eqref{eq:Lgeneral} supplies its asymptotic closed form:
\begin{equation}
K_{11}^{\rm th}(\ell)
= (N+1)\Bigl[\frac{k\,L(k;k,\ell)}{D}
+\sum_{m=1}^{k-1}\frac{\mathcal N_m\,L(k;m,\ell)\,m}{D^2}\Bigr],
\label{eq:K11th}
\end{equation}
with $\E\varepsilon^2=N+1$ in the Mehta normalization.  The measured-to-theory ratio converges to unity with $1/N$ corrections, and the $\ell=0$ bin, for which the fixed-depth kernel vanishes identically, carries the purely finite-$N$ Weingarten weight, exactly $-8/N$ [Eq.~\eqref{eq:k110closed} below, Table~S4]. At coefficient level the ladder equation reduces, for the leading $z^{-2}z'^{\,-2}$ coefficient of the deep-plane expansion, to the exact statement $K_{11}=\gamma$.  Each ladder insertion carries $(\bar{\mathcal R}^0)^2(\bar{\mathcal R}^0)'^{\,2}=O(\eta^{-4})$ and contributes only at strictly higher inverse-frequency order, so the leading coefficient is the vertex source at any ladder order.

The four steps are now executed at $k=2$, where the channel-pair space has three bins (diagonal, $\ell=1$, $\ell=0$) and the two-parameter ansatz
\begin{equation}
\Gamma^{(2)}_{ab}=\frac{\gamma_d\,\delta_{ab}
+\gamma_\ell\,\1(\ell_{ab}=1)}
{z^2z'^2(\bar{\mathcal R}_a^0)^2(\bar{\mathcal R}_b^0)'^{\,2}}
\label{eq:twoparam}
\end{equation}
leaves exactly one entry (the $\ell=0$ bin) unconstrained.  That entry is the object the diagnostic will examine. Steps~2--3 degenerate into the fluctuation dictionary itself.  The two exact projections are the diagonal ($A=B$) bin of the same-channel coefficient, whose value the Porter--Thomas ratio fixes through $(q-1)\E[\lambda_A^2]/D$, and the $\ell=1$ coefficient [Eq.~\eqref{eq:K11th}], and the dressed parameter system [Eq.~\eqref{eq:paramsys}] solves to $\gamma_d=1.08$, $\gamma_\ell=1.38$, within $1.4\%$ of the two bare projections, $(q-1)\E[\lambda_A^2]/D=1.09$ and $K_{11}^{\rm meas}(1)=1.39$ (the dressing terms are $O(\eta^{-4})$, $a_0\bar\kappa=7.4\times10^{-4}$ at $\eta=12$). Two parameters against two exact projections leave the system no freedom, so the solved values are a consistency check of the bookkeeping.  The predictive content is carried by the third entry. Step~4 then decides. The solved vertex generates the $\ell=0$ coefficient only through dressing of the $\ell\ge1$ sources, $K_{11}^{\rm pred}(0)=\gamma_d\,a_0W_2^{(\ell=0)}+\gamma_\ell\,a_0W_{\ell W}^{(\ell=0)}=0.006$ at $\eta=12$ ($0.001$ at $\eta=20$) against the measured $-0.573$, a deficit of two orders of magnitude. The $\ell=0$ coefficient is not among the constraints used to determine $\gamma_d,\gamma_\ell$, so the discrepancy is an out-of-constraint prediction. The decision criterion of Sec.~\ref{sec:closure:arch} proves the missing content irreducible rather than dressed.  The ansatz's $\ell=0$ dressing falls $0.006\to0.001$ between $\eta=12$ and $20$, following $(12/20)^4=0.13$, while the measured weight is identical at both frequencies. The self-consistency has therefore located the missing sector.

\emph{Completing the loop.} The missing content is resolved from the $m$-bin expansion. Resolving the coefficient by the shared-level count $m$ of the two eigenstates,
\begin{equation}
\begin{aligned}
K_{11}(\ell)&=\sum_{m=0}^{k}\mathcal N_m\,\mathrm{Cov}(m,\ell)\,\E[\lambda_A\lambda_B|m],\\
\mathrm{Cov}(m,\ell)&:=\E[p_A^ap_B^b\,|\,m,\ell]-\frac1{D^2},
\end{aligned}
\label{eq:mbin}
\end{equation}
with the projection weights spectrum-only and closed form, $\E[\lambda_A\lambda_B|m]=m\,\E\varepsilon^2+(k^2-m)\,\E\varepsilon_s\varepsilon_t$, $\E\varepsilon^2=N+1$, $\E\varepsilon_s\varepsilon_t=-1$ for the unconditioned (traceful) GOE, shows that the measured $\ell=0$ profile is pure finite-$N$ content requiring both finite-$N$ mechanisms: the Weingarten geometry of the two disjoint minors and the GOE level correlations (which give the $m=0$ bin the weight $-4$ rather than the zero of the independent-level approximation). The profile $\mathrm{Cov}(m,\ell)$ is a finite Weingarten object: the O($N$) Weingarten matrix of the orthogonal Haar integration \cite{CollinsSniady,Weingarten} over the pairings of the $4k$ index slots of the two channel minors, with Gram matrix $G(\pi,\tau)=N^{c(\pi\circ\tau)/2}$ ($c$ the cycle count), whose $p=2$ sector reproduces the classical one-row and two-row quartics exactly. For $k=2$ the evaluation closes [Eqs.~\eqref{eq:closedm}], and the $\lambda\otimes\lambda$ projection collapses the entire $\ell=0$ profile onto a single power:
\begin{equation}
\boxed{\;K_{11}(0)=\sum_m \mathcal N_m\,\mathrm{Cov}(m,0)\,\E[\lambda_A\lambda_B|m]
=-\frac{8}{N}\;}
\label{eq:k110closed}
\end{equation}
exactly, for all $N$.  The closed forms reproduce the measured profile to three digits, and Eq.~\eqref{eq:k110closed} gives $-0.571$, $-0.267$, $-0.133$ at $N=14,30,60$ against the measured $-0.573$, $-0.267$, $-0.14$ [Table~S4]. Promoting Eq.~\eqref{eq:mbin} from the covariance to the vertex [Eq.~\eqref{eq:wansatz}], licensed by the coefficient-level identity $K_{11}=\gamma$, is the exact completion of the leading deep-plane $K_{11}$ sector of the vertex at $k=2$, and the loop of Sec.~\ref{sec:closure:arch} is closed at that level.  Minimal ansatz $\to$ projected parameter system $\to$ unconstrained-entry residual $\to$ decision criterion (not ladder) $\to$ microscopic resolution ($m$-bins, level correlations) $\to$ finite-$N$ Weingarten completion $\to$ exact closure.  What remains is the bulk-regime vertex at $\eta=O(1)$.  The higher-$k$ and higher-resolvent sectors of the benchmark are carried out, along the direct route, by the replica-sector calculus of Secs.~\ref{sec:replica}--\ref{sec:kernel}, which organizes the same group integrals as a hierarchy of invariant pairing sectors with exact finite-$N$ weights, of which the $k=2$ sectors used here are the lowest entries. 

Three further results complete the section. First, the framework's fluctuation dictionary is verified directly on two-resolvent objects.  The diagonal bin
\begin{equation}
\mathcal D_{ab}(z,z'):=\sum_A\frac{\delta p_A^a\,\delta p_A^b}{(z-\lambda_A)(z'-\lambda_A)}
\label{eq:Dbin}
\end{equation}
factorizes exactly, $\E[\mathcal D_{ab}]=\Cov(p_a,p_b)\,\E[T(z,z')]$ with $T=\sum_A[(z-\lambda_A)(z'-\lambda_A)]^{-1}$ (no statistical input beyond $U\perp\Lambda$) and its projections reproduce the benchmark's closed forms term by term.  The same-channel ridge (the diagonal-bin projection carrying $q-1$) gives $q=5.19$ against the Selberg ratio $q_{\rm FF}=5.09$ at $(9,3)$, the channel-distance profile $D^2C_\ell=(-0.51,-0.16,0.71)$ against the exact $(-0.49,-0.15,0.70)$, and the integrated cross-channel weight $c$ against its closed form in both sectors [Table~S5]. Thus $q_{\rm FF}\neq3$ is not merely a one-point deviation from Porter--Thomas statistics.  It is the diagonal-bin projection of a structured multi-channel covariance field. Second, the two fluctuation branches are connected by a one-parameter family.  $H(\lambda)=H_{\rm FF}+\lambda W$ with $W$ a dense ETH-scaled GOE in the channel basis ($\E W_{ab}^2=1/D$) interpolates between the determinantal branch ($\lambda=0$) and the Wick branch (large $\lambda$), with the Porter--Thomas ratio flowing from $q_{\rm FF}=5.0$ toward the Haar value $3D/(D+2)=2.91$, the cross-channel weight $c$ flipping sign at $\lambda\simeq2$ toward the isotropy value, and the ladder crossing $\rho(\mathcal L)=1$ at $\lambda_c\simeq3.5$ [Table~S6; Table~S25 gives the full diagnostic set of the family, including the fluctuation ratio $\F$ and the participation ratio $N_{\rm ch}^{\rm eff}$]. This family reappears as the ``dense'' interaction class of Sec.~\ref{sec:rigidity}, where it is governed by an exact law rather than sampled. Third, the same projection logic extends upward and sideways [Sec.~\ref{app:closure}].  The three-resolvent analogue of the diagonal ridge identifies the projection carrying the third overlap cumulant $q_3$ (the same-state triple diagonal bin, with the lower orders protected by exact spectral symmetries) and a minimal solvable deformation $h(\varepsilon)=h_0+\varepsilon\,\mathrm{diag}(v)$ breaks $U\perp\Lambda$ and shows the energy-resolved dictionary $f(E),q(E),c(E)$ acquiring structure [Table~S7].

\subsection{Interaction deformation: exact moment rigidity of the deep-plane vertex}
\label{sec:rigidity}

\subsubsection{The moment-field identity}
\label{sec:rigidity:identity}

Consider a general Hamiltonian on the $k$-particle sector with an arbitrary orthonormal channel basis $\{\ket{\varphi_a}\}$, $a=1,\dots,D$, matrix elements $H_{ab}=\braket{\varphi_a|H|\varphi_b}$, diagonal $a_a=H_{aa}$, and coupling $V_{ab}=H_{ab}(1-\delta_{ab})$. Define the moment fields $s_r(a):=\sum_A\delta p_A^a\lambda_A^r$ with $\delta p_A^a=p_A^a-1/D$ the overlap fluctuation about its uniform channel value. The anchor of everything in this section is a model-independent identity.  By the spectral decomposition $\sum_Ap_A^a\lambda_A^r=\braket{\varphi_a|H^r|\varphi_a}=(H^r)_{aa}$ and $\sum_A\lambda_A^r=\Tr H^r$,
\begin{equation}
\boxed{\;
\begin{aligned}
&s_r(a)=\delta(H^r)_{aa}=(H^r)_{aa}-\frac{\Tr H^r}{D},\\
&K_{rs}(\ell)=\E[s_r(a)s_s(b)\,|\,\ell]
=\E\bigl[\,\delta(H^r)_{aa}\,\delta(H^s)_{bb}\,\big|\,\ell\,\bigr],
\end{aligned}
\;}
\label{eq:momentcov}
\end{equation}
the first line an identity for any Hamiltonian and any orthonormal channel basis, the second its moment-level consequence.  The entire deep-plane coefficient array of the two-resolvent covariance is computable from $H$ without eigenstates, in the ensembles of this article the fields $s_r$ are ensemble-centered ($U\perp\Lambda$ with $\E_U[p_A^a]=1/D$), so the second moments are covariances, while for a general ensemble they are the second cross-moments of the trace-subtracted diagonal moments, and the leading sector $K_{11}$ is the covariance of the channel-energy diagonal. Three structural facts follow. (i)~The leading coefficient is a Hamiltonian-moment covariance.  What the benchmark obtained through the full determinantal (Weingarten) machinery is, in general, a trace-subtraction identity. (ii)~Covariance conservation is manifest.  $\sum_bs_r(b)=\sum_A\lambda_A^r\sum_b\delta p_A^b=0$ identically, so $\sum_bK_{rs}(a,b)=0$ at every order. (iii)~The sector-basis question of the architecture reduces to a definite one.  How the channel-energy covariance $\E[\delta H\otimes\delta H]$ organizes by the channel geometry $\ell$, the $\ell$-distance decomposition for structured models, the isotropy pair $\{\delta_{ab},1-\delta_{ab}\}$ for channel-transitive ones.

In random free fermions the identity closes in one line. With $H_{aa}=\sum_{c\in C_a}h_{cc}$ and $\Tr H/D=(k/N)\sum_ch_{cc}$, $s_a=\sum_{c\in C_a}h_{cc}-(k/N)\sum_ch_{cc}$.  Since $\E h_{cc}h_{dd}=2\delta_{cd}$,
\begin{equation}
\boxed{\;K_{11}(\ell)=2\ell-\frac{2k^2}{N}\;}
\label{eq:ffclosed}
\end{equation}
exactly for every $N,k$.  At $k=2$ the $\ell=0$ weight is the $-8/N$ of Eq.~\eqref{eq:k110closed}, the finite-$N$ Weingarten completion of Sec.~\ref{sec:closure:solved}, and the two finite-$N$ mechanisms separated there (the Weingarten profile and the level correlations) are unified by the single trace-subtraction covariance $\E[(\sum_ch_{cc})^2]=2N$. The benchmark thereby becomes a consistency check of the general identity rather than a separate ingredient [Sec.~\ref{app:inter}]. For the pure dense reference ($H=\sigma\cdot\mathrm{GOE}(D)$, $\sigma^2=1/D$, product channels) the ensemble is channel-transitive, so the covariance has the isotropy form $\E[s_as_b]=x\delta_{ab}+y(1-\delta_{ab})$.  Conservation forces $y=-x/(D-1)$ (given transitivity, conservation alone fixes the ratio) and the channel-diagonal statistics fix
\begin{equation}
x=\E\bigl[(W_{aa}-\Tr W/D)^2\bigr]=2\sigma^2\Bigl(1-\frac1D\Bigr)
\label{eq:wick}
\end{equation}
exactly.  The isotropic deep-plane vertex closes with a single parameter, a diagonal-energy variance, not a fit. The same substitution closes the three-resolvent coefficients as channel-energy three-point cumulants, $K_{rst}=\E[s_r(a)s_s(b)s_t(c)]$.  In random free fermions $K_{111}\equiv0$ exactly (the third cumulant of Gaussian channel energies) and $K_{211}$ closes in piecewise Wick form [Eqs.~\eqref{eq:K111}--\eqref{eq:threept}].

\subsubsection{The deformation law and the rigidity law}
\label{sec:rigidity:law}

Consider the deformation family
\begin{equation}
H(\lambda)=H_0+\lambda W,
\label{eq:family}
\end{equation}
with $H_0=H_{\rm FF}$ the free-fermion channel Hamiltonian and $W$ a centered perturbation ($\E W_{ab}=0$), statistically independent of $H_0$, with law symmetric under $W\to-W$ (any centered Gaussian). Three exact statements organize the response.

\emph{Evenness.} The joint law of $H(\lambda)$ equals that of $H(-\lambda)$, so every ensemble-level observable is an even function of $\lambda$.  $q(\lambda)$, $c(\lambda)$, $\F(\lambda)$, $\rho(\bar W)(\lambda)$, $\rho(\mathcal L)(\lambda)$, $\Gamma^{(2)}(\lambda)$, $K_{11}(\ell;\lambda)$. Evenness alone does not imply analyticity.  Wherever an observable admits a $\lambda$-expansion its odd coefficients vanish and the first interaction correction is $O(\lambda^2)$, while the eigenstate-overlap observables, though even, have a singular regular-perturbative expansion at $\lambda=0$ (Sec.~\ref{sec:bulkb}).

\emph{Coupling splitting.} In expectation the squared channel coupling splits as
\begin{equation}
\bar W(\lambda)_{ab}=\E|V_{ab}^0+\lambda W_{ab}|^2
=\bar W_{0,ab}+\lambda^2\,\E|W_{ab}|^2,
\label{eq:wsplit}
\end{equation}
the linear term vanishing by centering and independence (for disjoint-support perturbations the splitting holds elementwise, per realization). Since the second term is entrywise nonnegative, $\rho(\bar W)(\lambda)$ is nondecreasing in $|\lambda|$.  An independent centered interaction can only enhance the channel-graph spectral radius. For the dense ETH-scaled perturbation the enhancement is additive to the quoted digits, $\rho(\bar W)=\rho(\bar W_0)+\lambda^2(D-1)/D$, with $\rho(\bar W_0)=k(N-k)$ exactly. Whether the ladder crosses below unity is then a competition between this connectivity growth and the spectral broadening, which shrinks $|\bar r^0|^4$ [Eq.~\eqref{eq:rhol}].

\emph{Leading deep-plane rigidity.} From the identity [Eq.~\eqref{eq:momentcov}],
\begin{equation}
s_a(\lambda)=s_a(0)+\lambda\,w_a,\qquad
w_a:=W_{aa}-\frac{\Tr W}{D},
\label{eq:srigid}
\end{equation}
exactly, per realization, for any $H_0$ and $W$, no statistical assumption enters this step. Consequently the exact two-point law is
\begin{equation}
\begin{aligned}
K_{11}^{ab}(\lambda)&=K_{11}^{ab}(0)+\lambda\,\Cov(s_a^{(0)},w_b)\\
&\quad+\lambda\,\Cov(w_a,s_b^{(0)})+\lambda^2\,\Cov(w_a,w_b),
\end{aligned}
\label{eq:generalrigid}
\end{equation}
with $s_a^{(0)}=s_a(0)$.  The coefficient is exactly quadratic in $\lambda$, and the linear term is present in general [a correlated counterexample is exhibited in Eq.~\eqref{eq:counter}]. For $W$ statistically independent of $H_0$ the cross covariances vanish (both fields are centered), and
\begin{equation}
\boxed{\;K_{11}(\ell;\lambda)
=K_{11}(\ell;0)+\lambda^2\,\E[w_aw_b\,|\,\ell].\;}
\label{eq:rigidity}
\end{equation}
The assumption hierarchy is therefore: algebra (no assumptions), then independence (the linear term drops), then a centered symmetric law (evenness of every observable), then diagonal-free $W$ (exact invariance). The interaction enters the leading deep-plane coefficient only through its channel diagonal, at order $\lambda^2$, through the covariance of $w$, no higher terms, no eigenstate dependence.

\subsubsection{Interaction geometries: closed forms and grids}
\label{sec:rigidity:classes}

Three deformation families carry the comparison, and each acts on the
deep-plane kernel in a different way.  Pair hopping leaves the
off-diagonal structure intact and shifts $K_{11}$ by a channel-diagonal
term [Eq.~\eqref{eq:W2b}]; a dense isotropic layer adds the layer of
Eq.~\eqref{eq:denselayer} without an $\ell$ dependence; and
density-density interactions deform the profile nonlinearly, with the
$O(\lambda^2)$ shift quadratic in the channel distance
[Eq.~\eqref{eq:ddlaw}].  The three cases, the table of their measured
coefficients, and the fit that separates the amplitudes from the
profile are collected in Sec.~\ref{app:derivations}.

\subsubsection{The rigidity ladder and higher-resolvent constraints}
\label{sec:rigidity:ladder}

\begin{figure}[t]
\includegraphics[width=0.92\columnwidth]{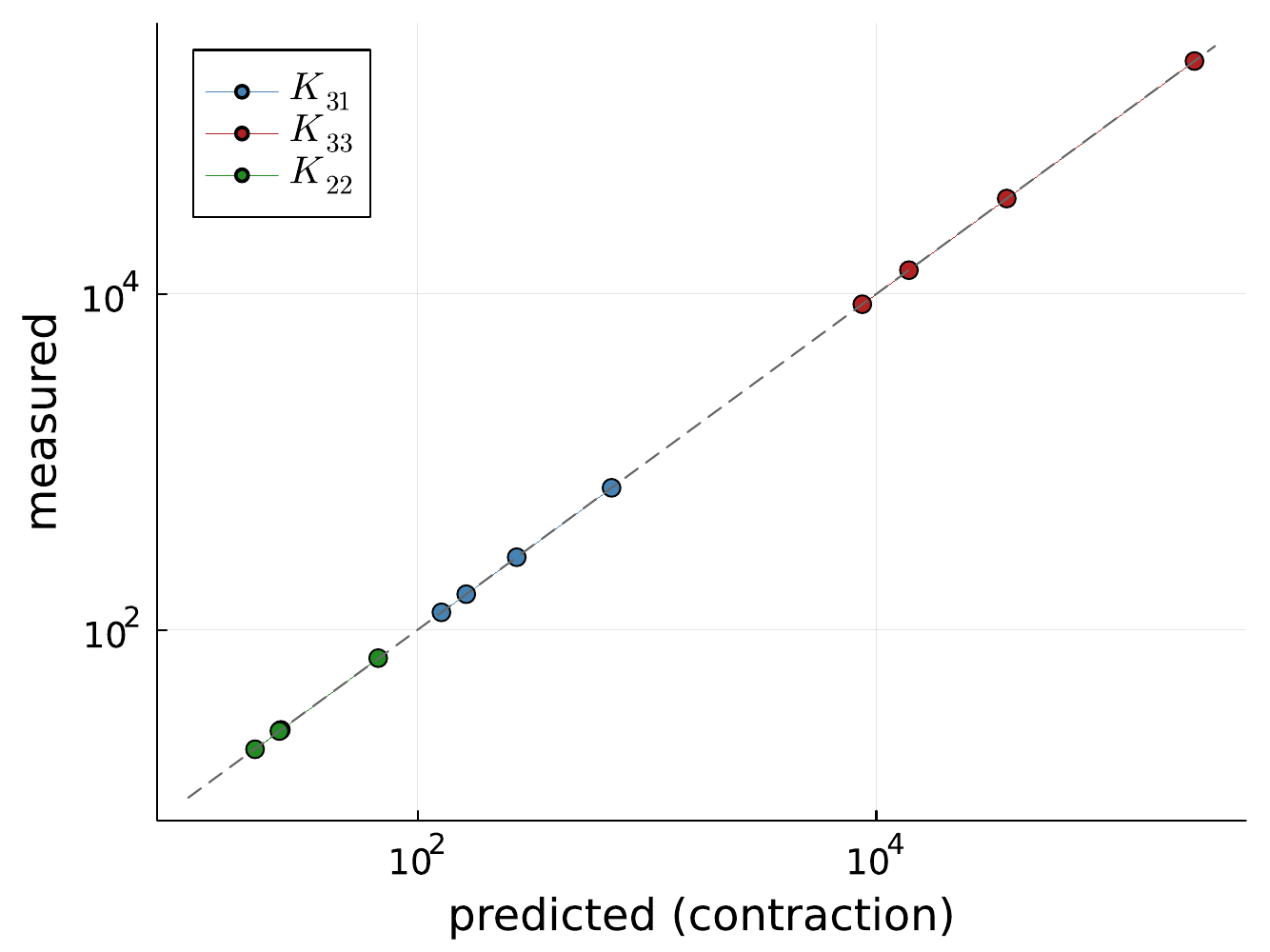}
\caption{Hierarchy validation of the rigidity ladder
[Eqs.~\eqref{eq:ladder}--\eqref{eq:ladderpb}], pair-hopping at $(8,4)$.
Measured vs contraction prediction for $K_{31},K_{33},K_{22}$ at
$\lambda=0,0.5,1,2$ (log-log, the dashed line is $y=x$).  All points
lie within a percent of the diagonal.  $K_{11},K_{21}$ are exactly
invariant, $K_{32}$ lies within the sampling error of the small
$C_{32}^4$ coefficient, $K_{111}\equiv0$ exactly, $K_{211}$ closes in
Wick form to an rms of $1.1\%$, and the correlated counterexample
[Eq.~\eqref{eq:generalrigid}] satisfies
$K_{11}(\lambda)/K_{11}(0)=(1+\alpha\lambda)^2$ per realization to
machine precision.}
\label{fig:ladder}
\end{figure}

Eq.~\eqref{eq:rigidity} is the $(r,s)=(1,1)$ entry of a complete hierarchy over the deep-plane moment sectors. Since $s_r(a;\lambda)=\delta(H^r)_{aa}(\lambda)$ and $H(\lambda)=H_0+\lambda W$, each moment field expands exactly, $s_r(\lambda)=s_r(0)+\sum_{j=1}^{r}\lambda^j A_{rj}$, with $A_{rj}$ the $j$-$W$-factor block of the channel-diagonal moment: pure matrix algebra, valid for any $W$. Every $K_{rs}(\lambda)$ is therefore a polynomial in $\lambda$ of degree at most $r+s$, whose top coefficient $[\lambda^{r+s}]K_{rs}=\E[\delta(W^r)_{aa}\delta(W^s)_{bb}]$ is fixed by $W$ alone. For a centered $W\perp H_0$ with $W\to-W$-symmetric law, every term with an odd total number of $W$ factors vanishes in expectation, so the coefficients obey the exact even finite polynomial (the rigidity ladder)
\begin{equation}
\boxed{\;
\begin{aligned}
&K_{rs}(\ell;\lambda)
=K_{rs}(\ell;0)+\sum_{j=1}^{\lfloor(r+s)/2\rfloor}
\lambda^{2j}\,C_{rs}^{2j}(\ell),\\
&C_{rs}^{2j}=\sum_{j_1+j_2=2j}\E[A_{rj_1}\otimes A_{sj_2}],
\end{aligned}
\;}
\label{eq:ladder}
\end{equation}
where for the Gaussian $W$ used here the interior coefficients are closed-form Wick contractions of the interaction's two-point function against $H_0$-blocks. For the pair-hopping interaction two simplifications are exact.  $A_{11}=w=0$, and $A_{21}\equiv0$ elementwise, the supports of $H_0$ and $W$ are disjoint channel-pair sets, so $\mathrm{diag}(H_0W)=0$, which gives $K_{11},K_{21}$ exactly invariant and explicit closed lines for $K_{31},K_{22},K_{32},K_{33}$ [Eqs.~\eqref{eq:ladderpb}]. Figure~\ref{fig:ladder} validates every line (the full $\lambda$-grid in Table~S8(d)), the closed lines reproduce the measured coefficients across the grid to within the sampling accuracy. Two structural facts complete the picture. First, the interaction corrections inherit the $\ell$-geometry of the free vertex exactly, whatever their functional form.  By the channel-pair selection rule of Sec.~\ref{sec:bench:twopoint}, every $C_{rs}^{2j}$ is a function of the pair through $\ell$ alone, here $C_{31}^2(\ell)$ is linear in $\ell$ to the measured digits and vanishes at $\ell=k/2$, measured $(-287.0,-143.5,0,+143.5)$ for $\ell=0,\dots,3$ at $(8,4)$, the same profile as $K_{11}=2\ell-2k^2/N$, so for this family the interaction renormalizes the amplitudes of the determinantal geometry but not the geometry itself. Second, the three-resolvent sector closes in the same way.  $K_{111}\equiv0$ exactly at every $\lambda$ [Eq.~\eqref{eq:K111}], $K_{211}$ closes in piecewise Wick form to an rms of $1.1\%$, and the pair-hopping rigidity extends to it [Eq.~\eqref{eq:threept}].

The bulk vertex admits the same bookkeeping, now evaluated rather than merely structured. With $\psi_A,\lambda_A$ the free-fermion eigen-data and $W_{AB}=\braket{\psi_A|W|\psi_B}$, the Rayleigh--Schr\"odinger expansions of $\delta\lambda_A$ and $\delta\psi_A$ (self-tested against finite differences to $O(\lambda^3)$, Sec.~\ref{app:inter}) determine the resolvent blocks $R^1,R^2$ and the self-energy blocks $G^1,G^2$ in closed form, and with $|V(\lambda)|^2=|V_0|^2+\lambda^2|W|^2$ elementwise the $O(\lambda^2)$ coefficient of the bulk vertex is the explicit ensemble quantity $\Gamma_2$ of Eq.~\eqref{eq:Gamma2}. The verification succeeds where the expectation is observable.  $|\mathcal R_a|\ge c(\eta)>0$ (the imaginary part of the channel resolvent is sign-definite), so $\E[GG]$ is finite at every $\eta$, and the finite-$\lambda$ ratio $[\Gamma^{(2)}(\lambda)-\Gamma^{(2)}(0)]/\lambda^2$ is consistent with $\Gamma_2$ at $\eta\gtrsim4$, normalized overlap $0.60$ ($\eta=4$), $0.78$ ($\eta=6$), $0.85$ ($\eta=6$--$8$), while below $\eta\simeq2$--$3$ the ratio is not $\lambda^2$-consistent, dominated by the near-degenerate level-pair content that makes the overlap kernel singular to ordinary perturbation theory (Sec.~\ref{sec:bulkb}). There is therefore an important distinction between the finite-$\eta$ resolvent vertex and its eigenstate-resolved kernel representation.  A finite broadening $\eta>0$ regularizes the resolvent observable itself, but not the individual-eigenstate perturbation denominators, which the near-degenerate level pairs make singular.

\subsection{Bulk versus deep plane: the nonperturbative boundary}
\label{sec:bulkb}

\subsubsection{Bulk flow, decoupling, and the hierarchy-level statement}
\label{sec:bulk:flow}

\begin{figure*}[t]
\includegraphics[width=0.90\textwidth]{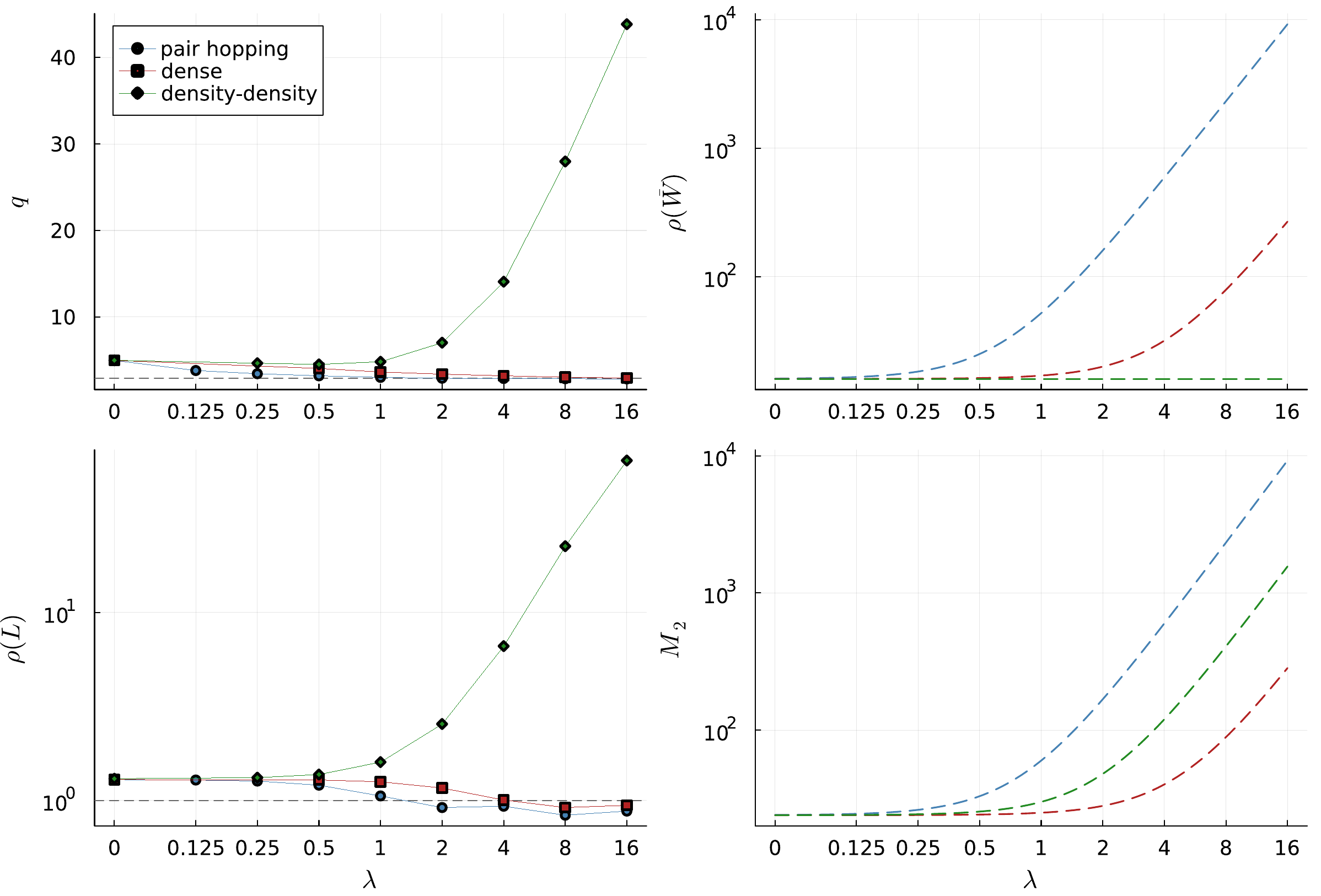}
\caption{Bulk diagnostics of the three interaction deformations at
$(8,4)$, $z=0.5+0.8i$ (pair-hopping and dense at $n_{\rm real}=400$,
density-density at $n_{\rm real}=800$, logarithmic $\lambda$-axes with
the $\lambda=0$ anchors at the left edge, full grids in
Tables~S9 and S8).
(a)~Porter--Thomas ratio $q$ (markers, Wick target $2.917$), 
pair-hopping saturates at $2.84$, dense recovers the random-matrix
crossover, density-density flows away as the overlap intensities
concentrate.  (b)~Channel-graph radius $\rho(\bar W)$ (exact laws,
lines), pair-hopping follows $\rho(\bar W_0+\lambda^2\bar W_2)$
elementwise, dense follows $k(N-k)+\lambda^2(D-1)/D$, density-density
is exactly frozen at $16$.  (c)~Ladder radius $\rho(\mathcal L)$
(markers, boundary at unity), crossings at $\lambda\in(1,2)$
(pair-hopping), $\lambda_c\simeq3.5$ (dense), and already $\lambda<1$
(density-density).  (d)~Second moment $M_2$ (exact quadratic laws, lines).}
\label{fig:bulk}
\end{figure*}

Figure~\ref{fig:bulk} records the flow of the robust bulk diagnostics for the three families (full $\lambda$ grids in Tables~S9 and S8). Four observations. (i)~The Porter--Thomas ratio flows from $q_{\rm FF}=5.0$ toward (but not to) the Wick value $3D/(D+2)=2.917$ \cite{PorterThomas}.  The pair-hopping family saturates at $2.84$ (the pure pair-hopping value), the dense family recovers the random-matrix crossover of Sec.~\ref{sec:closure:solved} ($4.94\to2.95$), and the density-density family flows away as the intensities concentrate ($q\to43.8$). (ii)~The cross-channel weight $c$ changes sign (between $\lambda=0.25$ and $0.5$ for pair-hopping, near $\lambda=2$ for the dense family) and overshoots the isotropy value $-0.028$ to $-0.043$.  The fluctuation ratio $\F$ falls from $17.0$ to $0.75$, far below the Wick value $2.03$, the pair-hopping ensemble realizes Wick-like one-point ratios without the Wick two-point structure. (iii)~The channel-graph radius follows the exact elementwise law $\rho(\bar W)=\rho(\bar W_0+\lambda^2\bar W_2)$ to all digits, and the second moments follow their exact quadratic laws. (iv)~The ladder crosses the propagation boundary $\rho(\mathcal L)=1$ between $\lambda=1$ and $2$ for pair-hopping (dense, $\lambda_c\simeq3.5$) and settles below unity, while the density-density family (whose channel graph is frozen) drives the ladder supercritical already between $\lambda=0$ and $1$, as its diagonal aligns the spectrum with the channel energies. The exact vertex decomposition of Eq.~\eqref{eq:vertexsplit} holds to $10^{-15}$ at every $\lambda$ [Sec.~\ref{app:inter}].

The contrast is complete.  The deep-plane sector measures the Hamiltonian's channel-diagonal geometry (which channels carry which energies) while the bulk sector measures the eigenstate-overlap statistics \cite{Mirlin}. The pair-hopping interaction is invisible to the former at every $\lambda$ and strongly visible to the latter (Fig.~\ref{fig:deeplaw} vs Fig.~\ref{fig:bulk}).  The pure pair-hopping reference ($H=W_{\rm 2b}$, $w_a\equiv0$) is the cleanest demonstration, its deep-plane sector vanishes identically while its bulk diagnostics are finite, the two Level-II sectors are carried by different structures of $H$, not by successive orders of a single expansion. This makes the hierarchy-level statement sharp: in the pair-hopping family $q$ approaches the Wick value to within $3\%$ while $K_{11}(\ell)$ stays exactly at the free-fermion profile $2\ell-4$. A natural order parameter for the deep-plane sector is the normalized distance to the Wick vertex,
\begin{equation}
\begin{aligned}
I_2(\lambda)&=\frac{\|\{K_{11}(\ell;\lambda)\}-\{K_{11}^{\rm iso}\}\|}{\|\{K_{11}\}\|},\\
\{K_{11}^{\rm iso}\}&=\gamma_d\Bigl[\delta_{\ell,k}-\frac{1-\delta_{\ell,k}}{D-1}\Bigr],
\end{aligned}
\label{eq:I2}
\end{equation}
over the sector-0 bins.  It is exactly constant for the pair-hopping family and flows toward the isotropy value for the dense family, where only the diagonal bin is Wick-like, the cross-distance geometry is never reached at finite $\lambda$ because $K_{11}(\ell;\lambda)-K_{11}(\ell;0)=O(\lambda^2/D)$ there.

The four-step architecture transfers to the interacting model with one simplification and one confirmation, the architecture transfers, the closure does not.  The free-fermion closure is replaced by the exact moment closure of Sec.~\ref{sec:rigidity:identity}. The simplification is that the deep-plane sector basis is now known in closed form.  The sector amplitudes $\gamma_\ell$ of the vertex ansatz are the channel-energy covariances [Eq.~\eqref{eq:ffclosed}], so Steps~1--2 of the architecture have analytic input, and the parameter system degenerates to its own projections exactly as in the free-fermion case (the dressing terms are $O(\eta^{-4})$ and the coefficient-level equation is $K_{11}=\gamma$ at any ladder order). The confirmation is that the residual diagnostic still does its job. At $\lambda=2$ (pair-hopping) the two-parameter ansatz $\{\delta_{ab},\1(\ell_{ab}=1)\}$ reproduces its two constrained bins by construction, and the unconstrained bins $\ell=2,3$ carry residuals of magnitude $1.38$, the diagnostic locates them as missing sectors, while the moment identity independently supplies their exact values ($0$ and $2$).  The conservation-fixed isotropy ansatz performs worse ($1.65$ over the cross bins), so at this $\lambda$ the vertex is organized by the $\ell$-geometry, not by isotropy, and the architecture selects the correct sector basis by its residual [Table~S10(c)]. At $(14,2)$ the same protocol reproduces the benchmark scenario exactly.  The unconstrained bin is $\ell=0$ and its residual $0.562$ is the finite-$N$ weight $-8/14=-0.571$, the Weingarten content of Sec.~\ref{sec:closure:solved}, now closed by Eq.~\eqref{eq:ffclosed}.

\subsubsection{The free-fermion spectral kernel}
\label{sec:bulk:kernel}

The bulk completion is the promotion of the $m$-bin resolution of Sec.~\ref{sec:closure:solved} from moments to spectral kernels. With $\mathrm{Cov}(m,\ell)=\E[p_A^ap_B^b\,|\,m,\ell]-1/D^2$ the benchmark's two-point overlap kernel and
\begin{equation}
\hat{\mathcal G}_2(z,z';m)
=\frac1{\mathcal N_m}\sum_{\substack{A,B\\|A\cap B|=m}}
\frac1{(z-\lambda_A)(z'-\lambda_B)}
\label{eq:G2}
\end{equation}
the exact two-eigenvalue resolvent average, the ansatz reads
\begin{equation}
\begin{aligned}
\Cov^{\rm pred}_{ab}(z,z')
&=\sum_{m=0}^{k}\mathcal N_m\,\mathrm{Cov}(m,\ell_{ab})\,\hat{\mathcal G}_2(z,z';m)\\
&\quad+\E\bigl[(\bar S_0-r^0)(\bar S_0'-r^{0\prime})\bigr],
\end{aligned}
\label{eq:kernelansatz}
\end{equation}
where $\bar S_0=\Tr[(z-H)^{-1}]/D$ is the empirical spectral Stieltjes and the final term restores the collective trace channel [Eq.~\eqref{eq:exactz}], so Eq.~\eqref{eq:kernelansatz} predicts the full covariance, which exceeds $\mathcal C^{(2)}_{ab}$ by that trace term. For random free fermions this is exact at every $\eta$, not only in the deep plane.  The $(m,\ell)$ two-point kernel theorem of Sec.~\ref{sec:benchmark} ($\E[p_A^ap_B^b]$ depends only on $m,\ell$) promotes identically to the resolvent level, and $\hat{\mathcal G}_2$ is computed per realization without any expansion (norm ratios $0.96$, $1.00$, $1.00$ at $\eta=0.8,2,4$, Fig.~\ref{fig:obstr} and Table~S10(b)). Its deep-plane limit is the moment collapse $\hat{\mathcal G}_2(m)\to\E[\lambda_A\lambda_B|m]/(z^2z'^{\,2})$, i.e.\ the $m$-bin resolution of Sec.~\ref{sec:closure:solved} and the closed form $K_{11}=2\ell-2k^2/N$, the deep-plane scalars are the leading moments of the bulk spectral functions. The two regimes are now characterized by coordinates.  In the deep plane the coefficient-level equation $K_{11}=\gamma$ is exact, the ladder is $\eta^{-4}$-inert, and a finite-dimensional degenerate parameter system closes.  In the bulk the equation is a genuine functional fixed point whose solution is carried by one spectral function $\hat{\mathcal G}_2(z,z';m)$ per $m$-sector, with the ladder relevant where $\rho(\mathcal L)>1$ and the Feshbach expansion uncontrolled in most realizations where $\mathrm{frac}_{>0.9}\approx1$ [Table~S10(a)].

\subsubsection{The nonperturbative overlap-kernel response}
\label{sec:bulk:nonpert}

\begin{figure*}[t]
\includegraphics[width=0.85\textwidth]{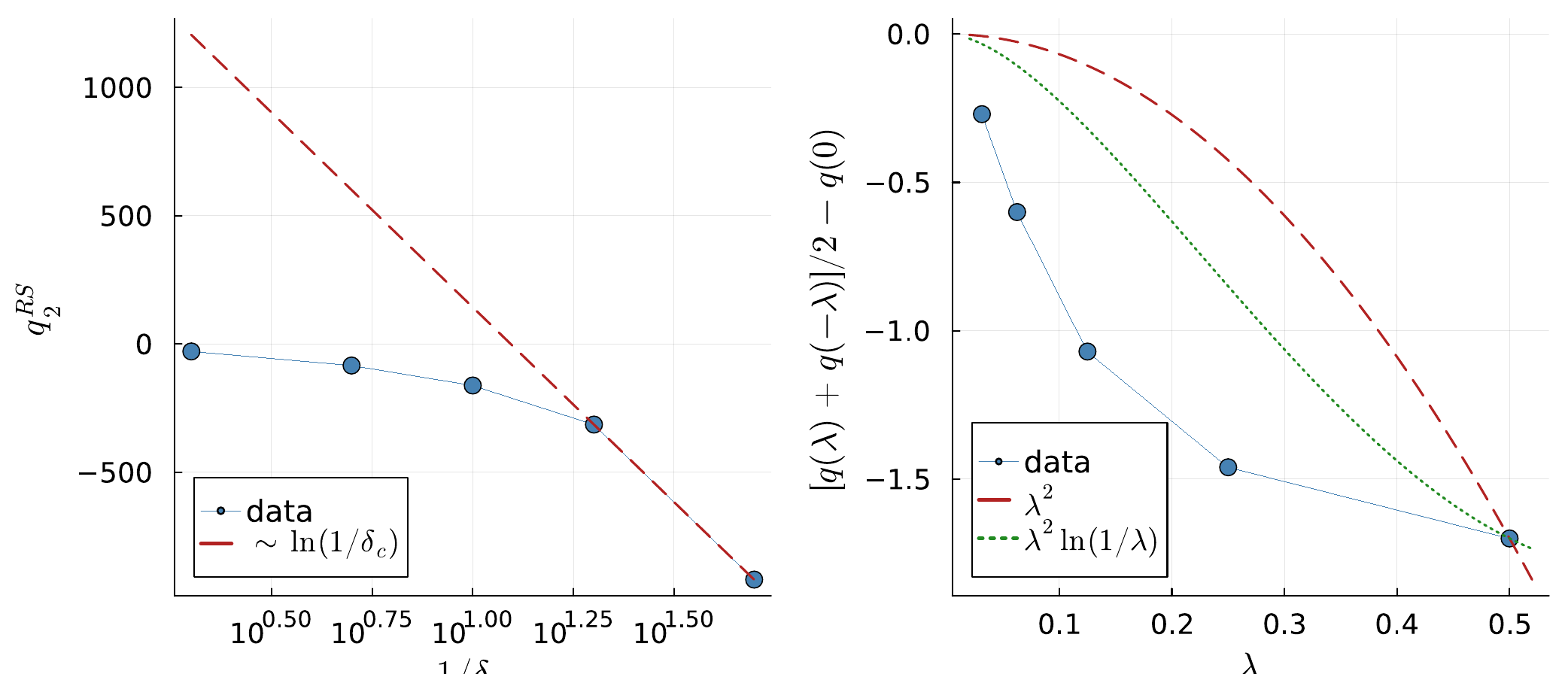}
\caption{The nonperturbative bulk obstruction at $(8,4)$.  Left: the
would-be $O(\lambda^2)$ coefficient of the overlap kernel,
$q_2^{\rm RS}$, with level pairs closer than $\delta_c$ excluded
($n_{\rm real}=400$), grows logarithmically with $1/\delta_c$
(reference line through the two smallest cutoffs).  Right.  The measured
even part $\tfrac12[q(\lambda)+q(-\lambda)]-q(0)$ ($q(0)=4.905$,
$n_{\rm real}=1600$) is reproduced by neither the $\lambda^2$ nor the
$\lambda^2\ln(1/\lambda)$ form (reference curves matched at
$\lambda=0.5$).  The failure coordinates of Sec.~\ref{sec:bulk:kernel}
and the kernel-closure ratios are in Table~S10.}
\label{fig:obstr}
\end{figure*}

For the interacting model the same ansatz requires the interacting two-point kernel $\mathrm{Cov}^{(\lambda)}(m,\ell)$ in place of the benchmark kernel.  Its leading-moment projection is fixed by the rigidity law [Eq.~\eqref{eq:rigidity}], and its bulk completion is precisely the open object. This object cannot be deformed perturbatively from $\lambda=0$.  Its would-be $O(\lambda^2)$ coefficient, evaluated by the Rayleigh--Schr\"odinger machinery of Eq.~\eqref{eq:Gamma2}, diverges, in the GOE-based benchmark considered here, the near-degenerate level pairs of the many-body spectrum give $\E[1/(\lambda_A-\lambda_B)^2]$ a logarithmic divergence, and the corresponding second-order eigenvector contributions have a logarithmically divergent second moment. The logarithm follows from the spacing distribution.  The second-order denominators contribute $\E[1/(\lambda_A-\lambda_B)^2]=\int\dd s\,P(s)/s^2$, and with the linear small-spacing repulsion $P(s)\propto s$ ($\beta=1$) of the GOE-like many-body spectrum this integral behaves as $\ln(1/\delta_c)$ under a spacing cutoff $\delta_c$, against $\delta_c^{-1}$ for uncorrelated levels ($\beta=0$) and a finite limit for $\beta>1$. The cutoff scan below is therefore the verification, not the source, of the logarithm. Fig.~\ref{fig:obstr} and Table~S18 exhibit the divergence numerically.  Excluding level pairs with spacing below a cutoff $\delta_c$ makes the coefficient grow like $\ln(1/\delta_c)$ ($-28.7\to-918.2$ as $\delta_c=0.5\to0.02$). The measured regularized response is correspondingly non-quadratic.  At $n_{\rm real}=1600$ the even part $\tfrac12[q(\lambda)+q(-\lambda)]-q(0)$ reads $-0.27,-0.60,-1.07,-1.46,-1.70$ at $\lambda=0.03125,0.0625,0.125,0.25,0.5$, dominated by the same near-degenerate pairs, and neither a $\lambda^2$ nor a $\lambda^2\ln(1/\lambda)$ fit reproduces it (the latter's residuals $\pm0.3$--$0.4$ exceed the signal at the smallest $\lambda$).  The $(m,\ell)$-resolved kernel bins show the same flat non-power response ($\Delta E\approx-(3$--$8)\times10^{-5}$ over the same range). This is the sharp form of the contrast organizing this article.  The channel-diagonal moment fields are exactly polynomial in $\lambda$ (matrix algebra, no eigen-decomposition [Eqs.~\eqref{eq:ladder}--\eqref{eq:rigidity}]), while the eigenstate-overlap observables have a singular regular-perturbative expansion at $\lambda=0$ (the deep-plane sector closes exactly because it never meets the level crossings, and the bulk kernel remains open because its deformation is singular to ordinary perturbation theory) consistent with the general distinction between regular analytic perturbation theory and the singular perturbation of eigenvectors and eigenspaces \cite{Kato}, the logarithmic divergence here being derived from the near-degenerate pairs rather than imported. In physical terms, the mixing parameter $\lambda|W_{AB}|/|\lambda_A-\lambda_B|$ is already of order unity on near-degenerate pairs at arbitrarily small $\lambda$, the many-body spacing being exponentially small in the large-system regime.  The obstruction is dense-spectrum eigenstate mixing.  The resolvent hierarchy itself is exact off shell. 

The main open problem is single.  Construct the regularized, energy-resolved interacting overlap kernel targeted by the measured response of Table~S18, $\mathrm{Cov}^{(\lambda)}(m,\ell;E,E')$. Three routes implement it. (i)~\emph{Energy-resolved stochastic flow.} View the kernel as a many-body analogue of the energy-resolved eigenfunction correlations studied in random and disordered systems \cite{Mirlin}, with a parameter-dependent stochastic or Dyson-type flow \cite{Dyson1962} as one route and field-theoretic treatments of the nonlinear sigma-model class as another \cite{Efetov}.  The logarithmic divergence is to be confirmed in unfolded-spacing scaling with growing system size. (ii)~\emph{Near-degenerate resummation.} The logarithmic sensitivity of the would-be $O(\lambda^2)$ coefficient signals that non-degenerate eigenstate perturbation theory fails uniformly in a many-body spectrum whose level spacing is exponentially small in the large-system regime.  A near-degenerate block resummation, in the sense of degenerate perturbation theory \cite{Kato}, is the corresponding mathematical step. (iii)~\emph{Higher-resolvent closure.} Replace the open interacting overlap kernel by the next level of the resolvent hierarchy itself.  Promote $\Gamma^{(2)}\to\Gamma^{(3)},\Gamma^{(4)},\ldots$, a hierarchy organized by resolvent multiplicity rather than perturbative order \cite{HOETH}, instead of imposing a random-matrix closure a priori.  The free-fermion benchmark sectors of this promotion are supplied by the replica-sector calculus of Secs.~\ref{sec:replica}--\ref{sec:kernel}. On any model with a fixed channel basis (a weakly interacting lattice fermion system included) the deep-plane rigidity law turns the sector-basis question into a diagonal-statistics computation [Eq.~\eqref{eq:rigidity}].

\subsection{The sector-resolved interaction response}
\label{sec:interact}

The open object left by the two-resolvent program of
Sec.~\ref{sec:bulk:nonpert} is the interacting two-point kernel
$\Cov^{(\lambda)}(m,\ell)=\E[p_n^ap_m^b\,|\,m,\ell]$ of the deformed
family $H=H_0+\lambda W$.  Three sector-resolved data points are added
here at the main size $(8,4)$ of Sec.~\ref{sec:benchmark}: the exact
$\lambda=0$ anchor, the sector localization of the logarithmic
divergence, and the measured sector-resolved response.

\emph{The anchor.}  At $\lambda=0$ the kernel is the two-copy sector
sum of the $d=8$ sector algebra (the $22$-class reduced algebra behind
the $q_4$ projection of Sec.~\ref{sec:q3}), and its fifteen sectors
close as exact rationals organized by three exact identities.  The
complementary-minor identity $\det U[A,C]=\pm\det
U[A^{\mathsf c},C^{\mathsf c}]$ of orthogonal matrices maps each sector
to its reflection,
\begin{equation}
\begin{aligned}
\mathcal E(m,\ell)&=E(k-m,k-\ell)\qquad(N=2k),
\end{aligned}
\label{eq:refl}
\end{equation}
visible in Table~S15 as the corner degeneracy
$\mathcal E(0,0)=\mathcal E(4,4)$ and the symmetry about $(2,2)$.  The
normalization $\sum_n p_n^a=1$ gives the exact row and column sum
rules $\sum_m\mathcal N_m\mathcal E(m,\ell)=\sum_\ell\mathcal
N_\ell\mathcal E(m,\ell)=1$ with
$\mathcal N_m=\binom{N}{m}\binom{N-m}{k-m}\binom{N-k}{k-m}$, and the
same-eigenstate slice admits the exact finite-$N$ closed form
\begin{equation}
\begin{aligned}
D^2\mathcal E(k,\ell)&=\frac{(N-k+1)(N-k+2)}{(N+1)(N+2)}\,[1+L(k,\ell)],
\end{aligned}
\label{eq:dress}
\end{equation}
with the benchmark kernel $L(k,\ell)$ of
Eq.~\eqref{eq:Lgeneral}.  At $(8,4)$ the dressing factor is $\tfrac13$:
the coincident slice sits at one third of its fixed-depth limit, and
the measured $\lambda=0$ bins reproduce Table~S15 to seven
digits.

\emph{Where the divergence lives.}  The would-be $O(\lambda^2)$
coefficient $C_2(m,\ell)$ of the kernel grows logarithmically in every
sector, $C_2(m,\ell;\delta_c)=a(m,\ell)+b(m,\ell)\ln(1/\delta_c)$,
with a slope that is strongly sector-localized: the disjoint corner and
its coincident twin carry the divergence, the first-neighbour rows are
an order of magnitude quieter and the mid-overlap row two orders of
magnitude quieter, the slope sharing the sign of the coefficient in all
fifteen sectors (full tables in Sec.~\ref{app:interact}).

\emph{The measured response.}  Against the exact anchor the regularized
response is subquadratic in every sector, with a sector-dependent
effective exponent, and the interaction lifts the anchor degeneracy
$\mathcal E(0,0)=\mathcal E(4,4)$ (fitted values in
Sec.~\ref{app:interact}).  The picture is therefore an exact
rational anchor with reflection symmetry and sum rules, a logarithmic
divergence concentrated in the two complementary corners, and a
subquadratic response whose effective exponent varies by sector.  The
anchor is exact; the finite-$\lambda$ response is a sector-resolved
diagnostic of the open problem, not a nonperturbative solution, and it
does not close the deformed kernel $K_r^{(0)}(s)\to K_r^{(\lambda)}(s)$.
What it demonstrates is that the free-fermion sector coordinates remain
a usable basis after the interaction is introduced.

This answers the three questions posed at the start of the section.
The exact sector structure fixes the whole $\lambda=0$ kernel,
including its reflection symmetry, sum rules and coincident-slice
closed form.  Beyond that, the coefficient of the deformation is not
determined by Hamiltonian moments alone: its logarithmic singularity is
the near-degenerate level pairing of Sec.~\ref{sec:bulk:nonpert}, whose
signs and weights are eigenstate resolved.  And the boundary is
testable, since the anchor and the sector-localized slope are
computable exactly while the response is measurable at fixed size.

\section{Discussion and conclusions}
\label{sec:disc}

The hierarchy thread---the replica-sector algebra, its exact assembly,
and its lift to resolvent vertices---organizes every finite-$r$
multi-resolvent correlation of random free fermions.  The benchmark
thread---the exact $r=2$ realization, its projected reconstruction, and
its interaction deformation---shows that the same sector geometry
carries the physics beyond the ETH envelope, from the moment-determined
deep plane to the eigenstate-resolved bulk.  
The boundary between what
Hamiltonian moments determine and what eigenstate information must
supply is thereby explicit for the levels and interaction classes
studied here.

The mathematical chain has a direct physical reading.  The weight
$p_n^a$ is the spectral weight of eigenstate $n$ in channel $a$, the
resolvent $\mathcal R_a$ its channel-resolved local density of states,
and $q_r$ its generalized participation moment.  The sector covariances
$\Cov_r$ are connected intensity correlations between channels and
eigenstates, $S^{(r)}$ their energy-resolved spectrum, and the Feshbach
vertices $\Gamma^{(r)}$ the irreducible hybridization fluctuations
behind them.  In a general system the sector coordinate $|A_i\cap
A_j|$ generalizes into connectivity invariants
$\mathcal I_r(a_1,\dots,a_r)$, and the moment--cumulant (M\"obius)
inversion keeps the partition-lattice structure of free probability
\cite{Lehner2004,CollinsSniady}.  The fluctuation hierarchy
is related in spirit to the Schwinger--Dyson and loop-equation
hierarchies \cite{MingoSpeicher}.
Alongside the spectral line $\mathcal K_r\to\mathcal C^{(r)}\to
\Gamma^{(r)}$, the same exact average carries the group-theoretic line
of Eq.~\eqref{eq:zonalK}: sector histogram
$T^{(r)}_c\to$ spherical transform $\mathcal S^{(r)}_\lambda\to$ zonal
kernel.  The two lines meet at $\mathcal K_r$, the group line
organizing its finite-$N$ content and the spectral line its physical
interpretation: higher-order ETH correlations are the real-axis
boundary values of the off-shell multi-resolvent fluctuation object.

The open problems and the routes toward them are stated in Sec.~\ref{sec:bulk:nonpert}.  The finite-size interaction results are exact for the centered Gaussian deformation families, and the decisive further test is a genuinely interacting model without the determinantal geometry---a chaotic spin chain or the embedded two-body ensembles (EGOE(2)) \cite{KotaBook,KotaTBRE}---while determining the hierarchy from a finite-dimensional constrained ansatz, one level removed from the construction here, is a separate problem whose first instance is Sec.~\ref{sec:closure}.  Natural continuations are a field-theoretic formulation, the unitary and symplectic classes \cite{CollinsSniady}, Krylov moment data \cite{Haydock1972,Haydock1980,Nandy2025}, and multi-time correlations \cite{Pappalardi2022,Pappalardi2025}.

\begin{acknowledgments}
This work is supported by the National Natural Science Foundation of
China under Grant No. 12305035.

All exact sector tables, closed forms and verification codes are
provided in the Zenodo reproducibility package of Ref.~\cite{CodeData}.  The script that produces or verifies each result is indexed in Table~\ref{tab:index}.
\end{acknowledgments}

\appendix

\section{Exact algebra and proofs}
\label{app:algebra}

The exact algebraic material is collected here, and Table~\ref{tab:exactmap} summarizes the scope.  Numerical tables marked S$k$ are the reproduction tables collected in Sec.~\ref{app:supp}.

\begin{table*}[htbp]
\caption{Scope and exactness map of the main objects of this article.}
\label{tab:exactmap}
\small
\begin{ruledtabular}
\begin{tabular}{ll}
object & status\\
\hline
replica-sector decomposition $\mathcal K_r$ (Sec.~\ref{sec:replica}) & exact, all finite $N$ and $r$\\
O($N$) Weingarten weights $w_c^{(r,k)}$ (Sec.~\ref{sec:weingarten}) & exact finite-$N$ closed forms\\
occupancy/transfer assembly (Sec.~\ref{sec:assembly}) & exact and constructive; realized at $r=3$ ($k=2,3,4$) and $r=4,5$ ($k=2$)\\
three-point kernel, $r=3$ (Sec.~\ref{sec:kernel}) & full finite-$N$ arrays ($16\times16$, $39\times39$, $81\times81$)\\
four-point content, $r=4,k=2$ (Sec.~\ref{sec:kernel4}) & representative channel-pattern slices\\
five-point diagonal, $r=5,k=2$ (Sec.~\ref{sec:kernel}) & complete diagonal sector table\\
three-resolvent cumulant $\mathcal C^{(3)}$ (Sec.~\ref{sec:vertices}) & exact sector representation; spectral dictionary exact\\
four-resolvent cumulant $\mathcal C^{(4)}$ (Sec.~\ref{sec:kernel4}) & exact deep-plane expansion; representative slices\\
Feshbach vertices $\Gamma^{(r)}$ (Sec.~\ref{sec:vertices}) & exact Feshbach relation; explicit leading deep-plane coefficients\\
interacting bulk kernel (Secs.~\ref{sec:rigidity}--\ref{sec:bulkb}) & open; exact $\lambda=0$ anchor and sector-localized response\\
thermodynamic limit & not established\\
\end{tabular}
\end{ruledtabular}
\end{table*}

\subsection{Properties of the multi-resolvent hierarchy}

the cumulants are the M\"obius coefficients of the standard
moment--cumulant inversion over the partition lattice $\Pi_r$ of the
$r$ resolvent arguments,
$\E[\delta\mathcal R_{a_1}\cdots\delta\mathcal R_{a_r}]
=\sum_{\pi\in\Pi_r}\prod_{B\in\pi}\mathcal C^{(|B|)}_{B}$
(blocks of size one vanish), so the hierarchy is graded by the number
of simultaneously correlated resolvent channels rather than by a
perturbative order.  The construction thereby separates three
levels, the raw moments $\mathcal M^{(r)}=\E[\delta\mathcal
R_{a_1}\cdots\delta\mathcal R_{a_r}]$, the connected content
$\mathcal C^{(r)}$, and the Feshbach-irreducible vertices
$\Gamma^{(r)}$ of Sec.~\ref{sec:gamma3}, related by two different
operations, the M\"obius inversion above and the Schur dressing
below, so that $\mathcal M^{(r)}\ne\mathcal
C^{(r)}\ne\Gamma^{(r)}$ in general.  With the arguments centered
($\E[\delta\mathcal R_a]=0$), $\mathcal M^{(r)}=\mathcal C^{(r)}$ for
$r\le3$, and the first place the two levels differ is
\begin{equation}
\mathcal M^{(4)}_{a_1a_2a_3a_4}=\mathcal C^{(4)}_{a_1a_2a_3a_4} +\mathcal C^{(2)}_{a_1a_2}\mathcal C^{(2)}_{a_3a_4}
+\mathcal C^{(2)}_{a_1a_3}\mathcal C^{(2)}_{a_2a_4}
+\mathcal C^{(2)}_{a_1a_4}\mathcal C^{(2)}_{a_2a_3},
\label{eq:M4}
\end{equation}
the three pairings of two-point cumulants being the first reducible
content.  Nor is the hierarchy a derivative hierarchy in
$z$.  The operator resolvent obeys $(z-H)^{-2}=-\partial_z(z-H)^{-1}$,
but the channel-resolved objects are projected quantities for which no
such reduction exists, $[\mathcal R_a(z)]^2=\sum_{nm}p_n^ap_m^a/
[(z-\lambda_n)(z-\lambda_m)]\ne\sum_np_n^a/(z-\lambda_n)^2=
-\partial_z\mathcal R_a(z)$, so the coincident-argument correlations
are genuinely new information: a correlation hierarchy between
different random channel weights rather than a Taylor or
energy-derivative hierarchy of a single resolvent.  In physical
terms the hierarchy measures the correlations of fluctuations of
channel-resolved spectral functions.  $\mathcal C^{(2)}$ is the
spectral-weight covariance of two channels, $\mathcal C^{(3)}$ the
connected three-channel spectral-weight correlation, and so on.  The
two-point
level of this hierarchy is the covariance governed by the ladder
equation $\mathcal C^{(2)}=\bar{\mathcal R}^2\bar{\mathcal R}^2[(1-
\mathcal L)^{-1}\Gamma^{(2)}+\cdots]$ with $\mathcal
L_{ab,cd}=\bar{\mathcal R}_a^2\bar{\mathcal R}_b^2|V_{ac}|^2|V_{bd}|^2$
and the self-energy vertices $\Gamma^{(2)},\Gamma^{(3)},
\ldots$ of the Feshbach expansion
\cite{Feshbach1958,HOETH} and Sec.~\ref{sec:closure:arch}.  The
$r$-resolvent cumulants $\mathcal C^{(r)}$ and the self-energy
cumulants $\Gamma^{(r)}$ provide two related representations of the
same fluctuation sector, connected by the exact geometric ladder
dressing [Eqs.~\eqref{eq:ladder3} and \eqref{eq:laddergen} below].  Throughout,
$\Gamma^{(r)}$ denotes the cumulant of the self-energy fluctuation
$\delta\cG$ in the Feshbach representation.  It plays the role of an
irreducible vertex with respect to the geometric ladder dressing, we
use the term irreducible because these cumulants enter the expansion
with respect to the geometric resolvent dressing, and they should not
be confused with the one-particle-irreducible vertices defined through
a Legendre effective action, with the free cumulants of free
probability, nor with the M\"obius (partition-lattice) irreducibility
of the moment--cumulant inversion above.  Schur irreducibility and
partition-lattice irreducibility are distinct operations.  Throughout,
``irreducible'' refers
exclusively to irreducibility with respect to this Feshbach geometric
dressing, not to one-particle irreducibility in the
Legendre-field-theoretic sense.

\subsection{The orthogonal Weingarten sector tables}
\label{app:wgtab}

The eleven coset classes $\mu\vdash6$ have, relative to any fixed
reference pairing $\pi_0$, the shell sizes
$|M_\mu|=\#\{\rho\in M_{12}:\mathrm{type}(\pi_0,\rho)=\mu\}$,
independent of $\pi_0$ and summing to $11!!=10395$.
$(1^6)\!:1$, $(2,1^4)\!:30$, $(2^2,1^2)\!:180$, $(2^3)\!:120$,
$(3,1^3)\!:160$, $(3,2,1)\!:960$, $(3^2)\!:640$, $(4,1^2)\!:720$,
$(4,2)\!:1440$, $(5,1)\!:2304$, $(6)\!:3840$.  The Weingarten values
$w_\mu(N)$ are the rational functions of Eq.~\eqref{eq:reduced}, each
of the form $\tilde n_\mu(N)/\mathfrak D_6(N)$ with the common
denominator
\begin{equation}
\mathfrak D_6(N)=N^2\prod_{j\in\{5,4,3,2,1,-1,-2,-3,-4,-6,-8,-10\}}(N-j),
\label{eq:D6}
\end{equation}
In lowest terms every denominator is a product of the linear factors
of $\mathfrak D_6$, with numerator degree at most eight.  All eleven
values satisfy the sphere identity
$\sum_\mu|M_\mu|w_\mu=1/[N(N+2)(N+4)(N+6)(N+8)(N+10)]$ exactly, and the
$d=3,4$ reductions against the full Gram inversions are exact.  At
$d=8$ ($2{,}027{,}025$ pairings, $22$ coset classes) the weights are
$w_\mu(N)=\tilde n_\mu(N)/\mathfrak D_8(N)$ with
$\mathfrak D_8(N)=N^2(N-1)^2\prod_{j\in S_8}(N-j)$,
$S_8=\{7,6,5,4,3,2,-1,-2,-3,-4,-5,-6,-8,-10,-12,-14\}$, and numerators
of degree at most $12$; all twenty-two satisfy
$\sum_\mu|M_\mu|w_\mu=1/[N(N+2)\cdots(N+14)]$ exactly, as verified in
integer arithmetic on the complete table.  The diagonal sector sum
reproduces $\E[p^4]=q_4/D^4$ at $N=8,14,30$ exactly, and the same
weights close the full two-copy kernel $\mathcal E(m,\ell)$, fifteen
exact rationals at $(8,4)$ with the complementary-minor reflection
$\mathcal E(m,\ell)=\mathcal E(4-m,4-\ell)$ and the finite-$N$ dressing
of Eq.~\eqref{eq:dress}, the $\lambda=0$ anchor of the interacting
kernel of Sec.~\ref{sec:interact}.  At $d=9$ ($34{,}459{,}425$
streamed pairings, thirty coset classes) the structure is the same,
with $\mathfrak D_9$ of Eq.~\eqref{eq:D9}, numerator degree at most
$14$, and all thirty satisfying
$\sum_\mu|M_\mu|w_\mu=1/[N(N+2)\cdots(N+16)]$ exactly.  The individual
rational functions of the three tables are reproduction data: they are
collected as Tables~S11, S12 and S13 of Sec.~\ref{app:supp} and
archived as the per-depth files \texttt{data/wg\_depth6.json},
\texttt{data/wg\_depth8.json} and \texttt{data/wg\_depth9.json} of
Ref.~\cite{CodeData}.

The machine-readable data of this appendix are distributed in the Zenodo package of Ref.~\cite{CodeData}, whose layout is described in Sec.~\ref{app:models}: the portable tables \texttt{data/C1\_data.json} (with all integers emitted as strings for arbitrary precision), the Julia-native \texttt{data/C1\_data.jls}, and the per-depth tables \texttt{data/wg\_depth6.json}, \texttt{data/wg\_depth8.json} and \texttt{data/wg\_depth9.json}.  The per-depth tables are regenerated
from the printed LaTeX by the build script listed in Table~\ref{tab:index}, which
verifies the sphere identity
$\sum_\mu|M_\mu|w_\mu(N)=1/\prod_{i=0}^{d-1}(N+2i)$ in integer arithmetic at several $N$, and which repairs two typographical defects
of the earlier printing (a dropped factor $N$ in the constant part of
the $w_{1^9}$ numerator, printed here in its verified form, and a
missing plus sign in the $w_{5,2,1^2}$ numerator); a load-back script reconstructs every table of the paper exactly
in integer arithmetic from the package.

\subsubsection{Zonal tables and the depth-twelve cross-checks}

The zonal spherical tables behind Eq.~\eqref{eq:zonal} are computed
independently from the Murnaghan--Nakayama rule and verified against
the reduced engine at every class, with exact rational agreement at
all tested $N$.  The $d=4$ table (rows $\lambda$, columns $\mu\vdash4$ in the order $4,31,2^2,21^2,1^4$) is archived as Table~S14 of Sec.~\ref{app:supp}.
  At depth twelve Eq.~\eqref{eq:zonal} produces the
full-cycle class
\begin{equation}
\Wg^{O(N)}(12)=\frac{1}{24!}\sum_{\lambda\vdash 12}
\frac{f^{2\lambda}\,C_\lambda\,z_{24}\,\tilde J_\lambda[12]}{Z_\lambda(1^N)},
C_\lambda=\frac{Z_\lambda(1^{25})}{\tilde J_\lambda(1^{25})},
\label{eq:zonal12}
\end{equation}
with $z_{24}=24$ and the monomial-normalized Jack polynomials
$\tilde J_\lambda$ (leading monomial coefficient one).  The 77 weights
pass three independent checks.  (i) the identical route reproduces the
depth-eight table above exactly.  (ii) the sphere identity
$\sum_\mu|M_\mu|w_\mu(N)=1/\prod_{i=0}^{11}(N+2i)$ holds exactly at
$N=13,15,17$.  (iii) at $N\to\infty$ every weight reproduces the
signed-Catalan asymptotics of Eq.~\eqref{eq:cfree} in all 77
classes.  As an independent, engine-free cross-check, the full-cycle
weight $w_{12}(N)$ reproduces the closed form of Ref.~\cite{CM2009}
(their Eq.~(6.3), with the $\omega$-normalization
$1/[2^{d-1}(d-1)!]$ made explicit)
\begin{equation}
\begin{aligned}
\Wg^{O(N)}(12)
&=\frac{1}{23!!}\sum_{\substack{\lambda\vdash 12\\ \lambda_3\le 1}}
\frac{f^{2\lambda}}{2^{11}\,11!\,N}\\
&\quad\times\prod_{\substack{(i,j)\in\lambda\\ (i,j)\neq(1,1)}}
\frac{2j-i-1}{N+2j-i-1},
\end{aligned}
\label{eq:fcw}
\end{equation}
exactly in integer arithmetic at $N=13,15,17,19,21$.  The same
comparison at depths $4,6,8,9$ against the reduced engine fixes the
normalization exactly.

\subsubsection{The sector table at $k=3$}

The sector algebra is independent of $k$.  At $k=3$ only the slot
lists change (the valid pairings now run over $\sigma_r,\tau_r\in
S_3$ per channel, eighteen slots in all) and the kernel is built
from the $d=9$ sector algebra, whose thirty coset classes are the
partitions of nine.  The depth-nine table is computed by streaming
the $17!!$ pairings of $M_{18}$.  The reduced Gram
closes with the common denominator
\begin{equation}
\mathfrak D_9=N^2(N-1)^2(N+2)^2\prod_{j\in S}(N-j),
\label{eq:D9}
\end{equation}
where $S$ collects the even negatives $-4,-6,\dots,-16$, the odd
negatives $-1,-3,-5$, and the positives $2,3,4,5,6,7,8$.  The degree
is $23$, and the thirty weights are collected
in the machine-readable data of Sec.~\ref{app:wgtab}.  The sphere identity
$\sum_\mu|M_\mu|w_\mu=1/\prod_{i=0}^8(N+2i)$ holds exactly in
rational arithmetic at every $N$ tested.  The $k=3$ triple geometry
has
thirty-nine patterns per side (the additional overlap values
$\ell=3$ and $t=3$ open the coincident sectors), and the full
$39\times39$ array follows from the same sector sum, evaluated by
a streaming classification of the pairing pairs (twenty-four
threads, checkpointed).  At $(14,3)$ the diagonal sectors read
\begin{equation}
\begin{aligned}
\E[p_1p_2p_3]\big|_{3333}&=\frac{5}{2376192},\\
\E[p_1p_2p_3]\big|_{2222}&=\frac{3}{31116800},\\
\E[p_1p_2p_3]\big|_{1111}&=\frac{26863}{917447731200},\\
\E[p_1p_2p_3]\big|_{0000}&=\frac{152687}{4994993203200},
\end{aligned}
\label{eq:k3diag}
\end{equation}
and the four consistency identities of this section anchor the
table.  The Selberg identity gives the fully coincident sector
$\E[p^3]=q_3/D^3$ with the $k=3$ value $q_3=41405/408$ at
$N=14$, reproduced by the computed sector exactly
($5/2376192$), and independently by direct Monte Carlo at $(14,3)$
to three significant digits.  The two-point
marginal $\sum_{C_3}\E[p_1p_2p_3]=\E[p_1p_2]=q_2/D^2=1/20384$ closes
over the coincident row.  Covariance conservation
$\sum_{C_3}\E[\delta p_1\delta p_2\delta p_3]=0$, with the two-point
terms supplied by the $k=3$ two-point kernel of the same $d=6$
machinery, holds exactly.  And the table is invariant under replica
and transpose exchange.  The sector theorem therefore holds at
$k=3$ with the identical algebra, only the valid-pairing lists
growing, as anticipated.

The large-$N$ limit of the same kernel closes in closed form
for every determinant depth, as the second general statement of the
construction:
\begin{quote}
\noindent\emph{Proposition (large-$N$ identity-sector limit).}
For the Gaussian one-body ensemble, with $k$ and $r$ fixed and the
overlap geometry fixed, the scaled $r$-point kernel
converges,
$F_r(A_1,\dots,A_r;C_1,\dots,C_r):=\lim_{N\to\infty}N^{kr}\,
\E[p_1\cdots p_r]=\sum_{\pi\in M_{2kr}}R(\pi;A)\,R(\pi;C)$, with
$R(\pi;A)$ the row-side determinant sum defined below.
\end{quote}
\noindent\emph{Proof.}  In the pairing basis the Gram matrix obeys
$G(\pi,\pi)=N^{kr}$ and $G(\pi,\pi')\le N^{kr-1}$ for $\pi\ne\pi'$
(distinct pairings merge at least one union component), so the
association-scheme inverse obeys
$\Wg(\pi^{-1}\pi')=N^{-kr}\delta_{\pi,\pi'}+O(N^{-(kr+1)})$.  Only the
identity class survives at the leading order $N^{-kr}$.  Inserting the
leading term into Eq.~\eqref{eq:weingarten} leaves the free double sum
\begin{align}
F_r(A_1,\dots,A_r;C_1,\dots,C_r)
:=\lim_{N\to\infty}N^{kr}\,\E[p_1\cdots p_r]\notag\\
=\sum_{\pi\in M_{2kr}}R(\pi;A)\,R(\pi;C),
\label{eq:freer}
\end{align}
where $R(\pi;A)=\sum_{\sigma_t\in S_k}\mathrm{sgn}(\prod_t\sigma_t)
\prod_{\alpha}\delta(A_{t}[\,\cdot\,],A_{\pi(\alpha)}[\,\cdot\,])$
is the row-side determinant sum of the pairing, the $t$-sum running
over the $r$ determinants.  The subleading coset classes carry the
signed-Catalan weights
\begin{equation}
\lim_{N\to\infty}N^{d+|\mu|-\ell(\mu)}\,\Wg_\mu(N)
=\prod_{j\in\mu}(-1)^{j-1}\mathrm{Cat}_{j-1},
\label{eq:cfree}
\end{equation}
the product of signed Catalan numbers over the class parts, extracted
from the exact $d=4,6,8,9$ tables (all 68 classes).  Here $\mu\vdash d$
is a partition of $d=kr$, so $|\mu|=d$ and the exponent is
$d+|\mu|-\ell(\mu)=2d-\ell(\mu)$, the standard orthogonal Weingarten
scaling.  Equivalently,
\begin{multline}
\Wg_\mu(N)
=N^{-kr-\Delta(\mu)}
\Bigl[\prod_{j\in\mu}(-1)^{j-1}\mathrm{Cat}_{j-1}\Bigr]
\bigl[1+o(1)\bigr],\\
\Delta(\mu)=kr-\ell(\mu),
\label{eq:Wgdefect}
\end{multline}
the suppression graded by the coset defect $\Delta$.  The identity
class ($\Delta=0$) is the leading layer, $\Delta=1$ the first
contraction layer, $\Delta=2$ the second, and so on, one extra
inverse power per missing part.  These are two distinct gradings of
the hierarchy.  The resolvent multiplicity $r$ fixes which order of
spectral correlation is being resolved, whereas $\Delta$ ranks the
finite-$N$ suppression of the contraction sectors at that fixed
order.  The $\Delta=0$ layer of Eq.~\eqref{eq:Wgdefect} is the
identity sector behind the Gaussian-overlap limit of the Proposition
above, and the $\Delta\ge1$ layers are its finite-$N$ corrections.  The
signed-Catalan weights of the subleading classes are not part of the
Gaussian-overlap limit itself.
Equation~\eqref{eq:freer} reproduces the
$N\to\infty$ limits of every exact table in hand.  The two-point kernel
in all $9$ entries (the coincident corner
$F_2=q_2(2,\infty)\,2^2=24$), the three-point kernel in all $256$
$k=2$ entries (the coincident corner $F_3=720=q_3(2,\infty)\,2^3$), the
$k=3$ three-point table in every entry whose label blocks are small
enough to enumerate (zero mismatches, the corner
$75600=q_3(3,\infty)\,6^3$ from the closed form), and the four-point
$k=2$ table with zero mismatches (the corner
$F_4=q_4(2,\infty)\,2^4=2520\cdot16=40320$ from the closed form).  At
$k=4$ the three-point anchors follow.
$F=13824=(4!)^3$ for three pairwise disjoint channels (independent
minors) and $F=14515200=q_3(4,\infty)\,(4!)^3=1050\cdot24^3$ at the
fully coincident corner, with the two-point anchor
$q_2(4,\infty)(4!)^2=8640$ from the same limit.  With the depth-twelve
zonal expansion of Sec.~\ref{sec:weingarten} the finite-$N$ $k=4$
three-point kernel is assembled by the identical sector sum.  The
$81\times81$ pattern array (the three-set-four overlap classes
$(m_{12},m_{13},m_{23},t)$ per side) is built from the $77$
depth-twelve Weingarten weights of Eq.~\eqref{eq:zonal12} as
$\E[p_1p_2p_3]=\sum_c T_c\,w_c(N)$, the sector table
$T_c=\sum_{\pi,\pi'}R[\pi]C[\pi']\,\delta_{\mathrm{type}(\pi\pi'),c}$
carried by the signed support maps
$R[\pi]=\sum_{\sigma\in S_4^3}\mathrm{sgn}(\sigma)\,\delta_\Delta(\pi;\sigma)$,
with $\delta_\Delta(\pi;\sigma)$ the indicator that the
$\sigma$-assigned row labels coincide within every pair of $\pi$.
The anchors of the $k=4$ table hold exactly.  The fully coincident
corner follows the Selberg closed form $\E[p^3]=q_3(4,N)/C(N,4)^3$
($1/2212210$ at $N=13$, the identical sector sum reproduces the
Selberg corners at $k=2,3$ exactly), with the transpose
symmetry $T_c(pa,pb)=T_c(pb,pa)$ exact on
every computed sector.  The $N\to\infty$ limit of the table closes
completely.  The free kernel of Eq.~\eqref{eq:freer} equals the
identity-class sector, $F(pa,pb)=T_{[1^{12}]}$, and this identity now
fills the whole $81\times81$ free table (including the large-label
blocks left open by the direct enumeration) through the signed-support
dot product $F=\sum_\pi R[\pi]C[\pi]$, reproducing the free anchors
$F=13824=(4!)^3$ (disjoint), $F=14515200=q_3(4,\infty)(4!)^3$
(coincident), and $F=207360=q_2(4,\infty)(4!)^3$ (two-shared), and the
free limit of every finite-$N$ sector computed.  The machine-readable
tables are included in the Zenodo package of
Ref.~\cite{CodeData} (\texttt{kernel4pt3\_data.jls}).

Beyond the free table, the finite-$N$ sectors that the support maps cannot
reach directly (the $t=3,4$ pairs of the three-set-four classes,
whose direct signed supports are intractably large) are closed
by the occupancy-matrix algebra.  The slot-incidence matrix
factorizes as $m(\sigma,\tau)=M_1(\sigma)+M_2(\tau)$, where $M_1$ is fixed
by the $\sigma$-slots and $M_2$ by the $\tau$-slots, so the signed census
$A_m=\sum_{\sigma,\tau}\mathrm{sgn}(\sigma\tau)\,\delta_{m,M_1(\sigma)+M_2(\tau)}$
is the convolution of two $13824$-term histograms (packed as three-bit-lane
integers, mirror symmetry removing half of the pairs), and the sector table
is $T_c=\sum_m A_m\,N_c(m)$ with
$N_c(m)=\#\{(\pi,\pi'):\mathrm{type}(\pi\pi')=c\}$.  Each matrix then splits
over the connected components of its support graph, and on a connected block
the counts follow from a finite-state transfer matrix.  Processing the block
row by row, the state is the multiset of open alternating chains
$\{(c_1,c_2,a)\}$ (the open $\pi'$-tails at columns $c_1,c_2$ of a chain
holding $a$ pairs) and
$\mathcal F_{r+1}(s')=\sum_{s\to s'}w(s\to s')\,\mathcal F_r(s)$,
where the transitions open, extend, merge or close chains, a closure of age
$a$ emitting a cycle of part $a+1$, and the cycle-type generating function
travels with the transfer.  The DP reproduces brute-force pairings
exactly in integer arithmetic on the largest brute-force blocks, and the
full assembly passes the transpose, free-limit and $\sum_c T_c=0$ gates.
The computationally demanding block then reproduces the Selberg anchor through the engine
itself, $\E[p^3](4444,4444)=q_3(4,N)/C(N,4)^3$ exactly at $N=13,15,17$
($1/2212210$, $1/11992344$, $1/52581816$), and the remaining
computationally demanding diagonal
sectors follow in closed rational form at $N=13$.
$\E(3333,3333)=1/53093040$ and
$\E(3343,3343)=\E(3433,3433)=\E(4333,4333)=1/16591575$
(the three $S_3$-permuted members of the $[3,3,4,3]$ family degenerate, as
they must), with free limits
$F=14515200$, $368640$ and $1451520$ respectively.

The same construction lifts the hierarchy one rung further.  As a
higher-multiplicity realization we apply it to $r=5$, $k=2$, where the
$k=2$ five-point kernel is assembled on the depth-ten zonal weights
(\texttt{data/wg10.json}, generated by the identical Jack route and
verified by the sphere identity, the signed-Catalan free limit in all
$42$ classes, and the full-cycle closed form of
Eq.~\eqref{eq:fcw} at depths $n=10$).  The complete diagonal sector
table, all $1750$ five-set overlap classes (the $S_5$-orbits of the
five-$2$-set Venn structures, the kernel being replica-symmetric
across each orbit), is computed by the occupancy assembly of
Eqs.~\eqref{eq:censusgen}--\eqref{eq:chaindp}: the $32\times32$ signed census, the component decomposition and the chain transfer matrix,
whose row/column pairing lists extend to the ten-slot cells that the
fully coincident classes produce.  The occupancy route reproduces the
Weingarten sector engine exactly in integer arithmetic on the disjoint and
four-coincident-plus-one-disjoint classes, the conservation gate
$\sum_c T_c=0$ and the replica symmetry hold class by class, the
corner follows the Selberg form
$\E[p^5]=q_5(2,N)/C(N,2)^5$ exactly ($1/92378$ at $N=11$, $1/352716$
at $N=13$, $1/1144066$ at $N=15$), the free anchors close as
$F=32=(2!)^5$ (disjoint), $F=80640=q_4(2,\infty)2^5$
(four-coincident-plus-one-disjoint), $F=241920$
(four-coincident-plus-one-sharing) and $F=3628800=q_5(2,\infty)2^5$
(coincident), and the adjacent diagonal sectors record the finite-$N$
values $\E=1/1100385$ and $49/37413090$ at $N=11$, an explicit
finite-$N$ example that disjoint supports do not imply factorization
of the corresponding principal-minor moments,
$\E[x^8y^2]/\E[x^8]\E[y^2]=1.215$ for the disjoint pair at $(2,11)$,
the sector-level echo of the independent-factorization failure.

\subsubsection{The $r\to r-1$ hierarchy: sector tables and the closed chain}
\label{sec:marginal}

The levels of the hierarchy are not independent.  Each is the exact
marginal of the next.  The overlap matrix
$p_a(n)=|\det U[A,C]|^2$ is doubly stochastic, so summing the last
channel collapses the $r$-point kernel,
\begin{equation}
\sum_{a_r}\mathcal K_r(A;C)
=\mathcal K_{r-1}(A_1,\dots,A_{r-1};C_1,\dots,C_{r-1}),
\label{eq:marginal}
\end{equation}
which holds for every last mode set $C_r$, because
$\sum_{a_r}p_{a_r}(n_r)=1$ per realization.  The marginal is exact at
every $N$.  At the sector level the channel sum decomposes into the
pattern classes of $(C_1,\dots,C_r)$ extending $P_{C\setminus r}$,
with their pattern multiplicities,
\begin{equation}
\sum_{P_C\supset P_{C\setminus r}}\mathrm{mult}(P_C)\,
\mathcal K_r(P_A,P_C)
=\mathcal K_{r-1}(P_{A\setminus r},P_{C\setminus r}),
\label{eq:margsec}
\end{equation}
for every $N$ and every full eigenstate pattern $P_A\supset P_{A\setminus r}$ (the left-hand side being independent of the extension), with $\mathcal K_r$ read through its sector expansion
Eq.~\eqref{eq:sectgen}.  The marginalization is an identity of the
kernel functions.  It is not an identity of the raw coefficient
vectors $T^{(r)}_c$, the depth-$2kr$ and depth-$2k(r-1)$ Weingarten
bases are distinct algebras, and the recursion is carried by the
weight functions $w^{(r)}_c(N)$.  Equation~\eqref{eq:marginal} is verified
exactly in integer arithmetic where the closed forms exist.  At $r=5$ the four coincident
channels with the coincident states sum over the last channel to
$1/24310=q_4(2,11)/D^4$ (the three last-channel classes weighted
$36+18+1$ by their exact sectors, and the same value with the last
state replaced by the disjoint pair $\{3,4\}$, confirming the
$C_r$-independence).  At $r=4$ the analogous sum closes to
$1/18564=q_3(2,14)/D^3$ with the same independence.  The recursion
together with the doubly stochastic identity therefore provides an
exact consistency hierarchy linking all $r$.  Every higher-rank kernel
is constrained by the marginal identities of the level below, and its
connected counterpart is the cumulant
expansion whose sector form is the $\sum_c T_c=0$ conservation gate of
the tables above.

The same bookkeeping answers the natural complexity question.  The
naive pairing sum runs over $(2kr-1)!!$ pairings, while the compressed
assembly organizes the computation into the pattern classes, the
signed occupancy census of $k!^{\,r}\times k!^{\,r}$ terms (packed as
integer-lane histograms, mirror symmetry removing half), and the
component-wise chain transfer matrix.  The sector invariance is
general.  The assembly realizes it constructively at any finite
multiplicity, the explicit realization cost being controlled by the
growth of the pattern and sector spaces, not by the pairing
enumeration.  At
$(r,k)=(3,4)$ the connected blocks take hours, and the diagonal
five-point table minutes, on the same transfer matrix.  In the
division of labor between the two sections, Sec.~\ref{sec:weingarten}
determines the universal group-theoretic weight layer, the O($N$)
Weingarten weights and the zonal machinery, shared by every ensemble
with Haar-distributed overlap geometry, while Sec.~\ref{sec:kernel}
determines the model-dependent occupancy layer, the channel-pattern
census specific to the Slater determinantal ensemble, evaluated in
full at $r=3$.

\subsection{The three-point kernel: sector tables}
\label{app:kernel}

Table~S16 lists the exact $D^3\Cov_3$ sector table at
$N=14$ (the $16\times16$ array of Eq.~\eqref{eq:cov3}, row index
$(m_{12},m_{13},m_{23},t)$, column index
$(\ell_{12},\ell_{13},\ell_{23},t_\ell)$), and Table~S17
the $N\to\infty$ (Gaussian-overlap) limits.  The closed forms of
all $256$ sectors are rational functions of $N$ with denominators
composed of factors $(N+j)$, $j\in\{-10,\dots,5\}$, times $N^2$
(the data files of Ref.~\cite{CodeData} contain the full table).  The
representative entries of Eq.~\eqref{eq:diag3} are the diagonal
sectors of Table~S16.

\subsubsection{The zonal expansion}

The solution of Eq.~\eqref{eq:reduced} also has the Fourier-type closed
form of Collins--Matsumoto \cite{CM2009}.  For every class
$\mu\vdash d$,
\begin{equation}
\Wg^{O(N)}(\mu)
=\frac{1}{(2d-1)!!}\sum_{\lambda\vdash d}
\frac{f^{2\lambda}\,\omega^{\lambda}_\mu}{Z_\lambda(1^N)},
\label{eq:zonal}
\end{equation}
where $Z_\lambda(1^N)=\prod_{(i,j)\in\lambda}(N+2j-i-1)$,
where $f^{2\lambda}=(2d)!/\prod_{\square\in 2\lambda}\mathrm{hook}
(\square)$ is the dimension of the $S_{2d}$-irrep $2\lambda$, and
\begin{equation}
\omega^{\lambda}_\mu
=\frac{1}{2^d d!}\sum_{\zeta\in H_d}\chi^{2\lambda}(\sigma_\mu\,\zeta),
\qquad \sigma_\mu\ \text{of cycle type}\ 2\mu,
\label{eq:omega}
\end{equation}
is the zonal spherical function of the Gelfand pair
$(S_{2d},H_d)$ \cite{CM2009}.  Since $Z_\lambda(1^N)$ and
$f^{2\lambda}$ are elementary, the whole finite-$N$ content of the
sector calculus sits in the $N$-independent zonal table
$\omega^\lambda_\mu$.  The $N$-dependence is the fixed, factored
denominator structure $\prod_{\square}(N+\mathrm{content})$, one
linear factor per cell of $\lambda$.  For the identity class $\mu=1^d$,
$\omega^{\lambda}_{1^d}=1$ for every $\lambda$, so its weight is the pure
spherical sum
$\Wg^{O(N)}(1^d)=\frac{1}{(2d-1)!!}\sum_{\lambda\vdash d}f^{2\lambda}/Z_\lambda(1^N)$.
The identity sector is exactly the sum over spherical representations.
Inserting Eq.~\eqref{eq:zonal} into the sector sum of
Eq.~\eqref{eq:sectgen} thereby decomposes the kernel itself into the
same spherical basis.  With the $N$-independent sector amplitude
$\mathcal S^{(r)}_\lambda(\mathcal P_A,\mathcal P_C)=\sum_c
T^{(r)}_c\,\omega^{\lambda}_c$,
\begin{equation}
\begin{aligned}
&\mathcal K_r(A_1,\dots,A_r;C_1,\dots,C_r)\\
&\quad=\frac{2^{kr}(kr)!}{(2kr)!}
\sum_{\lambda\vdash kr}\frac{f^{2\lambda}\,
\mathcal S^{(r)}_\lambda(\mathcal P_A,\mathcal P_C)}
{Z_\lambda(1^N)},
\end{aligned}
\label{eq:zonalK}
\end{equation}
separating the kernel exactly into an $O(N)$ spectral weight
$f^{2\lambda}/Z_\lambda(1^N)$ and a microscopic overlap-geometry
amplitude $\mathcal S^{(r)}_\lambda$ carried by the zonal table and
the sector counts.  For fixed $r$ and $k$ the representation-theoretic
weights $f^{2\lambda}/Z_\lambda(1^N)$ are universal, fixed by the
group construction $(O(N),\wedge^k)$ and by $N$ alone, so the
factorization is universal within the $O(N),\wedge^k$
construction, not across ensembles, whereas the spherical amplitudes
$\mathcal S^{(r)}_\lambda$ encode the model-specific overlap geometry.
$T^{(r)}_c$ is the microscopic sector histogram and
$\mathcal S^{(r)}_\lambda$ its spherical transform.  We verified
Eq.~\eqref{eq:zonal} against the reduced engine of
Eq.~\eqref{eq:reduced} with $\omega^\lambda_\mu$ computed
independently from the Murnaghan--Nakayama rule.  $d=1,2,4,6$, every
class, exact rational agreement at all tested $N$.  The zonal tables of depths four and six are distributed as machine-readable data (Sec.~\ref{app:wgtab}).  At depth twelve the same formula produces
all $77$ Weingarten weights behind the $k=4$ three-point kernel of
Sec.~\ref{sec:kernel}, with no Gram inversion, the
$N$-dependence entering only through the factored denominators
$Z_\lambda(1^N)$, and the full-cycle weight reproduces the closed
form of Ref.~\cite{CM2009} in exact integer arithmetic.  The explicit
formulas and the three independent checks (the depth-eight
reproduction, the sphere identity, and the signed-Catalan asymptotics
of Eq.~\eqref{eq:cfree}) are collected in Sec.~\ref{app:wgtab}.
  It closes the depth-twelve sector algebra
behind the $k=4$ three-point kernel of Sec.~\ref{sec:kernel}.

\subsubsection{The three-copy sector sum}

For $r=3$, $k=2$ each squared minor contributes four slots
(two permutation copies of two rows), and
Eq.~\eqref{eq:weingarten} with $d=6$ gives the three-point moment as
\begin{equation}
\E[p_1p_2p_3]
=\sum_{\sigma,\tau}\mathrm{sgn}(\sigma\tau)
\sum_{\pi,\pi'}\Delta_\pi(R)\Delta_{\pi'}(C)\,
w_{\mathrm{type}(\pi\pi')}(N),
\label{eq:sectorsum}
\end{equation}
with the pairings $\pi,\pi'$ running over $M_{12}$, the permutations
$\sigma,\tau\in S_2^{3}$, the shorthand
$R\equiv R(\sigma,\tau)$, $C\equiv C(\sigma,\tau)$ (the pairings are
involutions, $\pi^{-1}=\pi$), where
$R(\sigma,\tau)$ and
$C(\sigma,\tau)$ are the $12$-tuples of row
(level) and column (mode) labels of the slots, with slot
$(i,j;1)$ carrying $(A_i(\sigma_i(j)),C_i(j))$ and slot $(i,j;2)$
carrying $(A_i(j),C_i(\tau_i(j)))$.  Equation~\eqref{eq:sectorsum} is
evaluated exactly.  For each of the sixteen row patterns and sixteen
column patterns the valid pairings are collected ($\Delta_\pi\neq0$
requires every $\pi$-edge to join equal labels, so the valid sets have
size $3^{m_{12}+m_{13}+m_{23}-3t}\,15^{t}$ at most), and the
type-weighted double sum is carried out over the $10395\times10395$
type matrix once and for all.  The result is the sector table of
Sec.~\ref{sec:kernel}.

\section{Numerical protocol and independent checks}
\label{app:numerics}

Numerical tables marked S$k$ are the reproduction tables collected in Sec.~\ref{app:supp}.  Models, sampling, error bars, and finite-size settings are stated here, together with the checks that do not appear in the main text; the implementation files and raw output behind each table are listed in Sec.~\ref{app:models} and deposited in the data package.

\subsection{Numerical protocol}
\label{app:num}

The Monte Carlo draws real GOE one-body Hamiltonians $h$ in the Mehta
normalization ($\E h_{aa}^2=2$, $\E h_{ab}^2=1$), diagonalizes them to
obtain the Haar one-body eigenvector matrix $U$ and levels $E_a$, and
builds the many-body levels $\lambda_A=\sum_{a\in A}E_a$ and the exact
overlaps $p_A^C=|\det U[A,C]|^2$ for all $D=\binom Nk$ label and mode
sets (Julia, MKL backend).  All quantities in Sec.~\ref{sec:bulk} are
bin-averaged over the sixteen channel-triple patterns.  The direct
cumulant $\mathcal C^{(3)}$ uses the exact flat-envelope centering
$\E_U[p]=1/D$, i.e.\ $\delta\mathcal R_a=\mathcal R_a-\frac1D
\Tr(z-H)^{-1}$ per realization.  The sector reconstruction uses
Eq.~\eqref{eq:bulk3} with $\hat{\cG}_3$ computed per realization from
the many-body spectrum and, at $(14,2)$, the closed-form kernel
table.  At $(8,4)$ the sector values $\Cov_3$ are measured in the same
ensemble, so the check tests the sector dependence of
$\E[\delta p\delta p\delta p]$ together with the $U\perp\Lambda$
factorization.  The deep-plane coefficients use the moment fields
$s_r=\delta(H^r)_{aa}$ in the channel basis with the channel
Hamiltonian built by second quantization and checked elementwise.
With bin-averaging over the channel-triple patterns, $K_{111}$ is
consistent with its exact zero within the bin noise (machine
precision at $(8,4)$), and $K_{211}$ agrees with the piecewise Wick
closed form to $2\%$ on the diagonal and disjoint bins at $(14,2)$
($21.28$ vs $21.70$ and $0.934$ vs $0.954$, $n_{\rm real}=2000$) and
to $1.3\%$ rms at $(8,4)$ ($27.14$ vs $27.43$ on the diagonal bin)
(Sec.~\ref{sec:q3}).  At $(8,4)$ the
deep-plane projection of the sector decomposition,
$\sum_s\Cov_3(s)\E[\lambda_n^2\lambda_m\lambda_l|s]$, reproduces the
closed-form $K_{211}$ for each representative channel triple to
$0.3$--$5\%$.  The
fully coincident sector $\E[(\delta p)^3]$ is measured over all
$D^2$ label--mode pairs and reproduces the Selberg value
$K_3=q_3-3q_2+2$: $28.856$ against $28.943$ at $(14,2)$ and
$36.32$ against $36.49$ at $(8,4)$, both at $n_{\rm real}=2000$.

The Sec.~\ref{sec:spectral} checks at $(14,2)$.  The class moment sums
of a fixed realization reproduce Eq.~\eqref{eq:momdict} to max
relative deviation $1.8\times10^{-14}$ over the $25$ class-moment
combinations.  Eq.~\eqref{eq:disp3} holds on an $8192$-point grid with
rms $4.2\times10^{-3}$ of the peak real part (direct cumulant of the
diagonal channel triple at $\eta=0.04$) and rms $1.3\times10^{-2}$ for
the sector reconstruction, the difference being the discretization of
its sharper pole structure.  With the exact $16\times16$ kernel table
of Sec.~\ref{sec:kernel} and the eigenstates classified by their
one-body pair pattern, the sector reconstruction of
Eq.~\eqref{eq:bulk3} at the probe point of Table~\ref{tab:bulk}
reproduces the direct cumulant to $4.3\%$ on the diagonal pattern and
$2.0$--$2.3\%$ on the single-overlap and disjoint bins
($n_{\rm real}=2000$, exact channel spectrum), and the moment sum
$\sum_s N_s\Cov_3(s)M_s(2,1,1)$ reproduces the closed-form $K_{211}$
of Sec.~\ref{sec:q3} to $0.9\%$ on the diagonal channel triple and
$0.6$--$0.9\%$ over all four representative triples, the total mass
of Eq.~\eqref{eq:mass0} vanishing identically as it must.

The checks of Sec.~\ref{sec:gamma3} reuse the same ensemble.  The
self-energy $\cG_a=z-a_a-1/\mathcal R_a$ is computed per realization
at $z=i\eta$ for $\eta=128$--$1024$.  Its ensemble mean is estimated
over all channels and realizations (two passes).  The ladder identity
Eq.~\eqref{eq:ladder3} is evaluated per realization set on the
coincident channel triple with the mean-field resolvent $\bar{\mathcal
R}_a^0=1/(z-a_a-\bar\cG_a)$.  The leading-coefficient check compares
$z\cG_a$ with $v_a=\sum_{c\in A,q\notin A}h_{cq}^2$.  And the
$h$-side moments use the moment-field identity
$s_r=\delta(H^r)_{aa}$ in closed form without diagonalization, with
the channel-graph traces $\Tr H^2$, $\Tr H^3$ (including all
mode-triple and star triangles of the Johnson graph) verified against
the eigenvalue sums to $5\times10^{-11}$.  The exact Wick values of
Eqs.~\eqref{eq:Gamma3}, \eqref{eq:Gamma4} and \eqref{eq:K2211} come
from the monomial Wick engine, which evaluates
arbitrary products of the
independent diagonal ($\E d_c^2=2$) and off-diagonal ($\E
h_{cq}^2=1$) Gaussian fields exactly in rational arithmetic.

The ladder bookkeeping of Sec.~\ref{sec:gamma3} is checked at $(14,2)$ with $n_{\rm real}=10^5$: the identity Eq.~\eqref{eq:ladder3} holds per realization set to $10^{-32}$ for $\eta=128$--$1024$, and the leading-coefficient identification Eq.~\eqref{eq:Gdeep} holds to
$\E|z\cG_a-v_a|/\E|v_a|=1.2\times10^{-3}$ at $\eta=1024$, decaying as
$1/\eta$.  The vertex moments approach their exact limits from the
h-side ($10^6$ realizations, moment-field identity, no
diagonalization), $\E[(\delta v)^3]=192.38$ against $8k(N-k)=192$ and
$\kappa_4(\delta v)=1180.5$ against $1152$, and from the z-side,
$\E[(z\,\delta\cG)^2]=48.08$ against $\E[(\delta v)^2]=48$,
$\E[(z\,\delta\cG)^3]=193.96$ against $192$,
$\E[(z\,\delta\cG)^4]-3\E[(z\,\delta\cG)^2]^2=1215$ against $1152$,
$\E[(z^3\delta\Delta)^3]=193.87$ against $192$, with the inversion
control $\E[(z^3\delta\Delta)^3-(z\,\delta\cG)^3]=-0.09$ against zero
at $\eta=1024$.  The diagonal channel sum closes the projection
$\sum_a\Gamma^{(3)}_{aaa}\,z_1z_2z_3\to8Dk(N-k)=17472$, measured
$17491$ ($10^6$ realizations).

The four-point Wick tables.  The exact values of
Eq.~\eqref{eq:K2211} at $(14,2)$ are
\begin{equation}
\begin{aligned}
&K_{2211}=\frac{19040256}{57967},\ \frac{5245664}{57967}, -\frac{185664}{57967},\ -\frac{98304}{57967},\\
&K_{3111}=\frac{1049472}{4459},\ \frac{175200}{4459}, -\frac{33120}{4459},\ -\frac{6912}{4459},
\end{aligned}
\label{eq:KtabA}
\end{equation}
for the coincident quadruple, the pair-sharing quadruple
$([12],[13],[12],[13])$, the generic quadruple $([12],[13],[24],[56])$
and the disjoint quadruple, respectively (Monte Carlo $329.7\pm1$ and
$233\pm9$ at $n_{\rm real}=10^6$ for the coincident values, consistent
within the heavy-tailed sampling error of the cubic fields), and at
$N=8,10,12,16$ the coincident values are
$K_{2211}=11304/49,\;2760704/10125,\;36800/121,\;78253/225$ and
$K_{3111}=1098/7,\;23808/125,\;7120/33,\;5019/20$.  At $(14,2)$ the
pair terms of Eq.~\eqref{eq:deep4} carry $71.2$ and $95.0$ of the two
totals, so the pair terms are substantial.  The four-point
pattern space of Sec.~\ref{sec:kernel4} holds $1657$ restricted-growth
classes of four 2-sets.

\noindent\parbox{\linewidth}{\raggedright
The entry script and the data file behind each result are listed in Table~\ref{tab:index}.\par}

The computational scales quoted in the main text.  The $k=3$ table
streams the $17!!=34{,}459{,}425$ pairings of $M_{18}$ ($2.0\times10^{12}$
pairing-pair classifications, twenty-four threads, checkpointed).  The
free-limit checks of Sec.~\ref{sec:kernel} cover $936$ ($k=3$) and
$6340$ (four-point $k=2$) entries with zero mismatches.  The occupancy
assembly of Sec.~\ref{sec:assembly} handles the $t=3,4$ classes whose
direct signed supports hold up to $\sim5\times10^{8}$ pairings, and the
chain DP is verified against brute force on blocks up to
$2.56\times10^{9}$ pairs.  In the independent-factorization comparison
of Sec.~\ref{sec:annealed} the exact sector tables span, in units
$\E\cdot D^r$, $[1/3,5]$, $[0.4510,40.593]$ and $[0.2160,544.375]$ at
$r=2,3,4$.  Column-resolved at $r=4$.  Coincident column
$[0.2160,544.375]$ with the coincident row at $q_4=544.375$, generic
and permuted-generic columns identical by replica symmetry
$[0.3862,2.2873]$, disjoint column $[0.2160,1.3367]$.

\subsubsection{Numerical verification}

We verify Eq.~\eqref{eq:bulk3} in Monte Carlo over the same GOE ensemble.  At $(14,2)$, where the kernel of Sec.~\ref{sec:kernel} is exact, both sides are compared bin by bin over all sixteen channel-triple patterns at
$(z_1,z_2,z_3)=(0.5+0.8i,-0.4+0.9i,0.2+1.1i)$, $n_{\rm real}=2000$: the imaginary parts, which dominate the cumulant at these frequencies, agree on the diagonal, single-overlap, and disjoint bins within the Monte Carlo noise, and the sector values of the measured overlap triple product reproduce the closed-form kernel table (Table~\ref{tab:bulk}; protocol and full numbers in Sec.~\ref{app:num}).  At $(8,4)$, the main size of Sec.~\ref{sec:benchmark}, the sector values are measured in the same ensemble, and the deep-plane projection of Eq.~\eqref{eq:bulk3}, $\sum_s\Cov_3(s)\E[\lambda_n^2\lambda_m\lambda_l|s]$, reproduces the closed-form $K_{211}$ on every representative channel triple within sampling error ($n_{\rm real}=2000$), confirming that beyond $k=2$ the three-point overlap statistics depend only on the triple geometry; at $k=4$ the pattern dependence of the sector theorem is measured rather than closed, the $d=12$ pairing algebra being beyond direct enumeration.

\begin{table}[tbp]
\caption{Bulk three-resolvent cumulant at $(14,2)$, $n_{\rm real}=2000$:
direct bin-averaged $\mathcal C^{(3)}$ against the sector
reconstruction of Eq.~\eqref{eq:bulk3} with the exact closed-form
kernel, for three representative channel-triple patterns.  The
imaginary parts dominate at these frequencies; the (small) real parts
agree within their noise.}
\label{tab:bulk}
\centering
\footnotesize
\begin{ruledtabular}
\begin{tabular}{cccc}
\hline\hline
$\ell$-pattern & direct (Im) & sector (Im) & ratio\\
\hline
$(2,2,2,2)$ & $1.1541\times10^{-4}$ & $1.1552\times10^{-4}$ & $0.999$\\
$(2,1,1,1)$ & $4.136\times10^{-5}$ & $4.266\times10^{-5}$ & $0.969$\\
$(0,0,0,0)$ & $6.384\times10^{-6}$ & $6.666\times10^{-6}$ & $0.958$\\
\hline
\end{tabular}
\end{ruledtabular}
\end{table}

\subsection{Benchmark details and verification}
\label{app:bench}

\subsubsection{Minor statistics: Jacobi, Selberg, and the tails}

The squared overlap is $p=\det(U[A,C]U[A,C]')$, a product of the eigenvalues $x_1,\dots,x_k$ of the Gram matrix. By the classical theorem of Muirhead on submatrices of Haar orthogonal matrices \cite{Muirhead}, the matrix $U[A,C]U[A,C]'$ has the matrix-variate beta distribution $\mathrm{Beta}_1(k/2,(N-k)/2)$ for $N\ge2k$, and the joint density of its eigenvalues is the Jacobi ensemble of index $\beta=1$,
\begin{equation}
P(x_1,\dots,x_k)\propto
\prod_{i<j}(x_i-x_j)\prod_{i=1}^{k}x_i^{-1/2}(1-x_i)^{(N-2k-1)/2},
\label{eq:jacobi}
\end{equation}
with exponents $a_0=-1/2$ and $b_0=(N-2k-1)/2$.  For $k>N/2$ the complementary-minor identity $|\det U[A,C]|=|\det U[\bar A,\bar C]|$ maps the statistics onto the $(N-k)$-minor sector, so Eq.~\eqref{eq:jacobi} and the moment formulas below, evaluated at $k\mapsto N-k$, cover every occupation $0\le k\le N$. The product moments follow from Selberg's integral \cite{Forrester},
\begin{equation}
M_k(t):=\E\prod_{i=1}^{k}x_i^t
=\prod_{j=1}^{k}
\frac{\Gamma\bigl(\frac{j+2t}{2}\bigr)}{\Gamma\bigl(\frac j2\bigr)}
\frac{\Gamma\bigl(\frac{N-k+j}{2}\bigr)}
{\Gamma\bigl(\frac{N-k+j+2t}{2}\bigr)}.
\label{eq:selberg}
\end{equation}
The Selberg moments of Eq.~\eqref{eq:selberg} telescope at $t=1$ to the exact flat envelope
\begin{equation}
M_k(1)=\frac{k!\,(N-k)!}{N!}=\frac1D,
\label{eq:flat}
\end{equation}
and for integer $t$ collapse to the rational product
\begin{equation}
M_k(t)=\prod_{j=1}^{k}\prod_{i=0}^{t-1}\frac{j+2i}{N-k+j+2i}.
\label{eq:Mkint}
\end{equation}
In the fixed-depth limit the normalized moments reduce to
\begin{equation}
\begin{aligned}
q_r\to\prod_{j=1}^{k}\prod_{i=1}^{r-1}\frac{j+2i}{j}, \quad q_2=\frac{(k+1)(k+2)}2,\\
q_3=\frac{(k+1)^2(k+2)^2(k+3)(k+4)}{48},
\end{aligned}
\label{eq:qrlimit}
\end{equation}
the moment sequence of $|\det G|^{2r}/\E[|\det G|^2]^r$ for a $k\times k$ matrix of independent standard Gaussians (equivalently that of a product of independent Gamma variables $Y_j\sim\Gamma(j/2,2/j)$), so the fixed-depth limit of the Haar minor is exactly the statistic of a Gaussian minor.  The non-Porter--Thomas character is the determinant product structure itself, present already for the Gaussian matrix. For $k\le2$ the moment sequence is moment-determinate and the convergence holds in distribution. The normalized cumulants are $K_r:=\kappa_r/\E[p]^r$ with
\begin{equation}
K_2=q_2-1=\frac{k(k+3)}2,\qquad
K_3=q_3-3q_2+2,
\label{eq:Kr}
\end{equation}
against the Porter--Thomas values $K_r^{\rm PT}=2^{r-1}(r-1)!$, so the ratios $K_2/K_2^{\rm PT}=k(k+3)/4$ and $K_3/K_3^{\rm PT}=(q_3-3q_2+2)/8$ equal $2.5$ and $9.3$ at $k=2$ and $7.0$ and $126$ at $k=4$.  
The single-channel density is known exactly at both ends. The Mellin transform $M_k(t)=\E p^t$ has its rightmost pole at $t=-\tfrac12$, contributed by the single factor $\Gamma(\tfrac12+t)$ of $j=1$, simple and isolated (the remaining numerator poles sit at $t\le-1$ and the denominator zeros at $t\le-(N-k+1)/2\le-1$). Inverting a Mellin transform with a simple pole at $t=-s$ gives $P(p)\sim Cp^{-s}$ without logarithmic enhancement, and the residue gives
\begin{equation}
\begin{aligned}
P(p)&\sim C_k\,p^{-1/2}\quad(p\to0),\\
C_k&=\pi^{-1/2}\prod_{j=2}^{k}\frac{\Gamma\bigl(\frac{j-1}{2}\bigr)}{\Gamma\bigl(\frac j2\bigr)}
\prod_{j=1}^{k}\frac{\Gamma\bigl(\frac{N-k+j}{2}\bigr)}{\Gamma\bigl(\frac{N-k+j-1}{2}\bigr)},
\end{aligned}
\label{eq:tailsmall}
\end{equation}
while the corner scaling of the Jacobi density gives the power-law cutoff at maximal intensity
\begin{equation}
P(p)\propto(1-p)^{k(N-k)/2-1}\quad(p\to1),
\label{eq:taillarge}
\end{equation}
which for $N=5$, $k=2$ assembles into the exact closed form $P(p)=\tfrac32\,p^{-1/2}(1-\sqrt p)^2$. In the rescaled variable $\tilde p=Dp$, the small-intensity law reads $\tilde P(\tilde p)\sim(C_k/\sqrt D)\,\tilde p^{-1/2}$ against the constant small-amplitude density of the Porter--Thomas law.  The free-fermion distribution enhances small intensities algebraically, and its maximal intensity is cut off by a power rather than by an exponential.

\subsubsection{Dictionary closed forms}

Sector populations.  Writing the bath block as $U[A,\mathrm{bath}]$ and the system-mode column as $u_A=U[A,0]$,
\begin{equation}
\begin{aligned}
D_A^0&=\det\bigl(U[A,\mathrm{bath}]U[A,\mathrm{bath}]'\bigr)\\
&\quad=\det(I_k-u_Au_A')=1-\sum_{a\in A}U_{a0}^2,\\
D_A^1&=\sum_{a\in A}U_{a0}^2,
\end{aligned}
\label{eq:D0}
\end{equation}
by Cauchy--Binet and the rank-one determinant identity, hence the partition $D_A^0+D_A^1=1$ and the occupation interpretation $D_A^1=\braket{\psi_A|c_0^\dagger c_0|\psi_A}$. The ensemble moments follow from the standard second and fourth moments of a Haar-distributed row:
\begin{equation}
\E D_A^1=\nu,\qquad
\Var D_A^1=\frac{2\nu(1-\nu)}{N+2}.
\label{eq:occmom}
\end{equation}
With the dictionary decomposition
\begin{equation}
f_{ji}^2=D_{ji}+g_{ji},
\label{eq:decomp}
\end{equation}
the diagonal entries close as follows. For the diagonal baseline, the channel resolution of identity gives $\sum_Bp_B^{\mu i}=1$ for every $\mu$, hence $\sum_{B\ne A}\sum_\mu p_A^{\mu i}p_B^{\mu i}=D_A^i-\Pi_A^i$ with the channel purity $\Pi_A^i:=\sum_\mu(p_A^{\mu i})^2$ and $\E\Pi_A^i=d_B\,q_{\rm FF}/D^2$, so
\begin{equation}
D_{ii}=\frac{D\bigl(\E D_A^i-d_B\,q_{\rm FF}/D^2\bigr)}{D-1}
=\frac{d_B}{D}\Bigl[1+O\bigl(D^{-1}\bigr)\Bigr];
\label{eq:meansector}
\end{equation}
for the smooth function the exact complementarity identity gives $\sum_{B\ne A}|\sigma_{AB}^{ii}|^2=D_A^i(1-D_A^i)$, so
\begin{equation}
f_{ii}^2=\frac{D\,\nu(1-\nu)N}{(N+2)(D-1)},
\label{eq:f2mean}
\end{equation}
and the correlation correction follows by subtraction,
\begin{equation}
g_{ii}=-\frac{D}{D-1}\Bigl[(\E D_A^i)^2+\frac{2\nu(1-\nu)}{N+2}-\frac{d_B\,q_{\rm FF}}{D^2}\Bigr]<0,
\label{eq:gmeansector}
\end{equation} $g_{ii}<0$ is of order one: in the large-$D$ limit the diagonal baseline $d_B/D\to\E D_A^i$ overestimates the smooth envelope $f_{ii}^2\to\E D_A^i(1-\E D_A^i)$, the difference being a correlation deficit rather than an $O(1/D)$ correction. It is proportional to $\E(D_A^i)^2-\E\Pi_A^i>0$, precisely the quantity the projector sum rule and the cavity subtraction of the framework control, and all three entries are energy independent because $U\perp\Lambda$. A flat ETH envelope therefore does not imply the absence of multi-channel correlation:
\begin{equation}
f_{ji}^2=\mathrm{constant}\quad\not\Longrightarrow\quad g_{ji}=0.
\label{eq:counterexample}
\end{equation}

\emph{Fluctuation sector.} The single-channel variance is
\begin{equation}
\Var(p_A^a)=\frac{q_{\rm FF}-1}{D^2},
\label{eq:Vdg}
\end{equation}
the direct analogue of the Gaussian $(q-1)/D^2$ with the Slater ratio. The population variance is Eq.~\eqref{eq:occmom}, and its decomposition into same-channel and cross-channel contributions reproduces the diagonal-fluctuation identity of the framework term by term,
\begin{equation}
V_{\rm dg}=\underbrace{\sum_\mu\Var(p_A^{\mu i})}_{d_B(q_{\rm FF}-1)/D^2}
+\underbrace{\sum_{\mu\ne\nu}\Cov(p_A^{\mu i},p_A^{\nu i})}_{\bar C_i\,e^{-2S_S}},
\label{eq:fpVdg}
\end{equation}
with $e^{-S_S}=\E D_A^i$ and $\langle f^2\rangle_B=1$ for the flat channel envelope, which determines $\bar C_i$ in closed form. In sector $1$ the projector is the occupation operator of the system mode, $P_1=\sum_\mu\ket{\varphi_{\mu1}}\bra{\varphi_{\mu1}}=c_0^\dagger c_0$, so $\sigma_{AB}^{11}=\braket{\psi_A|c_0^\dagger c_0|\psi_B}$ vanishes unless $B=A$ or $B$ differs from $A$ by a single particle--hole hop, $B=A\setminus\{a\}\cup\{b\}$ with $a\in A$, $b\notin A$.  On a hop $|\sigma_{AB}^{11}|=|U_{0a}U_{0b}|$, and summing over the $k(N-k)=z$ hop partners and factorizing the double sum gives the complementarity identity
\begin{equation}
\sum_{B\ne A}\bigl(\sigma_{AB}^{11}\bigr)^2
=\Bigl(\sum_{a\in A}U_{0a}^2\Bigr)\Bigl(\sum_{b\notin A}U_{0b}^2\Bigr)
=D_A^1(1-D_A^1),
\label{eq:complementarity}
\end{equation}
realization by realization.  The off-diagonal matrix elements of the projected occupation are supported exactly on the particle--hole graph of the quadratic Hamiltonian. Hence
\begin{equation}
V_{\rm off}=\frac{\E[D_A^1(1-D_A^1)]}{D-1}=\frac{\nu(1-\nu)N}{(N+2)(D-1)}=\frac{f_{11}^2}{D},
\label{eq:Voff}
\end{equation}
and
\begin{equation}
\F=\frac{V_{\rm dg}}{V_{\rm off}}=\frac{2(D-1)}{N},
\label{eq:fratio}
\end{equation}
exact and proportional to $D-1$ at fixed $N$, against the Gaussian reference $\F\to2$.  The population retains an $O(N^{-1})$ projection-weight fluctuation that does not decrease with the Hilbert-space dimension, while the off-diagonal matrix elements are suppressed by $1/D$, the ratio compares two different mechanisms of smallness rather than a single fluctuation scale. The two-channel covariance within a sector is
\begin{equation}
c:=D^2\Cov(p_\mu,p_\nu)
=\frac{\dfrac{2D\nu}{N+2}+1-q_{\rm FF}}{D(1-\nu)-1},
\label{eq:cexact}
\end{equation}
whose leading form at extensive filling,
\begin{equation}
c=\frac{2\nu(1-\nu)}{N+2}\Bigl(\frac{N}{N-k}\Bigr)^2
\Bigl[1+O\Bigl(\frac{q_{\rm FF}}{D}\Bigr)\Bigr],
\label{eq:cpos}
\end{equation}
is of order $2/(N+2)$ at half filling. The sign has a transparent origin in the law of total covariance,
\begin{equation}
\begin{aligned}
\Cov(p_\mu,p_\nu)
&=\underbrace{\Cov\bigl(\E[p_\mu|D_A^0],\E[p_\nu|D_A^0]\bigr)}_{\text{population}}\\
&\quad+\underbrace{\E\bigl[\Cov(p_\mu,p_\nu|D_A^0)\bigr]}_{\text{fixed sum}},
\end{aligned}
\label{eq:totcov}
\end{equation}
with the population term dominating at half filling and the fixed-sum term in the chaotic reference, whose isotropically saturated sum rule gives $\bar C_i=-(q-1)(1-1/d_B)/(D-1)=-2/D+O(D^{-2})$ at $q=3$ \cite{HOETH}.  The free-fermion ensemble realizes the opposite sign at large determinant depth, so the sign of the channel covariance is a fingerprint of the channel-amplitude law rather than a universal feature. In the normalization of the framework, with $f^{\mu i}=1$,
\begin{equation}
\bar C_i:=\frac{1}{d_B^2}\sum_{\mu\ne\nu}f^{\mu i}f^{\nu i}
\E[\delta_A^{\mu i}\delta_A^{\nu i}]_{\rm c}
=\Bigl(1-\frac1{d_B}\Bigr)c,
\label{eq:Cbardef}
\end{equation}
evaluating to $0.1193$ at $N=14$, $k=7$.  $\bar C_i\neq0$ although the envelope is exactly flat.  In the free-fermion benchmark $\bar C_i$ arises from correlated Slater-minor fluctuations within the channel sector, and energy-resolved envelope structure is an additional possible source in interacting models, not a necessary one.

\subsubsection{Channel-distance profile and the kernel derivation}

Channel pairs are classified by $\ell=|C_\mu\cap C_\nu|$ with multiplicities
\begin{equation}
N_\ell=\binom{k-1}{\ell}\binom{N-k}{k-1-\ell},
\label{eq:ell}
\end{equation}
$\sum_{\ell<k-1}N_\ell=d_B-1$.  The multiplicity-weighted sum reproduces the integrated weight,
\begin{equation}
\begin{aligned}
&d_B\Bigl[\sum_{\ell<k-1}N_\ell\Bigl(C_\ell+\frac1{D^2}\Bigr)-(d_B-1)\frac1{D^2}\Bigr]\\
&\quad=\sum_{\mu\ne\nu}\Cov(p_\mu,p_\nu)=\bar C_i\,e^{-2S_S},
\end{aligned}
\label{eq:ellsum}
\end{equation}
numerically consistent with the closed form within the Monte Carlo error of the heavy-tailed estimator (direct estimates $2.64\times10^{-2}$ through the channel-distance sum and $2.81\times10^{-2}$ through the diagonal split at $N=14$, $k=7$, against the exact $2.9822\times10^{-2}$). The profile $C_\ell$ is exact for $k\le5$ by a finite Wick algorithm.  Expanding the first row in Laplace minors of the remaining $(k-1)$-row frame and using the Gaussian representation of the frame, the fourth moments of the Pl\"ucker coordinates close on Wick pairings whose normalization cancels the Wishart scale exactly, and the first-row average reduces to the sphere moments. The result is
\begin{equation}
D^2C_\ell=\frac{P_\ell(N)}{(k!/2)\,\binom{N+2}{2}},
\label{eq:Cellkk}
\end{equation}
with $k=3$.  $P_0=-9N$, $P_1=N^2-12N+2$, $P_2=\tfrac12(7N^2-39N+14)$.  $k=4$.  $P_0=-24(2N-1)$, $P_1=3(N^2-21N+14)$, $P_2=3(3N^2-31N+26)$, $P_3=24(N^2-7N+7)$.  $k=5$.  $P_0=-300(N-1)$, $P_1=12(N^2-32N+37)$, $P_2=33N^2-531N+696$, $P_3=75N^2-825N+1200$, $P_4=180N^2-1560N+2460$ (the $k=2$ case being Eq.~\eqref{eq:Cellk2}). Each $C_\ell$ changes sign exactly once, at a threshold $N_c(k,\ell)$ that grows as $\ell$ decreases (at the benchmark sizes the crossing sits near $\ell\simeq k/2$), while $C_0<0$ for every finite $N$.  Disjoint channels remain anticorrelated at any finite $N$ yet become independent in the Gaussian fixed-depth limit. The rational functions are reproduced exactly by the Wick engine and, independently, by direct sampling to within the Monte Carlo error.

\emph{The kernel $L(k;m,\ell)$.} Expanding both determinants along the $m$ shared rows,
\begin{equation}
\E[p_1p_2]=(k-m)!^{\,2}\sum_{X,Y}\mu_m(|X\cap Y|),
\label{eq:Lred}
\end{equation}
where the sum runs over $m$-subsets $X\subseteq C_1$, $Y\subseteq C_2$ and $\mu_m(j)=(m!)^2[1+L(m;m,j)]$ is the depth-$m$ pair moment of the shared-row block.  The count of subset pairs with $|X\cap Y|=j$ is
\begin{equation}
N(k,m,\ell,j)=\sum_{a=0}^{\min(\ell,m)}
\binom{\ell}{a}\binom{k-\ell}{m-a}\binom{a}{j}\binom{k-a}{m-j},
\label{eq:Ncount}
\end{equation}
which gives Eq.~\eqref{eq:Lgeneral}. The same-eigenstate closed form follows from the Bartlett decomposition of the $k\times(2k-\ell)$ Gaussian matrix $H$ formed by the two column sets.  Writing $H=QR$ with $Q\in O(k)$ and $R$ upper triangular, the entries $R_{ii}^2\sim\chi^2_{k-i+1}$ and $R_{ij}\sim\mathcal N(0,1)$ for $i<j$ are independent, and expanding $\det(R_C)^2$ over two permutations, the row contribution $\E[R_{ii}^2R_{i,c_{\sigma(i)}}R_{i,c_{\sigma'(i)}}]$ vanishes unless $\sigma(i)=\sigma'(i)$, leaving exactly
\begin{equation}
\E[p_1p_2]=\prod_{i=1}^{\ell}(k-i+1)(k-i+3)\,(k-\ell)!^{\,2},
\label{eq:Eppbartlett}
\end{equation}
equivalent to $L(k;\ell)=\ell(2k-\ell+3)/[(k-\ell+1)(k-\ell+2)]$. The formula reproduces every exact value obtained for $k\le5$ from the direct Wick expansion of the two-minor moment and the sampled values for $k\le7$ within the Monte Carlo error.  At fixed $\ell$ it decays as $1/k$ at large depth, so only overlapping channels retain a correlation in the thermodynamic limit, which is why the shallow-minor negative covariance (a finite-$N$ normalization effect) and the fixed-depth positive correlation can coexist.

\subsubsection{Energy resolution and the two-resolvent bridge}

The difference of the two many-body energies, $\lambda_B-\lambda_A=\sum_{j\in B\setminus A}\varepsilon_j-\sum_{i\in A\setminus B}\varepsilon_i$, is the sum of $2(k-m)$ one-body levels sampled from the empirical one-body density, whose leading large-$N$ difference density is the classical $2(k-m)$-fold convolution of the unit semicircle,
\begin{equation}
\rho_{\rm sc}^{*n}(\omega)=\frac1{2\pi}\int_{-\infty}^{\infty}
\Bigl[\frac{J_1(2t)}{t}\Bigr]^{n}e^{-it\omega}\,\dd t,
\qquad |\omega|<2n,
\label{eq:rhoconv}
\end{equation}
(the free convolution, which would give a semicircle of radius $2\sqrt n$, describes the eigenvalue density of a sum of matrices rather than of a sum of sampled eigenvalues of one matrix). With $\mathcal N_m=\binom{N}{m}\binom{N-m}{k-m}\binom{N-k}{k-m}$ (and $\mathcal N_k=D$), Eq.~\eqref{eq:Fomega} follows.  The prediction is confirmed by direct sampling at $N=60$, $k=2$, where the off-diagonal ridge follows the profile $\rho_{\rm sc}^{*2}$ bin by bin including the tails in $2\sqrt2<|\omega|<4$, the $\delta$-weight matches the finite-$N$ value of Eq.~\eqref{eq:Cellk2}, and the ridge height matches the finite-$N$ overlap moment measured independently ($X_N=0.445$ against the limit $L(2;1,1)=\tfrac12$), the remaining deviations being the level-repulsion dip at $\omega\simeq0$ and the eigenvalue-edge softening. In Eq.~\eqref{eq:Kab} the joint density of the two many-body energies for pairs sharing $m$ levels is
\begin{equation}
\rho_{AB}^{(m)}(E,E')=\int\rho_{\rm sc}^{*m}(s)\,\rho_{\rm sc}^{*(k-m)}(E-s)
\,\rho_{\rm sc}^{*(k-m)}(E'-s)\,\dd s,
\label{eq:rhoAB}
\end{equation}
and setting $E'=E+\omega$ and integrating over $E$ recovers Eq.~\eqref{eq:Fomega}.  The flat channel is
\begin{equation}
C_{\alpha\beta}^{\rm flat}(z,z'):=\sum_{A,B}\Bigl[\E\tfrac1{(z-\lambda_A)(z'-\lambda_B)}
-\E\tfrac1{z-\lambda_A}\,\E\tfrac1{z'-\lambda_B}\Bigr],
\label{eq:Cflat}
\end{equation}
carried by the spectral correlations of the one-body GOE spectrum. Finite-$N$ GOE spectral correlations modify $\rho_{AB}^{(m)}$ beyond the leading product-density approximation.  The central-frequency deviation observed at $N=60$ is the fingerprint of GOE level repulsion and provides a direct route to the finite-$N$ correlated spectral kernel, essential at the microscopic resolution $\omega\sim N^{-1}$.

\subsubsection{Relation to Mag\'an's analytic ETH example}
\label{sec:magan}

Random free fermions were established as an analytic example of eigenstate thermalization by Mag\'an \cite{Magan2016}.  For $H=\sum_{ij}h_{ij}c_i^\dagger c_j$ with $h$ drawn from the GOE, the multiparticle eigenstates satisfy ETH in the multiparticle sector, with correlations and entanglement entropies computed analytically. The structural statement underlying that analysis is that the one-body correlation matrix in an eigenstate,
\begin{equation}
C^{A}_{ij}:=\braket{\psi_A|c_i^\dagger c_j|\psi_A}
=\sum_{a\in A}U_{ia}U_{ja}
=\bigl(UP_AU'\bigr)_{ij},
\label{eq:Cij}
\end{equation}
separates into a thermal part that depends only on the macroscopic filling and a fluctuation part with GOE-like local statistics. The exact statements have three strengths. First, the sector population that the framework uses as its diagonal block is the diagonal element of the correlation matrix,
\begin{equation}
D_A^1=\braket{\psi_A|c_0^\dagger c_0|\psi_A}=C^{A}_{00}
=\sum_\mu p_A^{\mu 1},
\label{eq:DvsC}
\end{equation}
so the occupation observable is the diagonal element of the correlation matrix underlying the analytic ETH construction. Second, its variance is an exact channel-resolved projection.  Since $C^{A}_{00}$ is the sum over the channel family,
\begin{equation}
\Var C^{A}_{00}
=\underbrace{\sum_\mu\Var(p_A^{\mu 1})}_{V_{\rm same}}
+\underbrace{\sum_{\mu\ne\nu}\Cov(p_A^{\mu 1},p_A^{\nu 1})}_{V_{\rm cross}},
\label{eq:chanproj}
\end{equation}
which is the diagonal-fluctuation identity \eqref{eq:fpVdg} of the framework, carrying $q-1$ and $\bar C_i$ respectively. Third, the remaining moments of the correlation matrix are exact observable-level results, entering the framework as the dictionary correspondence recorded in Table~\ref{tab:status}. From the Haar four-point function,
\begin{equation}
\begin{aligned}
\Var C^{A}_{ii}&=\frac{2\nu(1-\nu)}{N+2},\\
\Var C^{A}_{ij}\Big|_{i\ne j}&=\frac{k(N-k)}{(N-1)N(N+2)},\\
\Cov\bigl(C^{A}_{ii},C^{A}_{jj}\bigr)\Big|_{i\ne j}&=-\frac{2\nu(1-\nu)}{(N-1)(N+2)},
\end{aligned}
\label{eq:Cijvar}
\end{equation}
the second of which gives the off-diagonal fluctuation variance of the GOE-like error structure, and the third of which is the trace sum rule implied by $\sum_iC^{A}_{ii}=k$,
\begin{equation}
N\,\Var C^{A}_{ii}+N(N-1)\,\Cov\bigl(C^{A}_{ii},C^{A}_{jj}\bigr)=0,
\label{eq:Ctrace}
\end{equation}
holding identically. $C^A$ is not itself a GOE matrix.  Its trace is fixed exactly, $\Tr C^A=k$, so the diagonal elements carry the nonzero covariance \eqref{eq:Ctrace} required to cancel their variance, in contrast to the independent diagonal entries of a GOE matrix.  The fluctuation statistics are GOE-like locally (in the variance ratio and in the vanishing of the off-diagonal mean) while the global projector constraint is exact, and it is this combination that the cavity subtraction of the framework is designed to restore. In the thermodynamic limit the three entries scale as $2\nu(1-\nu)/N:\nu(1-\nu)/N:2\nu(1-\nu)/N^2$, so the diagonal and off-diagonal fluctuations approach the GOE ratio $2:1$ while the diagonal--diagonal covariance is smaller by a further factor $1/N$ (numerically at $N=14$, $k=7$, $0.0310$, $0.0167$, $-0.00240$ against the predicted $0.03125$, $0.016827$, $-0.002404$). The two analyses answer different questions.  Mag\'an's result is that this ensemble thermalizes in the ETH sense with a smooth, energy-independent envelope; the present benchmark asks which microscopic overlap statistics realize that envelope, and finds they are not Porter--Thomas.  $q_{\rm FF}\neq3$, $\bar C_i\neq0$, and $g_{11}<0$ of order one, all with an exactly flat envelope. Table~S26 records the correspondence.

\subsubsection{Numerical checks, self-averaging, and status tables}

All checks are implemented in \texttt{verify\_ff\_hoeth.jl} and the companion scripts listed in Sec.~\ref{app:models}.  Two features of the scheme matter here: the channel sums are evaluated exactly for every sampled state---for a given $A$ all $\binom{N-1}{k}$ bath subsets are enumerated, so Monte Carlo enters only through the average over $A$ and over realizations---and the elementary hop formula for $\sigma_{AB}^{11}$ is cross-checked against the adjugate representation $\sigma_{AB}^{11}=u_A'\adj(M_{AB})u_B=-\det\left[\begin{smallmatrix}M_{AB}&u_A\\u_B'&0\end{smallmatrix}\right]$ evaluated by a hand-written LU factorization, whose augmented determinant makes the singular case ($\mathrm{rank}\,M_{AB}=|A\cap B|$, below $k-1$ off the hop graph) harmless.  The structural identities (partition, complementarity, the single-hop amplitude $|\sigma_{AB}^{11}|=|U_{0a}U_{0b}|$, and the agreement of the enumerated sector-1 channel sum with the adjugate value of $D_A^1$) hold to machine precision.  The statistical entries match their closed forms at the main benchmark point (Tables~S1 and S2) and along the half-filled line (Table~S3).  Four checks close the ledger: the moment hierarchy \eqref{eq:qr} against exact enumeration of all states and channels at $N=6$, $k=3$ and $N=8$, $k=4$ ($q_2,q_3,q_4$ to $0.4\%$ or better); the sign crossover of Eq.~\eqref{eq:cexact}, confirmed by direct sampling at $N=6,8,10$; the exact tails of the single-channel density, reproduced over three decades in $p$ at $N=6,8$, $k=2,3$, including the closed form $\tfrac32p^{-1/2}(1-\sqrt p)^2$ at $N=5$, $k=2$; and the Gaussian-determinant equivalence \eqref{eq:qrlimit}, verified by direct sampling at $k=2,3,4$.  The conditional mean $\E[p|U]=1/D$ holds realization by realization to $10^{-17}$, and the diagonal fluctuation identity \eqref{eq:fpVdg} term by term.

The measurement of $q_{\rm FF}$ must separate three objects: the ensemble value $q_{\rm ens}=q_{\rm FF}$ of Eq.~\eqref{eq:qff}, averaging over eigenstates, channels, and realizations; the single-realization value $q_U=\E[p^2|U]/\E[p|U]^2=D^2\E[p^2|U]$, since the conditional mean is exactly $U$-independent ($\E[p|U]=1/D$ by Cauchy--Binet, checked to $10^{-17}$); and the numerical estimate $\hat q$, which carries a sampling error on top of $q_U$.  The self-averaging ratio
\begin{equation}
\chi_q:=\frac{\Var_U(q_U)}{q_{\rm ens}^2}
\label{eq:chiq}
\end{equation}
decreases slowly at half filling, from $\chi_q=0.064$ at $N=6$ to $0.035$ at $N=16$, consistent with $\chi_q\propto N^{-0.5\pm0.1}$ over the accessible sizes (an exploratory trend, not used in any asymptotic claim), so the relative rms realization-to-realization fluctuation of $q_U$ stays near $20\%$ there, while at fixed determinant depth it decays close to $N^{-5/3}$ ($\chi_q=0.065,0.034,0.010,3.5\times10^{-3},7.0\times10^{-4}$ at $k=2$ for $N=6,10,20,40,100$).  Since $q_{\rm ens}$ grows with $N$ at fixed filling, absolute and relative fluctuations differ: the absolute variance $\Var_U(q_U)=\chi_q\,q_{\rm ens}^2$ grows from $0.8$ at $N=6$ to $6.2$ at $N=16$, so a single Hamiltonian reproduces the ensemble value absolutely at fixed depth but only relatively at extensive filling.  Table~\ref{tab:status} lists, for each quantity, whether it is an identity, an exact closed form, or a dictionary-level correspondence.

\begin{table*}[tbp]
\caption{Status of the quantities used in the benchmark; the last row is open.}
\label{tab:status}
\centering
\begin{ruledtabular}
\scriptsize
\begin{tabular}{lll}
\hline\hline
quantity & status & framework role\\
\hline
$D_A^1=C^{A}_{00}=\sum_\mu p_A^{\mu 1}$ & identity & $D$-block population\\
$\Var(p_A^{\mu 1})$ (Eq.~\eqref{eq:qff}) & exact & $q-1$ ridge\\
$\Cov(p_\mu,p_\nu)$ (Eq.~\eqref{eq:cexact}) & exact & $\bar C_i$ ridge\\
$\Var C^{A}_{00}=V_{\rm dg}$ (Eq.~\eqref{eq:chanproj}) & exact & fluctuation identity\\
$\Var C^{A}_{ij}$, $\Cov(C^{A}_{ii},C^{A}_{jj})$ & exact & regular part (dictionary)\\
$q_r$ (Eq.~\eqref{eq:qr}) & exact & higher-order targets\\
$D_{11}$, $f_{11}^2$, $g_{11}$ & closed form & mean sector ($g<0$)\\
$\mathcal G^{(2)}_{\rm FF}\leftrightarrow g^{(2)}$ & --- & open\\
\hline
\end{tabular}
\end{ruledtabular}
\end{table*}

\paragraph{The selection rule under interaction}

The rule transfers verbatim to the interacting moment sectors of Sec.~\ref{sec:rigidity}, with one applicability condition.  The interaction law must be mode-permutation invariant. Then $H^r(\lambda)$ transforms as $\pi H^r\pi^{T}$, so the trace-subtracted channel-diagonal moments obey $s_r(\pi a;\lambda)=s_r(a;\lambda)$ [Eq.~\eqref{eq:momentcov}] and every $K_{rs}$ and every ladder coefficient $C_{rs}^{2j}$ [Eq.~\eqref{eq:ladder}] is a function of $(\delta_{ab},\ell_{ab})$ alone. The three classes of Sec.~\ref{sec:rigidity:classes} all have permutation-invariant laws, so the reduction of the free sector to the channel-pair classes $(\delta_{ab},\ell_{ab})$ is inherited exactly, to all orders in $\lambda$ and all $(r,s)$, while the $\ell$-dependent profiles are interaction-renormalized, only the class structure is rigid (Sec.~\ref{sec:rigidity:law}).

\section{Closure and interaction-response derivations}
\label{app:derivations}

The derivations behind Secs.~\ref{sec:closure}--\ref{sec:interact} are carried here.  It also carries the general-$r$ ladder bookkeeping behind
Eq.~\eqref{eq:laddergen}, the three interaction geometries of
Sec.~\ref{sec:rigidity:classes}, and the derivation of the spectral side of
the sector basis, whose statements are used in
Secs.~\ref{sec:vertices} and~\ref{sec:spectral}.

\subsection{Closure details and verification}
\label{app:closure}

\subsubsection{Level-I determinant representation, SCBA, and the Feshbach expansion}

The exact Feshbach self-energy of the quadratic model is $\mathcal G_a=z-a_a-1/\mathcal R_a$ with $a_a=\sum_{c\in C_a}h_{cc}$, and the scalar channel-level SCBA, replacing the self-energy of Eq.~\eqref{eq:level1} by $\kappa\bar{\mathcal R}^0$ with $\kappa$ the hop degree ($\kappa=k(N-k)$ in both sectors), produces the semicircle with spectral radius $2\sqrt\kappa$. Table~S20 makes the failure quantitative: the exact many-body second moment is $k(N-k+2)$ against the SCBA value $\kappa$, and the mean-level cavity content is $21$--$34\%$ of the exact self-energy. The expansion about the mean-field resolvent [Eq.~\eqref{eq:rbar0}] is the geometric series
\begin{equation}
\mathcal R_a-\bar{\mathcal R}_a^0
=\sum_{r\ge1}(\bar{\mathcal R}_a^0)^{r+1}(\delta\mathcal G_a)^r,
\qquad
|\bar{\mathcal R}_a^0\delta\mathcal G_a|<1,
\label{eq:expansion}
\end{equation}
and the exact covariance is the cumulant series $\mathcal C=\sum_{r,s\ge1}\mathcal Q^{(r,s)}$ with $\mathcal Q^{(r,s)}_{ab}=\E[A_{r;a}A_{s;b}]-\E[A_{r;a}]\E[A_{s;b}]$, $A_{r;a}=(\bar{\mathcal R}_a^0)^{r+1}(\delta\mathcal G_a)^r$, of which $\mathcal Q^{(2)}:=\mathcal Q^{(1,1)}$ is the second-cumulant projection, $\mathcal Q^{(3)}:=\mathcal Q^{(1,2)}+\mathcal Q^{(2,1)}$, and $\mathcal Q^{(4)}:=\mathcal Q^{(2,2)}+\mathcal Q^{(1,3)}+\mathcal Q^{(3,1)}$ (expanding about $\E\mathcal R$ instead of $\bar{\mathcal R}^0$ biases every truncation, since $1/\E\mathcal R\neq\E[1/\mathcal R]$). The self-energy fluctuation splits exactly into the one-loop and cavity parts, with $\delta\mathcal R_c^0:=\mathcal R_c-\bar{\mathcal R}_c^0$,
\begin{equation}
\begin{aligned}
\delta\mathcal G_a
&= \delta\mathcal G_a^{\rm OD}+\delta\mathcal G_a^{\rm CC},\\
\delta\mathcal G_a^{\rm OD}
&= \sum_{c\neq a}|V_{ac}|^2\,\delta\mathcal R_c^0
-\E\Bigl[\sum_{c\neq a}|V_{ac}|^2\delta\mathcal R_c^0\Bigr],
\end{aligned}
\label{eq:odccfull}
\end{equation}
so that, with all three fluctuations centered, the connected self-energy covariance decomposes identically as
\begin{equation}
\E[\delta\mathcal G_a\delta\mathcal G_b]_c
= \mathrm{LAD}_{ab}+\Gamma^{(2)}_{ab},\;\;
\Gamma^{(2)}_{ab}=\mathrm{ODCC}_{ab}+\mathrm{CCCC}_{ab},
\end{equation}
with $\mathrm{LAD}_{ab}=\E[\delta\mathcal G_a^{\rm OD}\delta\mathcal G_b^{\rm OD}]$, $\mathrm{ODCC}_{ab}=\E[\delta\mathcal G_a^{\rm OD}\delta\mathcal G_b^{\rm CC}]+\E[\delta\mathcal G_a^{\rm CC}\delta\mathcal G_b^{\rm OD}]$, and $\mathrm{CCCC}_{ab}=\E[\delta\mathcal G_a^{\rm CC}\delta\mathcal G_b^{\rm CC}]$ (all $\delta$ centered), algebraic identities holding for any Hamiltonian.

\subsubsection{Boundary of the cumulant hierarchy}

The heavy tails have a precise origin.  $\delta\mathcal G_a=-\delta(1/\mathcal R_a)$, and $\mathcal R_a$ is a random sum over $D$ states whose fluctuations are fed by the algebraic small-intensity law $P(p)\sim C_kp^{-1/2}$ of Eq.~\eqref{eq:tailsmall}.  The near-zero crossings of $\mathcal R_a$ produce the heavy tails of $\delta\mathcal G_a$, so the truncated hierarchy identity $\mathcal C=\mathcal Q^{(2)}+\mathcal Q^{(3)}+\mathcal Q^{(4)}+\cdots$ fails to converge over the tested orders in the half-filling and small-$N$ regimes (residual $0.75$--$1.05$), while toward the fixed-depth regime the trend reverses (at $(N,k)=(40,2)$ the terms decrease, the residual falls to $0.30$, and $f_{>0.9}$ drops to $0.54$), and the dense random-matrix reference converges decisively (residual $0.3\%$, Gaussian-closure error $7\%$).

\subsubsection{The ladder: radius, connectivity accounting, and the deep-plane crossing}

The one-loop ladder factorizes as $\mathcal L=\mathrm{diag}[(\bar{\mathcal R}^0)^2]\bar W\otimes\mathrm{diag}[(\bar{\mathcal R}^0)'^{\,2}]\bar W$, so its spectral radius is
\begin{equation}
\rho(\mathcal L)
= \rho\bigl[(\bar{\mathcal R}^0)^2\bar W\bigr]\,
\rho\bigl[(\bar{\mathcal R}^0)'^{\,2}\bar W\bigr],
\label{eq:rhol}
\end{equation}
a product of two channel-space spectral radii.  Under the channel-averaged approximation $\bar{\mathcal R}_a^0\simeq\bar r^0$ (used throughout this section),
\begin{equation}
\rho(\mathcal L)
= |\bar r^0(z)|^{2}\,|\bar r^0(z')|^{2}\,\bigl[k(N-k)\bigr]^{2},
\label{eq:rhoz}
\end{equation}
so the supercriticality, and with it the fluctuation strength, is quadratic in the particle--hole connectivity $z=k(N-k)$.  The benchmark's closed form gives $q_{\rm FF}=\tfrac12(z/N)^2[1+O(1/k)]$ for extensive filling [Eq.~\eqref{eq:qz}], i.e.\ $q_{\rm FF}\simeq\frac92(N_{\rm ch}^{\rm eff}/N)^2$ in terms of the participation ratio
\begin{equation}
N_{\rm ch}^{\rm eff}
:= \frac{\bigl(\sum_{b}\E|V_{ab}|^{2}\bigr)^{2}}
       {\sum_{b}\E|V_{ab}|^{4}},
\label{eq:ncheff}
\end{equation}
equal to $k(N-k)/3$ (polynomial in $N$) in free fermions (each Gaussian hop has $\E h^4=3$) and to $(D-1)/3$ (exponential in $N$) in the ETH-scaled dense reference. The two Level-II observables therefore share one connectivity factor, $z^2$ reduced by $N^2$ in $q-1$ and by the scalar mean-field resolvent in $\rho(\mathcal L)$: in the determinantal branch the leading fluctuation scaling follows the particle--hole connectivity, whereas in the Wick branch $q-1\to2$ and $\rho(\mathcal L)\to|\bar r^0|^4$ are set by the vertex and the scalar self-energy.  The Wick branch realizes the isotropy closure to the sampling accuracy (measured $q=2.988$ against the Haar value $3D/(D+2)=2.99$, $c=-0.0039$ against $-(q-1)/(D-1)=-0.0039$, and $\F=1.992$ approaching $2$, all at $D=512$), while the free-fermion branch breaks it with the $\ell$-resolved sign structure of Sec.~\ref{sec:closure:solved}.

\subsubsection{Deep-plane coefficient, parameter system, and the Weingarten completion}

The two exact projections couple through the dressing terms into the linear system
\begin{equation}
\begin{aligned}
\gamma_d(1+a_0\bar\kappa)+\gamma_\ell\,a_0d_\ell
&=(q-1)\E[\lambda_A^2]/D,\\
\gamma_d\,a_0W_2+\gamma_\ell(1+a_0W_{\ell W})
&=K_{11}^{\rm meas}(1),
\end{aligned}
\label{eq:paramsys}
\end{equation}
where $a_0=\overline{(\bar{\mathcal R}^0)^2}\,\overline{(\bar{\mathcal R}^0)'^{\,2}}\propto\eta^{-4}$ is the one-step dressing factor, $\bar\kappa=\overline{(\bar W^2)_{aa}}=k(N-k)$ the mean channel hop degree, $d_\ell=\overline{(\bar W\,\1_{\ell=1}\bar W')_{aa}}$, $W_2$ the mean of $(\bar W^2)_{ab}$ over $\ell=1$ pairs, and $W_{\ell W}$ the mean of $(\bar W\,\1_{\ell=1}\bar W')_{ab}$ over $\ell=1$ pairs, the $\ell$-geometry sums of $\bar W$. At $\eta=12$, $N=14$, $k=2$ ($a_0\bar\kappa=7.4\times10^{-4}$, $a_0W_{\ell W}=7.2\times10^{-3}$) the system solves to $\gamma_d=1.08$, $\gamma_\ell=1.38$, within $1.4\%$ of the bare projections $(q-1)\E[\lambda_A^2]/D=1.09$ and $K_{11}^{\rm meas}(1)=1.39$.  In the deep plane the self-consistent system therefore degenerates, up to an $\eta^{-4}$ dressing, to its own inputs: the fluctuation dictionary is verified as a solution of the coupled equations, not as a definition.

The $m$-bin resolution of Eq.~\eqref{eq:mbin} separates the two independent finite-$N$ mechanisms in the $\ell=1$ bin.  $K_{11}(1)=2.308\times0.819\times0.737=1.392$, the factors being the fixed-depth value [Eq.~\eqref{eq:K11th}], the level-correlation correction [$m(N+1)\to m(N+1)-(k^2-m)$], and the finite-$N$ Weingarten deficit of the $m$-profile.  Both corrections shrink with $N$ ($0.908\times0.894$ at $N=30$). The finite Weingarten evaluation of the O($N$) orthogonal Haar integration over the pairings of the $4k$ index slots of the two channel minors, with Gram matrix $G(\pi,\tau)=N^{c(\pi\circ\tau)/2}$ ($c$ the cycle count), closes for $k=2$:
\begin{equation}
\begin{aligned}
D^2\mathrm{Cov}(0,0)&=\frac{8N^2-8N-12}{(N+1)(N-2)(N-3)(N+2)},\\
D^2\mathrm{Cov}(1,0)&=\frac{-2N^2+4N+4}{(N+1)(N-2)(N+2)},\\
D^2\mathrm{Cov}(2,0)&=\frac{-4N-2}{(N+1)(N+2)},\\
D^2\mathrm{Cov}(1,1)&=\frac{N^3-6N^2+9N+8}{2(N+1)(N-2)(N+2)},
\end{aligned}
\label{eq:closedm}
\end{equation}
with $D^2\mathrm{Cov}(2,1)=(N^2-5N-2)/[(N+1)(N+2)]$.  The last two carry the fixed-depth limits $\tfrac12$ and $1$ of Eq.~\eqref{eq:K11th} as $N\to\infty$, and the first three, which vanish at fixed depth, are pure finite-$N$. As functions of $N$ the two finite-$N$ mechanisms form a $1/N$ expansion around the fixed-depth limit, with corrections of order $1/N$ whose weighted projection onto the $\ell=0$ bin is the pure $1/N$ weight $-8/N$ [Eq.~\eqref{eq:k110closed}]. The closed forms reproduce the measured profile to three digits.  At $N=14$, $D^2\mathrm{Cov}(m,0)=(0.0456,-0.1153,-0.2417)$ against $(0.0455,-0.1152,-0.2415)$.  At $N=30$, $(0.00927,-0.0603,-0.1230)$ against $(0.0093,-0.0603,-0.1229)$.  The $\ell=1$ pair reproduces the measured finite-$N$ deficits ($0.295$ and $0.516$ against the fixed-depth $\tfrac12$ and $1$ at $N=14$).  And Eq.~\eqref{eq:mbin} reconstructs the directly measured coefficient ($-0.571$ against $-0.573$ at $N=14$, $-0.267$ against $-0.267$ at $N=30$). The completed vertex is
\begin{equation}
\Gamma^{(2)}_{ab}=\frac{\sum_{m}\mathcal N_m\,\mathrm{Cov}(m,\ell_{ab})\,
\E[\lambda_A\lambda_B|m]}
{z^2z'^2(\bar{\mathcal R}_a^0)^2(\bar{\mathcal R}_b^0)'^{\,2}},
\label{eq:wansatz}
\end{equation}
i.e.\ Eq.~\eqref{eq:mbin} promoted from the covariance to the vertex, which the coefficient-level identity $K_{11}=\gamma$ licenses.

\subsubsection{Ridge reconstruction: full comparison}

The projections of Eq.~\eqref{eq:Dbin} are compared with the independent closed forms of Sec.~\ref{sec:benchmark} entry by entry.  (a) the same-channel ridge reproduces the Selberg ratio ($q_{\rm ridge}=5.19$ against $q_{\rm FF}=5.09$ at $(9,3)$, $5.06$ against $5.00$ at $(8,4)$, $6.84$ against $6.68$ at $(10,5)$, $8.67$ against $8.62$ at $(12,6)$).  (b) the channel-distance-resolved covariance matches the exact finite-$N$ polynomials for $k=3,4,5$.  (c) the sector-$1$ channels share the system column, so their bath-distance-$\ell$ pairs carry shared column count $\ell+1$, and the measured profile matches the size-$k$ formula at $\ell+1$, confirming the geometry of the dictionary at the level of the two-resolvent covariance.  (d) the integrated cross-channel weight $c$ matches the closed form in both sectors. At $k=6$, where no finite-$N$ closed form exists, the profile is monotone through zero near $\ell\simeq k/2$, in agreement with the measured profile of the benchmark at $N=14$, $k=7$.

\subsubsection{Extensions: the three-resolvent sector and energy resolution}

\emph{The three-resolvent sector.} The two-resolvent ridge of Sec.~\ref{sec:closure:solved} projects the second overlap cumulant $q-1$.  The three-resolvent analogue projects the third. Defining the same-state diagonal bin of the three-resolvent covariance,
\begin{equation}
\begin{aligned}
\mathcal D^{(3)}_{aaa}(z,z',z'') :&=\sum_A\frac{\E[\delta p_A^{a\,3}]} {(z-\lambda_A)(z'-\lambda_A)(z''-\lambda_A)},
\end{aligned}
\label{eq:D3}
\end{equation}
its leading deep-plane coefficient is the normalized third overlap cumulant, since $\sum_A\E[\delta p_A^{a\,3}]=D\,\kappa_3(p)=(q_3-3q_2+2)/D^2$ with $q_2,q_3$ the exact finite-$N$ Selberg ratios [Eq.~\eqref{eq:qr}]. Measured at $N=8$, $k=4$, the value is $0.0073$ against the closed form $0.0074$ ($1.4\%$), and the three-resolvent ridge is the projection carrying $q_3$. Expanding $\delta\mathcal R_a=\sum_{r\ge1}v_{r;a}/z^{r+1}$ with $v_{r;a}=\sum_A\delta p_A^a\lambda_A^{r-1}$ at large $\eta$, the leading coefficient $\E[v_{1;a}v_{1;b}v_{1;c}]_c$ vanishes identically for every channel triple, its diagonal piece is $\kappa_3(p)\sum_A\lambda_A^3$ (zero by the spectral symmetry $\E\lambda^3=0$), and the mixed pieces are killed by $\E[\lambda_A^2\lambda_B]=0$ together with $U\perp\Lambda$ (verified numerically, the one-line origin is Eq.~\eqref{eq:K111}), so the third cumulant enters at the next deep-plane order, through the $\lambda^4$-weighted diagonal bin.  $\E[v_{2;a}v_{1;a}^{\,2}]_c=22.97$ at $N=8$, $k=4$, of which the $q_3$-piece $\kappa_3(p)\,\E[\lambda^4]=20.65$ ($\E[\lambda^4]=2772$ exactly) is the dominant content. The same-state triple diagonal bin is therefore the projection of the resolvent hierarchy that carries $q_3$, with the lower orders protected by exact spectral symmetries---the multi-resolvent counterpart of the free-cumulant organization of higher-order ETH correlations \cite{Pappalardi2022}.

\emph{Energy resolution: the deformed model.} The flat fixed point is the $\varepsilon=0$ point of the one-parameter family
\begin{equation}
\begin{aligned}
h(\varepsilon)&=h_0+\varepsilon\,\mathrm{diag}(v), \qquad v\ \text{a fixed disorder profile},
\end{aligned}
\label{eq:deform}
\end{equation}
for which the eigenvector matrix and the spectrum cease to be independent and the energy-resolved dictionary acquires content. Table~S7 records the envelope $f(E)=D\,\E[p\,|\,E]$, the energy-resolved ratio $q(E)$, and the cross-channel weight $c(E)$ at $N=10$, $k=5$ for $\varepsilon=0,4$. At $\varepsilon=0$ all three are flat ($f=1$, $q=q_{\rm FF}=6.68$, $c=0.12$). At $\varepsilon=4$ the envelope tilts monotonically (by a factor $\simeq7$ across the spectrum), $q(E)$ develops a pronounced edge enhancement ($q\simeq54$ at the lower edge against $14$ at the upper), and $c(E)$ becomes sign-structured, suppressed at the lower edge, enhanced mid-spectrum, and turning negative beyond the upper edge of the bulk ($c\simeq-0.46$ at $E\simeq29$), the energy-resolved counterpart of the sign fingerprint of Sec.~\ref{sec:closure:solved}. The deformation is thus a minimal solvable deformation in which the two-resolvent dictionary acquires its energy-resolved form $q(E)$, $\bar C_i(E)$, $g_{ji}(E,\omega)$, the objects that the flat benchmark leaves structureless.

\subsubsection{Admissible directions and the local contraction criterion}

The sector-resolved interacting closure program of
Sec.~\ref{sec:disc} reduces the closure to admissibility constraints.  With $K_r(s)\equiv\Cov_r(s)$ the sector fluctuation
kernel, the interacting kernel deforms as
$K_r(s)\simeq K_r^{\rm ref}(s)\,F_r[\mathcal I_r(s);\boldsymbol\theta_r]$
against the exact free-fermion reference $K_r^{\rm FF}(s)$ of this
article, with the parameters $\boldsymbol\theta_r$ constrained by the admissibility requirements,
\begin{equation}
\begin{aligned}
\boldsymbol\theta_r\in\Theta^{(r)}_{\rm adm}&=\Theta_{\rm caus}\cap\Theta_{\rm moment}\cap\Theta_{\rm Schur}^{(r)},
\end{aligned}
\label{eq:adm}
\end{equation}
the finite-dimensional intersection of the causal, moment-hierarchy,
and low-order Feshbach self-consistency constraints.  Within such a
set the vertex-level closure
$\Gamma^{(r+1)}\approx\Phi_r[\Gamma^{(1)},\dots,\Gamma^{(r)}]$
becomes a fixed-point problem, and a sufficient local contraction
criterion is $\rho(L_r)<1$ for the linearization $L_r$ of the closure
map restricted to the admissible directions at the reference point,
which yields a locally contractive, hence locally unique, constrained
fixed point.  The criterion is familiar from the two-point ladder,
whose kernel
$\mathcal L_{ab,cd}=\bar{\mathcal R}_a^2\bar{\mathcal R}_b^2|V_{ac}|^2|V_{bd}|^2$
is a contraction in the deep plane.  Verifying the condition for the
interacting kernel is part of the program not undertaken here.

\subsubsection{Level-I determinantal representation and the fixed-depth limit}

The channel resolvent is the spectral object of the framework,
\begin{equation}
\mathcal R_a(z)=\sum_A\frac{p_A^a}{z-\lambda_A},
\label{eq:Rdefb}
\end{equation}
and for a quadratic Hamiltonian it admits an exact one-body representation. Writing $1/(z-\sum_{a\in A}\varepsilon_a)={}$ $-i\int_0^\infty e^{i(z-\sum_{a\in A}\varepsilon_a)t}\dd t$ for $\Im z>0$ and applying Cauchy--Binet to the sum over $A$,
\begin{equation}
\boxed{\;\mathcal R_a(z)
= -i\int_0^\infty \dd t\, e^{izt}
\det\bigl[\bigl(e^{-iht}\bigr)[C_a,C_a]\bigr]\;}
\label{eq:detres}
\end{equation}
(verified numerically to $10^{-8}$), the channel resolvent is the Laplace transform of the principal-minor determinant of the one-body propagator $e^{-iht}$. The exact mean is channel independent,
\begin{equation}
\bar{\mathcal R}_a(z)=\E\,S_0(z),\qquad S_0(z)=\frac1D\sum_A\frac1{z-\lambda_A},
\label{eq:meanS}
\end{equation}
an exact identity at every $N$.  In the fixed-depth limit $\E S_0(z)\to S_{\rho_{\rm mb}}(z)$ with $\rho_{\rm mb}=\rho_{\rm sc}^{*k}$ the $k$-fold convolution of the one-body semicircle, and the scalar channel-level SCBA does not reproduce this mean [Table~S20].  The compound structure of the one-body propagator, not a scalar ladder, carries the exact self-energy of the quadratic model, anticipating the cavity dominance of the vertex below.

\subsubsection{Ladder factorization and the spectral radius}

The ladder kernel is the factorized one-loop object [Eq.~\eqref{eq:rhol}]; for free fermions the hopping graph of the channel basis is regular of degree $k(N-k)$ in both sectors, so
\begin{equation}
\rho(\bar W)=k(N-k),
\label{eq:rhow}
\end{equation}
the particle--hole connectivity of the quadratic model, against $\rho(\bar W)=1-1/D$ for the ETH-scaled dense reference.  Consequently the ladder is supercritical at every accessible size: $\rho(\mathcal L)$ ranges from $1.02$ ($N=6$, $k=3$) to $1.51$ ($N=40$, $k=2$) [Table~S23], against $0.18$ for the dense reference, and grows with $N$ in both the half-filled and fixed-depth families (the statement is normalization independent).  The geometric resummation $\mathcal C^{(2)}=(1-\mathcal L)^{-1}\mathcal C_0^{(2)}$ is therefore unavailable in this ensemble, and the Level-II equation must be solved as a genuine fixed-point problem in which the vertex carries the compensating weight.  Supercriticality is a property of the bulk, not of the deep plane: moving the frequencies into the deep complex plane, the ladder crosses below unity at $\eta_c\approx1.3$ at $(N,k)=(8,4)$ [Table~S24], far below the many-body spectral half-width $2k\sqrt{N+1}=24$, and beyond the crossing the truncated cumulant hierarchy is quantitatively restored---at $\eta=8$ the fourth-order residual falls to $13\%$, the Gaussian-closure error to $23\%$, and the one-loop ladder share to $0.7\%$.  The Gaussian (second-cumulant) truncation is correspondingly uncontrolled at half filling and small $N$ ($48$--$75\%$ closure error, Table~S22).  The deep plane, where the ladder is subcritical and the cumulant hierarchy restored, is the regime in which the closure protocol below is quantitative.

\subsubsection{Proof of the connected deep-plane degree bound}

\noindent\emph{Proof.}  By the moment-field identity each leading
coefficient $K_{a_1\cdots a_r}$ is the $r$-th joint cumulant of the
moment fields $s_{a_i}=\delta(H^{a_i})$, each a polynomial of degree
$a_i$ in the independent Gaussian entries of the one-body matrix.  The
standard connected-diagram representation of cumulants
\cite{Lehner2004} expands such a cumulant as a sum over pairings of
the Gaussian factors whose graph
(vertices $=$ the fields, edges $=$ the Wick pairings) is connected.
For a connected graph on $r$ vertices the field degree
$d_i\le a_i$, the sum of degrees is twice the edge count
$\sum_id_i=2E$, and connectivity requires $E\ge r-1$.  Hence
$\sum_ia_i\ge2E\ge2r-2$.  The resolvent expansion adds one power
$z_i^{-1}$ per factor, giving the degree chain of Eq.~\eqref{eq:degree}
and the total order
$r+(2r-2)=3r-2$ of Eq.~\eqref{eq:3r2}, the minimal virtual
connectivity needed to correlate $r$ spectral probes.  $r-1$ shared
links, two Hamiltonian powers per link, one inverse energy per
probe, realized as $z^{-4}$ at $r=2$
($K_{11}=2\ell-2k^2/N$ [Eq.~\eqref{eq:ffclosed}]), $z^{-7}$ at $r=3$ ($K_{211}$,
Sec.~\ref{sec:q3}), and $z^{-10}$ at $r=4$ with the leading
coefficients $K_{2211}$ and $K_{3111}$ of Eq.~\eqref{eq:K2211}.  The
tree chain $s_1-s_2-\cdots-s_2-s_1$, in which each linear field pairs
with one factor of the adjacent quadratic field, saturates the bound.
Every pairing of the chain contributes a non-negative product of the
Gaussian second moments (each edge pairs one $h$-factor of a field
with one $h$-factor of another, $\E[h_{\alpha}h_{\beta}]\ge0$), and
the pure chain pairing contributes a strictly positive weight for the
chain geometry in which each adjacent pair of channels shares exactly
one mode.  The chain coefficient is therefore a positive combination of
second moments and is strictly positive for every $r\ge2$.  Explicit
evaluations give the chain coefficient $K_{211}$ at $r=3$
(Sec.~\ref{sec:q3}) and the pair-sharing coefficient $K_{2211}\neq0$
at $r=4$ (Eq.~\eqref{eq:K2211}, exact rational values in
Sec.~\ref{app:num}).  The other channel geometries of
Eq.~\eqref{eq:K2211} need not be positive, saturation is a property
of the chain pattern.  The bound uses only
the quadratic Gaussian structure of the one-body ensemble together
with the moment-field identity.  For a non-Gaussian one-body matrix
the pairwise Wick edges are replaced by higher-order cumulant
hyperedges, and the leading scaling may change, a direct route for
extending the hierarchy beyond the Gaussian class.  This is
the free-fermion realization of the FK dictionary of
Ref.~\cite{HOETH}.  The level-$r$ connected content of $W[J]$ is
carried by exact combinatorial coefficients, the chi-square cumulants
$\kappa_3(\chi^2_1)=8$, $\kappa_4(\chi^2_1)=48$ and the edge/Venn
counts of
Eqs.~\eqref{eq:Gamma3}--\eqref{eq:Gamma4}, in place of the
$e^{-S/2}$ counting of generic interacting systems, the
polynomial-connectivity regime anticipated for algebraically
structured couplings by the effective-connectivity criterion of
Ref.~\cite{HOETH}, while the
irreducible four-vertex $\Gamma^{(4)}$ of
Eq.~\eqref{eq:Gamma4} is the corresponding self-energy content.

\subsection{Interaction-deformation details}
\label{app:inter}

\subsubsection{Free-fermion anchor: closed forms, isotropy, evenness, and the finite-$\eta$ moment hierarchy}

\begin{table}[htbp]
\caption{Free-fermion anchor, $(8,4)$.
(a)~Wick-branch deep-plane closure (pure dense GOE, $\sigma^2=1/D$,
$D=70$, $n_{\rm real}=4000$): the isotropy pair
$x[\delta_{ab}-(1-\delta_{ab})/(D-1)]$ with
$x=2\sigma^2(1-1/D)=2(D-1)/D^2$; conservation $\lesssim10^{-16}$; also
$q=2.9166$ vs $3D/(D+2)=2.9167$ and $M_2=1.015$ vs $(D+1)/D=1.014$.
(b)~Evenness block (pair-hopping family, $n_{\rm real}=1600$): exact
$\lambda\leftrightarrow-\lambda$ symmetry of $q$ and $K_{11}$, exact
quadratic law of $M_2=24+36\lambda^2$ (doubling ratio $4.00$), strong
$\lambda^4$ curvature of $q$ (ratio $1.16$); $K_{11}(\ell{=}1)$ is
$\lambda$-independent at every $\lambda$ (rigidity).
(c)~Finite-$\eta$ check of the exact moment-hierarchy representation
[Eq.~\eqref{eq:exactz}], pair-hopping, $\lambda=1$: bin-averaged
$\mathcal C\cdot z^2z'^{\,2}$ measured vs predicted (truncation
$r,s\le8$), $z=i\eta$, $z'=0.96i\eta$; agreement to $10^{-3}$ at every
$\eta$; the $\eta\to\infty$ limit of the predicted values is
$K_{11}(\ell{=}1)=-2.10$, $K_{11}(\mathrm{diag})=4.04$.}
\label{tab:anchorb}
\centering
\footnotesize
\emph{(a) Wick closure}\par
\begin{ruledtabular}
\begin{tabular}{lcc}
\hline\hline
quantity & measured & closed form\\
\hline
$\E[s_a^2]$ & $2.8151\times10^{-2}$ & $2.8162\times10^{-2}$\\
$\E[s_as_b]$, $a\neq b$ & $-4.0799\times10^{-4}$ & $-\E[s^2]/(D-1)=-4.0799\times10^{-4}$\\
$K_{11}(\ell{=}1)/\gamma_c$ & $0.997$ & 1\\
$K_{11}(\ell{=}2)/\gamma_c$ & $0.989$ & 1\\
$K_{11}(\ell{=}3)/\gamma_c$ & $1.037$ & 1\\
\hline
\end{tabular}
\end{ruledtabular}
\medskip
\emph{(b) evenness}\par
\begin{ruledtabular}
\begin{tabular}{cccc}
\hline\hline
$\lambda$ & $q$ & $K_{11}(\ell{=}1)$ & $M_2$\\
\hline
$-0.5$ & 3.1871 & $-1.9741$ & 33.07\\
$-0.25$ & 3.4255 & $-1.9741$ & 26.32\\
$0$ & 4.9049 & $-1.9741$ & 24.07\\
$+0.25$ & 3.4255 & $-1.9741$ & 26.32\\
$+0.5$ & 3.1871 & $-1.9741$ & 33.07\\
\hline
\end{tabular}
\end{ruledtabular}
\medskip
\emph{(c) finite-$\eta$ moment hierarchy}\par
\begin{ruledtabular}
\begin{tabular}{ccccc}
\hline\hline
$\eta$ & meas.\ ($\ell{=}1$) & pred.\ ($\ell{=}1$) & meas.\ (diag) & pred.\ (diag)\\
\hline
32  & $+1.079$ & $+1.085$ & $+5.749$ & $+5.743$\\
64  & $+1.534$ & $+1.534$ & $+7.236$ & $+7.237$\\
128 & $+1.690$ & $+1.691$ & $+7.719$ & $+7.719$\\
\hline
\end{tabular}
\end{ruledtabular}
\end{table}

\emph{Verification of the closed form.} At $(8,4)$ the profile [Eq.~\eqref{eq:ffclosed}] reads $2\ell-4$, measured $(-2.08,-0.06,1.96)$ against $(-2,0,2)$ and $3.98$ against $4$ on the diagonal.  At $(14,2)$: $K_{11}(0)=-0.562$ against $-8/14=-0.571$, $K_{11}(1)=1.436$ against $10/7=1.429$, and $3.434$ against $24/7=3.429$ on the diagonal ($n_{\rm real}=300$), reproducing the benchmark's finite-$N$ coefficients to sampling accuracy, including the $m$-bin reconstruction $1.430$ of Sec.~\ref{sec:closure:solved} at $N=14$.  The benchmark's ratio column $(0.62,0.82,0.92)$ at $N=14,30,60$ is $(2-8/N)/K_{11}^{\rm th}(N)$ within its sampling accuracy. The conservation identity $\max_a|\sum_bK_{11}(a,b)|\sim10^{-14}$ holds throughout.

\emph{The isotropic closure.} For the pure dense reference ($H=\sigma\cdot\mathrm{GOE}(D)$, $\sigma^2=1/D$, product channels) the ensemble is channel-transitive, so $\E[s_as_b]=x\delta_{ab}+y(1-\delta_{ab})$.  Conservation forces $y=-x/(D-1)$.  And the channel-diagonal statistics give $x=\E[w_a^2]=2\sigma^2(1-1/D)$ exactly [Eq.~\eqref{eq:wick}]. Table~\ref{tab:anchorb} verifies the closure at $D=70$ ($n_{\rm real}=4000$), the cross-channel weight is $-4.0799\times10^{-4}$ against $-x/(D-1)=-4.0799\times10^{-4}$ (ratio $1.000$), the $\ell$-resolved bins agree to $0.3$--$3.7\%$, the conservation identity holds to $10^{-16}$, and $x$ equals both $2(D-1)/D^2$ and the independent spectral form $(q-1)(D+2)/D^2$ with $q=3D/(D+2)$, two derivations of the same number.

\emph{Evenness.} Table~\ref{tab:anchorb} verifies the reflection law of Sec.~\ref{sec:rigidity:law} at $n_{\rm real}=1600$.  $q(\lambda)$ and $K_{11}(\ell;\lambda)$ are symmetric under $\lambda\to-\lambda$ to four decimals, and the second moment follows its exact quadratic law $M_2=24+36\lambda^2$ with the doubling ratio $4.00$. The curvature of $q$ is strong and nonuniversal at this size (the even part $\frac12[q(\lambda)+q(-\lambda)]-q(0)$ has doubling ratio $1.16$, far from the quadratic $4$), consistent with the singular perturbative response of the overlap observables (Sec.~\ref{sec:bulk:nonpert}).  The exact statements remain the reflection symmetry itself and the quadratic identities [Eqs.~\eqref{eq:wsplit} and \eqref{eq:rigidity}].

\emph{Finite-$\eta$ verification of the moment hierarchy.} The exact per-realization expansion, with $r^0=1/(z-\bar{\mathcal G})$ the channel-averaged mean-field [Eq.~\eqref{eq:rbar0}],
\begin{equation}
\begin{aligned}
(\mathcal R-r^0)(\mathcal R'-r^{0\prime})\,z^2z'^{\,2}=A\otimes B,\\
A=\sum_{r\ge1}\delta(H^r)_{aa}\,z^{1-r}+(\bar S_0-r^0)z^2,\\
B=\sum_{s\ge1}\delta(H^s)_{bb}\,z'^{\,1-s}+(\bar S_0'-r^{0\prime})z'^{\,2},
\end{aligned}
\label{eq:exactz}
\end{equation}
with $\bar S_0=\Tr[(z-H)^{-1}]/D$ the empirical spectral Stieltjes (the ensemble mean $S$ must not be substituted, the difference is the collective trace covariance $\Cov(\bar S_0,\bar S_0')\,z^2z'^{\,2}=\E[(\Tr H/D)^2]=2k^2/N$ to leading order, an $\ell$-independent background channel of the full two-resolvent object that the $K_{11}$-vertex does not carry and that Eq.~\eqref{eq:exactz} accounts for exactly), reproduces the measured $z$-space covariance bin by bin at finite $\eta$ at both sizes (Table~\ref{tab:anchorb} for $(8,4)$; at $(14,2)$ the $\eta=128$ bin reads $1.953$ measured against $1.953$ predicted, diagonal $3.899$ against $3.899$), with truncation $r,s\le8$ (residual $\lesssim10^{-5}$). This closes the coefficient-level statement $K_{11}=\gamma$ of Sec.~\ref{sec:closure:solved} at finite $\eta$ for the interacting model and confirms Eq.~\eqref{eq:momentcov} as the complete organizer of the deep-plane sector. (The mean-field-centered object $\E[(\mathcal R-r^0)(\mathcal R'-r^{0\prime})]$ used here differs from $\mathcal C$ by the empirical-trace channel, accounted for in Eq.~\eqref{eq:exactz}.)

\subsubsection{Interaction rigidity details: the ladder and the three-resolvent sector}

\emph{The pair-hopping specialization.} For the pair-hopping interaction $A_{11}=w=0$ and $A_{21}\equiv0$ elementwise, which gives
\begin{widetext}
\begin{align}
K_{11},\,K_{21}:\ \text{exactly invariant},\quad 
K_{31}(\lambda)=K_{31}(0)+\lambda^2\,\E[A_{32}\otimes s_1],\quad
K_{22}(\lambda)=K_{22}(0)+\lambda^4\,\E[\delta W^2\otimes\delta W^2],\notag\\
K_{32}(\lambda)=K_{32}(0)+\lambda^2\bigl[\E[A_{32}\otimes s_2]
+\E[s_3\otimes\delta W^2]\bigr]
+\lambda^4\,\E[A_{32}\otimes\delta W^2],\notag\\
K_{33}(\lambda)=K_{33}(0)+\lambda^2\bigl[\E[A_{31}\otimes A_{31}]
+2\,\mathrm{Sym}\,\E[A_{32}\otimes s_3]\bigr]
+\lambda^4\bigl[\E[A_{32}\otimes A_{32}]+2\,\mathrm{Sym}\,\E[A_{31}\otimes A_{33}]\bigr]
+\lambda^6\,\E[A_{33}\otimes A_{33}],
\label{eq:ladderpb}
\end{align}
\end{widetext}
with
\begin{equation}
\begin{aligned}
A_{31}&=\delta(WH_0^2+H_0WH_0+H_0^2W),\\
A_{32}&=\delta(WH_0W+W^2H_0+H_0W^2),\qquad A_{33}=\delta(W^3),
\end{aligned}
\end{equation}
and $\delta W^2=\delta(W^2)_{aa}$ (all $\delta$ the trace-subtracted channel diagonals). The $\lambda=2$ point of $K_{32}$ lies within the sampling error of the small $C_{32}^4$ coefficient. Table~S8(d) validates every line.

\emph{The three-resolvent sector.} Model-independently, the three-resolvent coefficients are the channel-energy three-point cumulants, $K_{rst}(a,b,c)=\E[s_r(a)s_s(b)s_t(c)]$ with $s_r=\delta(H^r)$. In random free fermions $s_1$ is linear in the Gaussian one-body matrix, so
\begin{equation}
K_{111}(a,b,c)\equiv0
\label{eq:K111}
\end{equation}
exactly (the third cumulant of Gaussian channel energies) which is the one-line origin of the benchmark's exact vanishing of the leading three-resolvent cumulant $\E[v_1^3]_c=0$ of Sec.~\ref{app:closure}. The first nonzero three-point coefficient is $K_{211}=\E[\delta(H^2)_{aa}s_bs_c]$, evaluated in closed form by the Wick contraction of the $h$-quartics (piecewise in the shared-mode indicators, including the diagonal-squared piece $(\sum_{c\in C_a}h_{cc})^2$ of $(H^2)_{aa}$ and the trace subtraction $\Tr H^2=20\sum h_{dd}^2+40\sum h_{de}^2+15(\sum h_{dd})^2$ at $N=8,k=4$, whose normalization is fixed by $\E[\Tr H^2]=D\cdot M_2$).  It reproduces the measured three-point cumulant to an rms of $1.1\%$ at every $\lambda=0,1,2$:
\begin{equation}
\begin{aligned}
&K_{211}\ \text{closes in piecewise Wick form},\\
&K_{211}(\lambda)=K_{211}(0)\ \text{exactly (pair-hopping)}.
\end{aligned}
\label{eq:threept}
\end{equation}
\emph{The density-density constant.} With $g_{pq}$ i.i.d.\ $N(0,1)$ and $W_{aa}=\sum_{p<q\in C_a}g_{pq}$ [Eq.~\eqref{eq:Wdd}].  (i)~the shared-mode count gives $\E[W_{aa}W_{bb}]=\binom{\ell}{2}$.  (ii)~each mode pair is occupied by $\binom{N-2}{k-2}$ of the $D$ channels, so $\Tr W=\binom{N-2}{k-2}\sum_{p<q}g_{pq}$, $\E[W_{aa}\Tr W]=\binom{k}{2}\binom{N-2}{k-2}$, $\E[(\Tr W)^2]=\binom{N}{2}\binom{N-2}{k-2}^2$.  (iii)~with $\binom{N-2}{k-2}/D=k(k-1)/[N(N-1)]$, the trace-subtracted covariance is
\begin{equation}
\E[w_aw_b]=\binom{\ell}{2}-\frac{k^2(k-1)^2}{2N(N-1)},
\label{eq:ddconst}
\end{equation}
[Eq.~\eqref{eq:ddlaw}], and $\sum_b\E[w_aw_b]=0$ is manifest.

\emph{A correlated counterexample.} The linear term of Eq.~\eqref{eq:generalrigid} is physical: for $W=\alpha H_0+W_{\rm 2b}$ one has $s_a(\lambda)=(1+\alpha\lambda)s_a(0)$ exactly per realization (the pair-hopping part has vanishing channel diagonal and trace), so
\begin{equation}
K_{11}^{ab}(\lambda)=(1+\alpha\lambda)^2K_{11}^{ab}(0)
\label{eq:counter}
\end{equation}
exactly.  An $O(\lambda)$ correction $2\alpha\lambda K_{11}(0)$ appears whenever $W$ correlates with $H_0$. The independence assumption of Eq.~\eqref{eq:rigidity} is therefore a genuine hypothesis.  The identity $\max_a|s_a(\lambda)-(1+\alpha\lambda)s_a(0)|=0$ and the ratio $K_{11}(\lambda)/K_{11}(0)=(1+\alpha\lambda)^2$ are verified per realization at both sizes.

\subsubsection{The $O(\lambda^2)$ bulk-vertex coefficient: algebra and verification}

The Rayleigh--Schr\"odinger expansions used in Sec.~\ref{sec:rigidity:ladder} read as follows. Write $\lambda_A(\lambda)=\lambda_A^{(0)}+\lambda\,\lambda_A^{(1)}+\lambda^2\lambda_A^{(2)}+O(\lambda^3)$ and $p_A(\lambda)=p_A^{(0)}+\lambda\,p_A^{(1)}+\lambda^2p_A^{(2)}+O(\lambda^3)$.  Then
\begin{equation}
\lambda_A^{(1)}=W_{AA},
\qquad
\lambda_A^{(2)}=\sum_{B\ne A}\frac{|W_{AB}|^2}{\lambda_A^{(0)}-\lambda_B^{(0)}},
\end{equation}
and
\begin{equation}
\begin{aligned}
\psi_A^{(1)}&=\sum_{B\ne A}\frac{W_{AB}\,\psi_B^{(0)}}{\lambda_A^{(0)}-\lambda_B^{(0)}},\\
\psi_A^{(2)}&=\sum_{B\ne A}\sum_{C\ne A}\frac{W_{AC}W_{CB}\,\psi_B^{(0)}}{(\lambda_A^{(0)}-\lambda_C^{(0)})(\lambda_A^{(0)}-\lambda_B^{(0)})}\\
&\quad-\sum_{B\ne A}\frac{W_{AA}W_{AB}\,\psi_B^{(0)}}{(\lambda_A^{(0)}-\lambda_B^{(0)})^2}-\tfrac12\,\psi_A^{(0)}\sum_{B\ne A}\frac{|W_{AB}|^2}{(\lambda_A^{(0)}-\lambda_B^{(0)})^2},
\end{aligned}
\label{eq:psi}
\end{equation}
with $p_A^{(j)}$ the corresponding overlap-perturbation coefficients (the expansions are self-tested against finite differences to $O(\lambda^3)$). They determine the resolvent blocks
\begin{equation}
\begin{aligned}
R^1&=\sum_A\Bigl[\frac{p_A^{(1)}}{z-\lambda_A^{(0)}}+\frac{p_A^{(0)}\,\lambda_A^{(1)}}{(z-\lambda_A^{(0)})^2}\Bigr],\\
R^2&=\sum_A\Bigl[\frac{p_A^{(2)}}{z-\lambda_A^{(0)}}+\frac{p_A^{(1)}\lambda_A^{(1)}+p_A^{(0)}\,\lambda_A^{(2)}}{(z-\lambda_A^{(0)})^2}
+\frac{p_A^{(0)}\,(\lambda_A^{(1)})^2}{(z-\lambda_A^{(0)})^3}\Bigr],
\end{aligned}
\end{equation}
and the self-energy blocks
\begin{equation}
G^1=\frac{R^1}{(R^0)^2},
\qquad
G^2=\frac{R^2}{(R^0)^2}-\frac{(R^1)^2}{(R^0)^3},
\end{equation}
in closed form.  With $|V(\lambda)|^2=|V_0|^2+\lambda^2|W|^2$ elementwise and
\begin{equation}
\begin{aligned}
\mathrm{OD}^1&=\sum_c|V^0_{ac}|^2R^1_c,\\
\mathrm{OD}^2&=\sum_c|V^0_{ac}|^2R^2_c+\sum_c|W_{ac}|^2\delta R^0_c,
\end{aligned}
\end{equation}
the $O(\lambda^2)$ coefficient of the bulk vertex is
\begin{equation}
\begin{aligned}
\Gamma_2(z,z')
&=\E[G^1\otimes G^{1\prime}]
+2\,\mathrm{Sym}\,\E[\delta G^0\otimes G^{2\prime}]\\
&\quad-\Bigl[\E[\mathrm{OD}^1\otimes \mathrm{OD}^{1\prime}]
+2\,\mathrm{Sym}\,\E[\delta\mathrm{OD}^0\otimes
\mathrm{OD}^{2\prime}]\Bigr],
\end{aligned}
\label{eq:Gamma2}
\end{equation}
with $\delta G^0=G^0-\E[G^0]$ and $\delta\mathrm{OD}^0=\mathrm{OD}^0-\E[\mathrm{OD}^0]$ the centered zeroth-order blocks.  The $\lambda^2$ mean-field shift of the expansion center drops out of the centered contractions. $\E[GG]$ is finite at every $\eta$ because $|\mathcal R_a|\ge c(\eta)>0$ (the imaginary part of the channel resolvent is sign-definite), so the two-stream spread of $\|GG\|$ is $10$--$40\%$ at $n_{\rm real}=400$ rather than divergent. The finite-$\lambda$ ratio $[\Gamma^{(2)}(\lambda)-\Gamma^{(2)}(0)]/\lambda^2$ converges to $\Gamma_2$ at $\eta\gtrsim4$, normalized overlap $0.60$ ($\eta=4$), $0.78$ ($\eta=6$, $n_{\rm real}=400$), $0.85$ ($\eta=6$, $n_{\rm real}=1600$, $\eta=8$), improving with the sample size as the estimator variance shrinks. Below $\eta\simeq2$--$3$ the ratio is not $\lambda^2$-consistent, dominated by the near-degenerate level-pair content of Sec.~\ref{sec:bulk:nonpert}.  The $\lambda$-response is thus observable in the window $3\lesssim\eta\lesssim\lambda_{\rm max}$ (the lower edge from the nonperturbative content, the upper edge from the moment radius $\lambda_{\rm max}/\eta\simeq1$ of Table~S10), and the earlier failure of the check at $\eta=0.8,2$ is this regime rather than a flaw of the machinery.

\subsubsection{Bulk machinery: failure coordinates, spectral kernel, and the projected system}

This subsection collects the bulk-closure diagnostics used in Sec.~\ref{sec:bulkb}: the failure coordinates and the free-fermion spectral-kernel check of Sec.~\ref{sec:bulk:kernel}, the projected-system residuals of Sec.~\ref{sec:bulk:flow} (Table~S10), and the full bulk-diagnostic grids summarized in Fig.~\ref{fig:bulk} (Table~S9).

The three failure coordinates are independent.  (i)~The ladder relevance $\rho(\mathcal L)$ against unity: above it the dressing resummation $(1-\mathcal L)^{-1}$ does not exist.  (ii)~The Feshbach-series validity, measured by the fraction of realizations with $\max_a|\bar{\mathcal R}_a^0\delta\mathcal G_a|>0.9$: above it the geometric expansion is uncontrolled.  (iii)~The moment-expansion radius $\lambda_{\rm max}/\eta$: above unity the $s$-field machinery (an expansion in $H^r/z^{r+1}$) diverges.  At $\lambda=2$, $\eta=0.8$ the ladder is already subcritical ($\rho(\mathcal L)=0.96$) while the Feshbach series is invalid in $99.5\%$ of realizations, so supercriticality is not necessary for non-closure, and the heavy tails of $\delta\mathcal G=-\delta(1/\mathcal R)$ act alone.  For the sector basis of Sec.~\ref{app:models}, the channel Hamming distance and the shared-mode count define the same basis, $d=|C_a\triangle C_b|=2(k-\ell)$ exactly for every channel pair, so the ``coupling-distance'' sectors coincide with the $\ell$-sectors; the measured bulk vertex is broadly spread over them ($41\%$ of $\|\Gamma^{(2)}\|_F^2$ at the middle bin, $\sim27\%$ at each neighbor at $\eta=0.8$, $\lambda=0$), so no scalar-per-$\ell$ basis compresses it.  What the scalar ansatz misses is therefore the frequency structure, not the spatial sector.

\subsubsection{Nonperturbative response data}

This subsection collects the nonperturbative-response data of Sec.~\ref{sec:bulk:nonpert}.  The cutoff scan of the would-be $O(\lambda^2)$ coefficient of the overlap kernel and the measured regularized response (Table~S18).

The same bookkeeping extends to arbitrary $r$ without any new input.
By the multilinearity of cumulants, the $r$-resolvent cumulant is
exactly
\begin{multline}
\mathcal C^{(r)}_{a_1\cdots a_r}
=\sum_{S\subseteq\{1,\dots,r\}}
\kappa\bigl[\{X_{a_i}\}_{i\in S};
\{\Delta_{a_j}\}_{j\notin S}\bigr],\\[2pt]
\Delta_a=\sum_{s\ge1}(\bar{\mathcal R}_a^0)^{s+1}(\delta\cG_a)^s,
\label{eq:laddergen}
\end{multline}
where the cumulant on the right takes the $X$-fields in the index set
$S$ and the ladder-dressed insertions in its complement.  The $r=3$
case is Eq.~\eqref{eq:ladder3} ($S=\{1,2\}$ contributes
$\kappa_3(X_a,X_b,\Delta_c)$, and so on).  The deep-plane ordering is
rigid.  $X_a=s_1(a)/z^2+O(z^{-3})$ and
$\Delta_a=\delta v_a/z^3+O(z^{-4})$, so the all-$X$ term carries the
leading $3r-2$ content (the $K$-kernel cumulants of
Sec.~\ref{sec:q3}, per the Proposition of Sec.~\ref{sec:gamma3}),
the mixed terms enter at successively deeper order with their leading
cross cumulants vanishing by the diagonal--off-diagonal independence
of $h$, and the pure vertex term is the single-rung insertion
\begin{equation}
\kappa_r(\Delta_{a_1},\dots,\Delta_{a_r})
=\Bigl[\prod_{i=1}^r(\bar{\mathcal R}_{a_i}^0)^2\Bigr]\,
\Gamma^{(r)}_{a_1\cdots a_r}\,\bigl[1+O(z^{-2})\bigr],
\label{eq:Deltagen}
\end{equation}
the general-$r$ form of Eq.~\eqref{eq:Delta3} (the $s\ge2$
rungs are relative $O(z^{-2})$ corrections, each extra
$\delta\cG$ factor costing one further deep-plane order).
Equation \eqref{eq:laddergen} is thus the exact relation between the
full $r$-resolvent cumulant $\mathcal C^{(r)}$ and the
Feshbach-irreducible vertices $\Gamma^{(r)}$ at every $r$, a Schur-type geometric dressing rather than an inversion.

Numerically, at $(14,2)$ the identity Eq.~\eqref{eq:ladder3} holds per
realization set to machine precision, the leading-coefficient
identification Eq.~\eqref{eq:Gdeep} closes as $1/\eta$, and the vertex
moments approach their exact limits from the h-side and the z-side
alike (full tabulation in Sec.~\ref{app:num}).  The diagonal
channel sum closes the vertex-level projection,
$\sum_a\Gamma^{(3)}_{aaa}\,z_1z_2z_3\to8Dk(N-k)$.

\subsubsection{Interaction geometries: the three families in detail}

Three classes of $W$ realize the three geometric possibilities of Eq.~\eqref{eq:rigidity}.  $w=0$ (frozen), $\Cov(w_a,w_b)\propto\delta_{ab}-1/D$ (isotropic layer), and $\Cov(w_a,w_b)=f(\ell)$ (structured polynomial).  The protocol is in Sec.~\ref{app:inter}.

\emph{(a) Random two-body pair-hopping interaction.}
\begin{equation}
W_{\rm 2b}=\sum_{\substack{p<q,\;r<s\\ \text{distinct}}}
g_{pqrs}\,\bigl(T_{pqrs}+T_{pqrs}^\dagger\bigr),
\qquad
T_{pqrs}=c_p^\dagger c_q^\dagger c_s c_r,
\label{eq:W2b}
\end{equation}
with i.i.d.\ $g_{pqrs}\sim N(0,1)$.  Its channel-basis matrix has zero diagonal ($W_{aa}=0$ exactly), support exactly the channel pairs differing in four modes, and $\E|W_{ab}|^2=1$ on every support edge, so the two-body channel graph is regular of degree
\begin{equation}
d_2=\binom k2\binom{N-k}{2},
\label{eq:d2}
\end{equation}
$\rho(\bar W_2)=d_2$, and the many-body second moment is exactly quadratic, $M_2(\lambda)=k(N-k+2)+\lambda^2d_2$.

\emph{(b) Dense ETH-scaled perturbation.} $W$ a dense GOE in the channel basis with $\E W_{ab}^2=1/D$, the Wick-branch reference of Refs.~\cite{HOETH} and of Sec.~\ref{sec:closure:solved}.

\emph{(c) Density-density interaction.}
\begin{equation}
W_{\rm dd}=\sum_{p<q} g_{pq}\,n_pn_q,
\label{eq:Wdd}
\end{equation}
with i.i.d.\ $g_{pq}\sim N(0,1)$ over all mode pairs.  Its channel-basis matrix is exactly diagonal, $W_{aa}=\sum_{p<q\in C_a}g_{pq}$.

\begin{figure*}[t]
\includegraphics[width=0.90\textwidth]{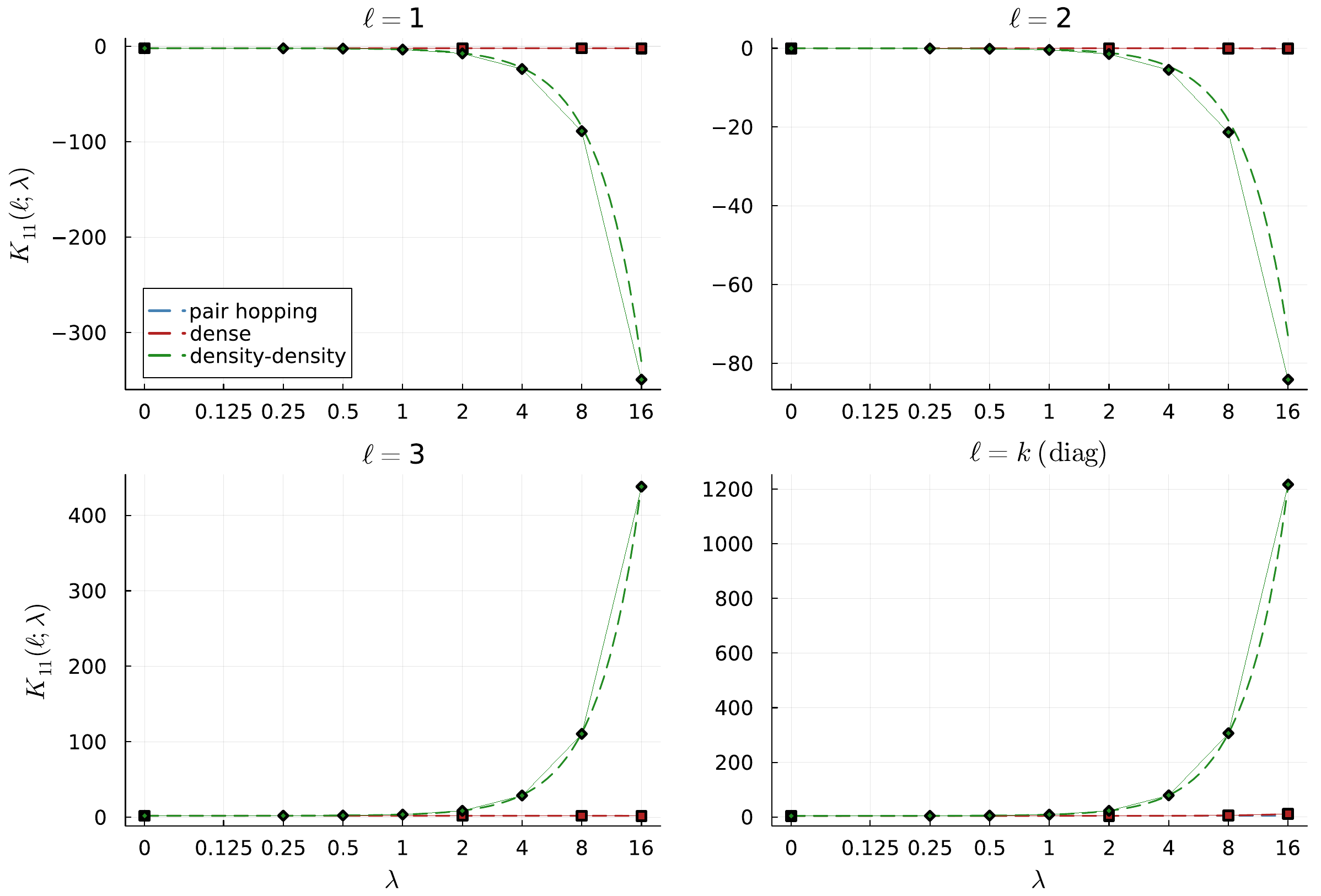}
\caption{The deep-plane deformation law [Eq.~\eqref{eq:rigidity}] for
the three interaction classes at $(8,4)$, per $\ell$-bin (small
multiples, logarithmic $\lambda$-axis with the $\lambda=0$ anchors
placed at the left edge, dashed, closed forms, markers, measured).
Pair-hopping is identically frozen at $2\ell-4$ at every $\lambda$
(measured to four decimals).  Dense follows $2\ell-4-\lambda^2\cdot2\sigma^2/D$ (cross
bins) and $4+\lambda^2\cdot2\sigma^2(1-1/D)$ (diag, measured diag at $\lambda=16$ agreeing with the
closed form to two significant digits).
Density-density follows $K_{11}(\ell;0)+\lambda^2[\binom{\ell}{2}-9/7]$
(the full grid in Table~S8).}
\label{fig:deeplaw}
\end{figure*}

Three corollaries of Eq.~\eqref{eq:rigidity}. (a)~The pair-hopping interaction has $W_{aa}=0$ and $\Tr W=0$, so $K_{11}(\ell;\lambda)=K_{11}(\ell;0)$ identically.  The deep-plane vertex sector is exactly invariant under the entire family, however strongly the bulk one-point statistics are deformed. Figure~\ref{fig:deeplaw} shows the profile identical to four decimals at every measured $\lambda\in\{0.5,1,2,4,8,16\}$ while $q$ flows from $5.0$ to $2.8$ [Table~S8(b)]. (b)~For the dense $W$, $\E[W_{aa}W_{bb}]=2\sigma^2\delta_{ab}$, $\E[W_{aa}\Tr W]=2\sigma^2$, $\E[(\Tr W)^2]=2\sigma^2D$, hence $\E[w_aw_b]=2\sigma^2(\delta_{ab}-1/D)$.  The interaction adds the isotropic layer
\begin{equation}
K_{11}(\ell;\lambda)=K_{11}(\ell;0)
+\lambda^2\,2\sigma^2\Bigl(\delta_{\ell,k}-\frac1D\Bigr)
\label{eq:denselayer}
\end{equation}
to the diagonal bin and a flat $-2\lambda^2\sigma^2/D$ to every cross-distance bin, the Wick vertex appearing as an additive isotropic layer on top of the frozen determinantal geometry, rather than replacing it.  Verified bin by bin [Table~S8(a)]. This family is the dense random-matrix crossover of Sec.~\ref{sec:closure:solved}, now governed by an exact law. (c)~For the density-density interaction any two channels share $\binom{\ell}{2}$ mode pairs, so $\E[W_{aa}W_{bb}]=\binom{\ell}{2}$.  The trace subtraction then shifts every bin by one constant [Eq.~\eqref{eq:ddconst}], and the rigidity law takes the closed form
\begin{equation}
\boxed{\;K_{11}(\ell;\lambda)=K_{11}(\ell;0)
+\lambda^2\Bigl[\binom{\ell}{2}-\frac{k^2(k-1)^2}{2N(N-1)}\Bigr]\;}
\label{eq:ddlaw}
\end{equation}
a pure quadratic polynomial in the channel geometry, with all coefficients fixed by $N,k$ and no fit.  Fig.~\ref{fig:deeplaw} verifies it at $(8,4)$, where the constant is $9/7$ [full grid in Table~S8(c)]. The three families are therefore not three separate models but three geometric realizations of one law [Eq.~\eqref{eq:rigidity}], summarized in Table~\ref{tab:summary}. The density-density family completes the contrast between the two sectors.  It deforms the deep plane while leaving the channel graph exactly frozen ($\rho(\bar W)$ is $\lambda$-independent, the off-diagonal being untouched), and its bulk one-point statistics flow away from the Wick value ($q$ grows to $43.8$, the intensities becoming increasingly concentrated), the mirror image, at the level of the diagnostics considered here, of the pair-hopping family, whose deep plane is frozen while its channel graph grows as $\lambda^2d_2$.

\begin{table*}[t]
\caption{Sector behavior of the three interaction classes at $(8,4)$:
the channel-diagonal structure, the exact deep-plane law
[Eq.~\eqref{eq:rigidity}], and the bulk status (evidence:
Figs.~\ref{fig:deeplaw}, \ref{fig:ladder}, \ref{fig:bulk},
\ref{fig:obstr}; full grids in Tables~S8 and
S10).}
\label{tab:summary}
\centering
\footnotesize
\begin{ruledtabular}
\begin{tabular}{llll}
\hline\hline
family & $W_{aa}$ & $\Delta K_{11}(\ell)/\lambda^2$ & bulk status\\
\hline
pair hopping & 0 & $0$ (frozen) & $q\to2.84$; crossing $\lambda\in(1,2)$\\
dense & isotropic & $2\sigma^2(\delta_{\ell,k}-1/D)$ & $q\to2.95$; $\lambda_c\simeq3.5$\\
density-density & structured & $\binom{\ell}{2}-\frac{k^2(k-1)^2}{2N(N-1)}$ & $q\to43.8$; supercritical at $\lambda<1$\\
\hline
\end{tabular}
\end{ruledtabular}
\end{table*}

\subsubsection{The spectral side of the sector basis}

\label{sec:spectral}

Equation~\eqref{eq:bulk3} has a spectral side.  To each eigenstate-triple
sector $s$ attach the exact multi-energy correlation density
\begin{equation}
\begin{aligned}
\rho_s^{(3)}(E_1,E_2,E_3)&=\frac1{N_s}\sum_{(n,m,l)\in s} \prod_{i=1}^{3}\delta(E_i-\lambda_{n_i}),
\end{aligned}
\label{eq:rhos}
\end{equation}
writing $n_1,n_2,n_3$ for $n,m,l$, so that $\hat\cG_3(z_1,z_2,z_3;s)
=\int\rho_s^{(3)}(E_1,E_2,E_3)\prod_{i=1}^{3}\frac{dE_i}{z_i-E_i}$, and
Eq.~\eqref{eq:bulk3} takes the exact spectral form
\begin{equation}
\begin{aligned}
\mathcal C^{(3)}_{abc}(z_1,z_2,z_3)&=\int\prod_{i=1}^{3}dE_i\; \frac{S^{(3)}(E_1,E_2,E_3)} {(z_1-E_1)(z_2-E_2)(z_3-E_3)}\\
&\quad ,
\end{aligned}
\label{eq:spectral3}
\end{equation}
with the multi-energy spectral cumulant
\begin{equation}
\begin{aligned}
S^{(3)}&=\sum_s N_s\,\Cov_3(s)\,\rho_s^{(3)},
\end{aligned}
\label{eq:S3def}
\end{equation}
so the sector covariance $\Cov_3(s)$ is the exact vertex content, and the bulk
cumulant is the threefold Stieltjes transform of $S^{(3)}$.  Thus
$S^{(r)}$ is the energy-resolved connected correlation density of the
spectral weights.  The resolvent hierarchy does not introduce a new
correlation object but is the Stieltjes/Green-function representation
of this underlying energy-resolved correlation spectrum, whose
Fourier/Laplace representation in the time domain connects it to the
corresponding multi-time
spectral-weight correlations.  The
construction promotes verbatim.  At multiplicity $r$ the $r$-variable
spectral cumulant is $S^{(r)}=\sum_s N_s\,\Cov_r(s)\,\rho_s^{(r)}$
with
$\rho_s^{(r)}=N_s^{-1}\sum_{(n_1,\dots,n_r)\in s}\prod_{i=1}^r\delta(E_i-\lambda_{n_i})$.
Hence the sector basis carries the energy-domain structure of the
hierarchy at every rung.

This identifies the dictionary with higher-order eigenstate
thermalization explicitly.  For fixed channels, $S^{(r)}$ is nothing
but the energy-resolved higher-order ETH correlation density,
\begin{multline}
P_r(E_1,\dots,E_r) :=\sum_{n_1,\dots,n_r}\Bigl\langle\prod_{j=1}^r\delta p_{n_j}^{a_j}\Bigr\rangle_c\\
\times\prod_{j=1}^r\delta(E_j-\lambda_{n_j})=S^{(r)}(E_1,\dots,E_r),
\label{eq:ETHdensity}
\end{multline}
so Eq.~\eqref{eq:spectral3} states
$\mathcal C^{(r)}=\mathcal S^{\otimes r}[P_r]$, and the boundary value
recovers the ETH density from the resolvent object,
\begin{multline}
P_r(E_1,\dots,E_r)=\frac{1}{\pi^r}\,\mathrm{Im}_1\cdots\mathrm{Im}_r\\
\times\mathcal C^{(r)}(E_1-\mathrm i0^+,\dots,E_r-\mathrm i0^+),
\label{eq:ETHinverse}
\end{multline}
with the convention of Eq.~\eqref{eq:disp3}, in the distributional
boundary-value sense.  Higher-order ETH is
therefore the on-shell projection of the multi-resolvent
hierarchy, and the resolvent representation is its off-shell
analytic generator.  The complex arguments
$z_j=E_j+\mathrm i\eta_j$ aggregate the spectral window
$|\lambda-E_j|\lesssim\eta_j$, so $\eta$ is a continuously tunable
spectral coarse-graining scale, and the joint energy arguments
$(E_1,\dots,E_r)$ carry the level correlations that a bare ETH moment
does not display.  The increment is explicit already at $r=2$.  The
diagonal ETH variance $V_{\rm ETH}(E,\Delta)=N_E^{-1}
\sum_{|\lambda_n-E|<\Delta}(\delta p_n^a)^2$ averages single-level
fluctuations only, while the resolvent covariance
\begin{equation}
\begin{aligned}
\E[|\delta\mathcal R_a(E+\mathrm i\eta)|^2]&=\sum_{n,m} \frac{\E[\delta p_n^a\delta p_m^a]} {(E+\mathrm i\eta-\lambda_n)(E-\mathrm i\eta-\lambda_m)}
\end{aligned}
\label{eq:crosslevel}
\end{equation}
carries the full cross-level covariance
($n\ne m$) together with the smoothing $\eta$.  Likewise the analytic
machinery has no bare-moment analogue.  The chain
$\delta p\to\delta\mathcal R\to\delta\cG$ of
Sec.~\ref{sec:gamma3} connects ETH fluctuations to Feshbach
self-energy fluctuations, and the deep-plane degree $3r-2$ of
Sec.~\ref{sec:gamma3} is an analytic scaling of the resolvent object
invisible in the bare ETH moments.  The multi-resolvent hierarchy is
thus not an alternative to higher-order ETH.  It is the analytic
generator whose on-shell boundary projection is higher-order ETH, and
the free-fermion sector algebra of this article provides a controlled
laboratory in which the full mapping between microscopic eigenstate
geometry and the analytic hierarchy is obtained exactly.  The
spectral cumulant is a signed measure of
vanishing total mass,
\begin{equation}
\begin{aligned}
\int S^{(3)}\prod_{i=1}^{3}dE_i&=\sum_s N_s\,\Cov_3(s)=0,
\end{aligned}
\label{eq:mass0}
\end{equation}
because the sector weights sum to zero,
$\sum_s N_s\Cov_3(s)=\E[(\sum_n\delta p_n^a)^3]=0$, the free gate of the
kernel.  Eq.~\eqref{eq:mass0} is verified exactly in rational
arithmetic for every channel pattern.

The spectral form makes the boundary structure of the cumulant
explicit.  $\mathcal C^{(3)}$ is analytic in each $z_i$ in the upper
half-plane, and for each variable the boundary value obeys the
dispersion relation.  Writing
$f_1(E_1)=\Im\mathcal C^{(3)}(E_1-\mathrm i0^+,z_2,z_3)$ with the
remaining variables held fixed,
\begin{equation}
\begin{aligned}
\Re\,\mathcal C^{(3)}(E_1-\mathrm i0^+,z_2,z_3)&=\frac1\pi\,\mathcal P\!\int\frac{f_1(E'_1)}{E_1-E'_1}\,dE'_1,
\end{aligned}
\label{eq:disp3}
\end{equation}
Equation~\eqref{eq:disp3} is
the three-point promotion of the single-resolvent boundary identity
$\mathcal R(\omega-\mathrm i0^+)/\pi=H[f](\omega)+\mathrm i f(\omega)$,
$f(\omega)=\Im\mathcal R(\omega-\mathrm i0^+)/\pi$, of the hierarchy's
two-point level \cite{HOETH}.  Iterating over the three variables gives
the full $2^3$ Hilbert dictionary $H_1H_2H_3$ relating the
real/imaginary boundary combinations of the cumulant, and the
$r$-point spectral cumulant carries the $r$-variable dictionary
$H_1\cdots H_r$.  The dispersion structure is closed under the sector
sum.  Each $\hat\cG_3(\cdot\,;s)$ is analytic in the upper half-plane in
every variable, so the sector basis carries the boundary-value
structure of the hierarchy level by level.

The deep plane of the spectral form is the moment dictionary.  The
sector moment sums factor over the coincidence classes of the triple
into the one-point power sums $p_j=\sum_n\lambda_n^j$.  For the
coincident class $n=m=l$,
$M_{(3)}(a,b,c)=p_{a+b+c}$.  For one coincident pair,
$M_{(2,1)}(a,b,c)=p_{a+b}\,p_c-p_{a+b+c}$ (orientations read
cyclically).  And for $n,m,l$ all distinct,
\begin{multline}
M_{(1,1,1)}(a,b,c)=p_ap_bp_c-p_{a+b}p_c-p_{a+c}p_b\\
-p_{b+c}p_a+2p_{a+b+c},
\label{eq:momdict}
\end{multline}
each identity exact per realization (each pattern sector belongs to
exactly one class).  The deep-plane coefficients of
Eq.~\eqref{eq:bulk3} are therefore the moments of $S^{(3)}$,
\begin{equation}
\begin{aligned}
K_{abc}(a,b,c)&=\int E_1^{a}E_2^{b}E_3^{c}\,S^{(3)}(E_1,E_2,E_3)\prod_{i=1}^{3}dE_i,
\end{aligned}
\label{eq:moments3}
\end{equation}
and the sector assembly
$K_{abc}(a,b,c)=\sum_s N_s\,\Cov_3(s)\,M_s(a,b,c)$ follows from
Eq.~\eqref{eq:S3def} since
$\int E_1^{a}E_2^{b}E_3^{c}\rho_s^{(3)}\prod dE_i=M_s(a,b,c)/N_s$.
The moments of $S^{(3)}$ are therefore the deep-plane coefficients.
$K_{211}$ is its $(2,1,1)$ moment and $K_{111}=0$ the vanishing
$(1,1,1)$ moment in the moment-field identification of
Eq.~\eqref{eq:momentcov}, the identification carrying the percent-level
accuracy verified below.  Equations
\eqref{eq:spectral3}--\eqref{eq:moments3} close the triangle
$\Gamma^{(3)}\leftrightarrow\mathcal C^{(3)}\leftrightarrow
S^{(3)}$.  The Feshbach ladder of Sec.~\ref{sec:gamma3} relates the
irreducible vertex $\Gamma^{(3)}$ to $\mathcal C^{(3)}$ (the leading
irreducible content being extracted in the deep plane), the spectral
side of $\mathcal C^{(3)}$ is
$S^{(3)}$, whose moments are the deep-plane coefficients, the ladder
structure of Ref.~\cite{HOETH} and the sector basis are the same
object in two representations.  A Lanczos/continued-fraction
resummation of the boundary values of $S^{(3)}$ would give the
sector-level self-energy closure of the same ladder.  This Krylov route
is beyond the present scope.

Numerically the spectral side is exact where it must be
(Sec.~\ref{app:num}), the class moment sums of a fixed realization
reproduce the power-sum factorizations of Eq.~\eqref{eq:momdict} to
machine precision, the boundary values of the exact channel spectrum
and of the sector reconstruction obey Eq.~\eqref{eq:disp3}, and the
sector reconstruction of Eq.~\eqref{eq:bulk3} at the probe point of
Table~\ref{tab:bulk} reproduces the direct cumulant and the
closed-form $K_{211}$ at the percent level, the total mass of
Eq.~\eqref{eq:mass0} vanishing identically as it must.  The decoupled
dictionary value and the exact per-realization closure
$s_2(a)s_1(b)s_1(c)$ agree with the closed form within the closure's
sampling error, bounding the spectral--channel correlation at the
percent level.  The bulk/deep-plane decoupling of
Sec.~\ref{sec:bulkb} extends to the spectral domain.

\subsection{Sector-resolved interaction response at $(8,4)$}
\label{app:interact}

Table~S19 collects the sector-resolved singular
response of the interacting two-point kernel of
Sec.~\ref{sec:interact} at $(8,4)$, $n_{\rm real}=800$.  The fitted
slope $b$ of the would-be $O(\lambda^2)$ coefficient against
$\ln(1/\delta_c)$ over $\delta_c=0.5,0.2,0.1,0.05,0.02$ (level pairs
below $\delta_c$ excluded), and the even part of the measured response
against the exact anchor,
$\Delta \mathcal E(m,\ell;\lambda)=\tfrac12[\mathcal E(\lambda)+\mathcal E(-\lambda)]-\mathcal E(0)$, at
$\lambda=2^{-4}$ and $2^{-1}$.  Over the fitted interval
$\lambda=2^{-6},\dots,2^{-1}$ the response grows by a factor
$4.2$--$6.2$ in the off-diagonal sectors and $8.5$--$10.1$ in the
coincident-eigenstate row, i.e.\ $\Delta \mathcal E\sim\lambda^{\alpha}$ with
an effective exponent $\alpha\approx0.45$ ($0.41$--$0.53$)
off-diagonal and $\alpha\approx0.63$ ($0.62$--$0.67$) coincident,
a factor $7$--$12$ below the cutoff-regularized quadratic estimate
$C_2(\delta_c)\lambda^2$ at any fixed $\delta_c$ (at $\delta_c=0.02$,
$\lambda=0.125$).  The $(1,1)$ bin reproduces the flat
$-(3$--$8)\times10^{-5}$ response quoted in Table~S18 of
Sec.~\ref{app:inter}, and
the Porter--Thomas even part
$\tfrac12[q(\lambda)+q(-\lambda)]-q(0)=-0.20,-0.39,-0.72,-1.18,-1.56,-1.80$
at $\lambda=2^{-6},\dots,2^{-1}$ matches the same reference within
statistics.  The slope anchors quoted in the main text are
$b(0,0)=-0.116$ and $b(4,4)=-0.054$ per $\ln(1/\delta_c)$, with
$|b|\lesssim0.016$ in the first-neighbor rows and $|b|\le0.005$ at
$m=2$.  At $\lambda=0.5$ the two complementary corners have moved
$-77\times10^{-5}$ and $-37\times10^{-5}$.

\subsection{The containment propositions}\label{app:containment}

\emph{Formal statements.}  The containment is stated as three propositions and a corollary, each proved with the equations already introduced (Secs.~\ref{sec:benchmark}--\ref{sec:bulkb}).

\emph{Proposition 1 (direct ETH objects are boundary projections).} Let
$\mathcal C_{ab}(z,z')=\Cov[R_a(z),R_b(z')]$. The energy-resolved ETH
fluctuation kernel $K_{\alpha\beta}(E,E')$ and its difference-resolved
form $F_{\alpha\beta}(\omega)$ are recovered from $\mathcal C_{ab}$ by
inverting the two-variable Cauchy transform [Eqs.~\eqref{eq:bridge}
and \eqref{eq:rescov}].  The double discontinuity of $\mathcal C_{ab}$
across the real axes returns the spectral-density covariance, the
direct kernel $K_{\alpha\beta}(E,E')$ plus the explicitly known
flat-weight part $C_{\alpha\beta}^{\rm flat}$ [Eq.~\eqref{eq:Cflat}], 
and setting $E'=E+\omega$ and integrating over $E$ returns
$F_{\alpha\beta}(\omega)$ [Sec.~\ref{app:bench}]. Direct two-point
ETH statistics are therefore derived observables of the two-resolvent
object.  Within the overlap-resolved fluctuation sector considered
here, the direct kernel is a boundary observable, while the resolvent
object retains the full analytic structure.

\emph{Proposition 2 (ETH moments are coefficients of one analytic
object).} Wherever the covariance admits the large-frequency expansion
$\mathcal C_{ab}(z,z')=\sum_{r,s}K_{rs}^{ab}/(z^{r+1}z'^{s+1})$
[Eq.~\eqref{eq:Krs}], its coefficients are the energy-weighted overlap
moments, $K_{rs}^{ab}=\sum_{A,B}\E[\delta p_A^a\delta p_B^b]\,
\lambda_A^r\lambda_B^s$.  The hierarchy of two-point ETH moments is the
coefficient array of a single analytic object, with the moment order
$(r,s)$ a subsidiary index below the resolvent multiplicity (at
multiplicity three the same statement organizes the three-resolvent
cumulants into $K_{rst}$ [Sec.~\ref{app:closure}]). Computing
moments directly computes coefficients without the object.

\emph{Proposition 3 (closure failure is sector-resolved).} Given the
exact ladder equation [Eq.~\eqref{eq:level2}], the projection residual
$\Delta_\alpha=P_\alpha[\mathcal C^{(2)}_{\rm
exact}]-P_\alpha[\mathcal C^{(2)}_{\rm ansatz}]$ splits into
ladder-generated and irreducible parts [Eq.~\eqref{eq:residual}],
separated by the $\eta^0$ versus $\eta^{-4}$ dressing scaling of
Sec.~\ref{sec:closure:solved}. A nonzero irreducible part is a
statement about a missing vertex sector, not a mismatch of one more
computed moment, a diagnostic that direct moment comparison does not
provide.

\emph{Corollary (the deep-plane/bulk split is structural).} The
moment-field identity [Eq.~\eqref{eq:momentcov}] holds for any
Hamiltonian without any resolvent input, and the bulk obstruction
[Sec.~\ref{sec:bulk:nonpert}] is the divergence of the
eigenstate-perturbation denominators themselves. The split between the
moment-determined deep plane and the eigenstate-resolved bulk is
therefore intrinsic to the fluctuation sector.  The resolvent
formulation exposes it rather than creates it.

\begin{table*}[htbp]
\caption{Result index: each main result, the script that produces or
verifies it, and the data file it reads or writes.  Paths are relative
to the reproducibility package of Ref.~\cite{CodeData}.}
\label{tab:index}
\small
\begin{ruledtabular}
\begin{tabular}{p{4.9cm}p{5.6cm}p{4.4cm}}
\hline\hline
result & entry point & data \\
\hline
exact replica-sector decomposition, general-$r$ assembly
(Sec.~\ref{sec:replica}, Sec.~\ref{sec:assembly})
& \texttt{occ\_assemble.jl}, \texttt{occ\_census.jl},
\texttt{occ\_dp.jl}, \texttt{wg\_cosets.jl}
& \texttt{data/wg\_depth6.json} \\
orthogonal Weingarten sector tables, depth six to nine
(Sec.~\ref{sec:weingarten}, App.~\ref{app:algebra})
& \texttt{wg\_cosets.jl}, \texttt{jack\_wg.jl}, \texttt{wg9.jl},
\texttt{omega\_gen.jl}, \texttt{data/build\_wg\_json.jl}
& \texttt{data/wg\_depth\{6,8,9\}.json}, \texttt{wg12.json} \\
two-resolvent benchmark: envelope, $q_r$, kernel $L(k;m,\ell)$
(Sec.~\ref{sec:benchmark}, App.~\ref{app:bench})
& \texttt{c1\_pack.jl}, \texttt{c1\_verify.jl}, \texttt{c1\_tex.jl}
& \texttt{data/C1\_data.json}, \texttt{.jls} \\
three-point kernel and $r\to r-1$ hierarchy
(Sec.~\ref{sec:kernel}, App.~\ref{app:algebra})
& \texttt{kernel3.jl}, \texttt{kernel3\_build.jl},
\texttt{kernel3\_eval.jl}, \texttt{kernel3k3\_build.jl},
\texttt{occ\_r4chk.jl}, \texttt{occ\_r5.jl}
& \texttt{kernel3\_out.txt} \\
cumulants and Feshbach vertices
(Sec.~\ref{sec:vertices}, App.~\ref{app:derivations})
& \texttt{three\_resolvent.jl}, \texttt{gamma3\_test.jl},
\texttt{spectral\_side.jl}, \texttt{wick\_exact2.jl}, \texttt{wick\_exact3.jl},
\texttt{fk\_check*.jl}
& \texttt{spectral\_side\_out.txt} \\
four-point kernel and four-resolvent content
(Sec.~\ref{sec:kernel4}, App.~\ref{app:algebra})
& \texttt{kernel4pt\_build.jl}, \texttt{kernel4pt\_bulk3.jl},
\texttt{kernel4pt\_*.jl}, \texttt{kernel4.jl}
& sector tables of App.~\ref{app:algebra} \\
interaction rigidity, bulk boundary and response
(Secs.~\ref{sec:closure}--\ref{sec:interact}, App.~\ref{app:derivations})
& \texttt{kernel\_lambda\_sectors.jl}, \texttt{anchor\_verify.jl},
\texttt{anchor\_Nscan.jl}, \texttt{kernel2k4\_anchor.jl},
\texttt{a4\_analysis.jl}, \texttt{a4\_paper\_tables.jl}
& response tables of App.~\ref{app:derivations} \\
\hline
\end{tabular}
\end{ruledtabular}
\end{table*}

\section{Model specification and reproducibility}
\label{app:repro}

The models are those of Sec.~\ref{sec:benchmark} and Secs.~\ref{sec:rigidity}--\ref{sec:bulkb}, with the sizes and sample counts stated in Sec.~\ref{app:numerics}.  The data directory of the deposited packages is documented in \texttt{data/README.md}, which fixes the JSON layout (denominator and numerator coefficient lists, class sizes, and the sphere-identity check).  Table~\ref{tab:index} maps each main result to its entry
point.

\subsection{Reproduction tables}
\label{app:supp}

The numerical tables that record the verification data are not typeset in this article.  They are archived in the reproducibility package of Ref.~\cite{CodeData}, both as the typeset supplement \texttt{data/tables/supp\_tables.pdf} and as the machine-readable files listed in Table~\ref{tab:index}.  The supplement carries its own table index, and three entries mark the groups: Table~S1 the two-point benchmark values (S1--S10 cover the benchmark, deformation and crossover grids), Table~S14 the zonal spherical table at $d=4$ (S11--S13 are the Weingarten rational functions of depths six, eight and nine), and Table~S20 the SCBA comparison (S15--S26 are the verification tables behind the closure, rigidity, response and reproduction sections, including the Mag\'an correspondence in Table~S26).
The division is by evidence type: the main text quotes the entries that each carry a conclusion on their own, stating the check they record; the supplement typesets the parameter grids behind them; and the package carries the full scans, the sector data, and the entry scripts of Table~\ref{tab:index}.
\subsection{Code and data availability}
\label{app:models}

All numerical scripts, the exact tables and the raw output are
deposited in the reproducibility package \cite{CodeData} (concept doi, 10.5281/zenodo.22949584; this version, 10.5281/zenodo.23726068), which carries the numerical material of both companion manuscripts.  Its layout is documented in \texttt{data/README.md}: \texttt{code/} holds the Julia drivers, \texttt{data/} the exact tables and exported kernel data, and \texttt{figures/} the generated figures.
Table~\ref{tab:index} maps each main result to its entry point, and the package README lists every file of code/ and data/, including the helper scripts behind each row.  Every
script runs under Julia with the MKL linear-algebra backend; the
sampling parameters of every table are listed in
Sec.~\ref{app:numerics}, and the models are those of
Sec.~\ref{sec:bench:dict} and Secs.~\ref{sec:rigidity}--\ref{sec:bulkb}.


\begin{thebibliography}{99}
\bibitem{Deutsch1991} J.~M. Deutsch, Quantum statistical mechanics in a
closed system, Phys. Rev. A \textbf{43}, 2046 (1991).

\bibitem{Srednicki1994} M.~Srednicki, Chaos and quantum thermalization,
Phys. Rev. E \textbf{50}, 888 (1994).

\bibitem{Dalessio2016} L.~D'Alessio, Y.~Kafri, A.~Polkovnikov, and
M.~Rigol, From quantum chaos and eigenstate thermalization to
statistical mechanics and thermodynamics, Adv. Phys. \textbf{65}, 239
(2016).

\bibitem{Wang2022} J.~Wang, M.~H. Lamann, J.~Richter, R.~Steinigeweg,
A.~Dymarsky, and J.~Gemmer, Eigenstate thermalization hypothesis and
its deviations from random-matrix theory beyond the thermalization
time, Phys. Rev. Lett. \textbf{128}, 180601 (2022).

\bibitem{Jafferis2023} D.~L. Jafferis, D.~K. Kolchmeyer,
B.~Mukhametzhanov, and J.~Sonner, Matrix models for eigenstate
thermalization, Phys. Rev. X \textbf{13}, 031033 (2023).

\bibitem{HOETH} Z.~Huang, A Multi-Resolvent Hierarchy for the ETH Smooth Function, arXiv:2607.19861 (2026).

\bibitem{Pappalardi2022} S.~Pappalardi, L.~Foini, and J.~Kurchan,
Eigenstate thermalization hypothesis and free probability,
Phys. Rev. Lett. \textbf{129}, 170603 (2022).

\bibitem{Mirlin} A.~D. Mirlin, Statistics of energy levels and
eigenfunctions in disordered systems, Phys. Rep. \textbf{326}, 259
(2000).

\bibitem{ErdosYau} L.~Erd\H{o}s and H.-T. Yau, \emph{A Dynamical
Approach to Random Matrix Theory}, Courant Lecture Notes
(AMS, Providence, 2017).

\bibitem{Tao2012} T.~Tao, \emph{Topics in Random Matrix Theory},
Graduate Studies in Mathematics Vol.~132 (AMS, Providence, 2012).

\bibitem{PottersBouchaud} M.~Potters and J.-P. Bouchaud, \emph{A First
Course in Random Matrix Theory} (Cambridge University Press,
Cambridge, 2021).

\bibitem{Pastur} L.~A. Pastur, Spectra of random self-adjoint
operators, Russ. Math. Surv. \textbf{28}, 1 (1973).

\bibitem{Mehta} M.~L. Mehta, \emph{Random Matrices}, 3rd ed.
(Elsevier, Amsterdam, 2004).

\bibitem{Magan2016} J.~M. Mag\'an, Random Free Fermions: An Analytical
Example of Eigenstate Thermalization, Phys. Rev. Lett. \textbf{116},
030401 (2016).

\bibitem{Weingarten} B.~Collins, S.~Matsumoto, and J.~Novak, The
Weingarten calculus, Not. Am. Math. Soc 69, no. 1 (2022).

\bibitem{CM2009} B.~Collins and S.~Matsumoto, On some properties of
orthogonal Weingarten functions, J. Math. Phys. \textbf{50}, 113516
(2009).

\bibitem{THF} X.~K. Guo, Z. Huang, On thermalization in random free fermions: Statistical origins and energy-dependent ETH structure, arXiv:2610.01414 (2026).

\bibitem{Alt1995} H.~Alt, H.-D.~Gr\"af, H.~L. Harney, R.~Hofferbert,
H.~Lengeler, A.~Richter, P.~Schardt, and H.~A. Weidenm\"uller,
Gaussian orthogonal ensemble statistics in a microwave stadium
billiard with chaotic dynamics: Porter--Thomas distribution and
algebraic decay of time correlations, Phys. Rev. Lett. \textbf{74},
62 (1995).

\bibitem{AbsilEdelmanKoev} P.-A. Absil, A. Edelman, and P. Koev, On
the largest principal angle between random subspaces, Linear Algebra
Appl. \textbf{414}, 288 (2006).

\bibitem{Feshbach1958} H.~Feshbach, Unified theory of nuclear
reactions, Ann. Phys. (N.Y.) \textbf{5}, 357 (1958).

\bibitem{Feshbach1962} H.~Feshbach, A unified theory of nuclear
reactions. II, Ann. Phys. (N.Y.) \textbf{19}, 287 (1962).

\bibitem{Lehner2004} F.~Lehner, Cumulants in noncommutative
probability theory I: Noncommutative exchangeability systems, Math. Z. 248, 67-100 (2004).

\bibitem{CollinsSniady} B.~Collins and P.~\'Sniady, Integration with
respect to the Haar measure on unitary, orthogonal and symplectic
group, Commun. Math. Phys. \textbf{264}, 773 (2006).

\bibitem{PorterThomas} C.~E. Porter and R.~G. Thomas, Fluctuations of
nuclear reaction widths, Phys. Rev. \textbf{104}, 483 (1956).

\bibitem{Kato} T.~Kato, \emph{Perturbation Theory for Linear
Operators}, 2nd ed. (Springer, Berlin, 1995).

\bibitem{Dyson1962} F.~J. Dyson, A Brownian-motion model for the
eigenvalues of a random matrix, J. Math. Phys. \textbf{3}, 1191 (1962).

\bibitem{Efetov} K.~B. Efetov, \emph{Supersymmetry in Disorder and
Chaos} (Cambridge University Press, Cambridge, 1997).

\bibitem{MingoSpeicher} J.~A. Mingo and R.~Speicher, Schwinger--Dyson
equations: classical and quantum, arXiv:1307.1806 (2013).

\bibitem{KotaBook} V.~K.~B. Kota, \emph{Embedded Random Matrix
Ensembles in Quantum Physics} (Springer, Cham, 2014).

\bibitem{KotaTBRE} V.~K.~B. Kota, A.~Rela\~no, J.~Retamosa, and
M.~Vyas, Thermalization in the two-body random ensemble,
J. Stat. Mech. (2011) P10028.

\bibitem{Haydock1972} R.~Haydock, V.~Heine, and M.~J. Kelly,
Electronic structure based on the local atomic environment for
tight-binding bands, J. Phys. C \textbf{5}, 2845 (1972).

\bibitem{Haydock1980} R.~Haydock, The recursive solution of the
Schr\"odinger equation, Comput. Phys. Commun. \textbf{20}, 11 (1980).

\bibitem{Nandy2025} P.~Nandy, A.~S. Matsoukas-Roubeas, P.~Mart\'inez-Azcona,
D.~K. Zhdanov, and A.~Dymarsky, Quantum dynamics in Krylov space:
Methods and applications, Phys. Rep. \textbf{1125}, 1 (2025).

\bibitem{Pappalardi2025} S.~Pappalardi, F.~Fritzsch, and T.~Prosen,
Full eigenstate thermalization via free cumulants in quantum lattice
systems, Phys. Rev. Lett. \textbf{134}, 140404 (2025).

\bibitem{CodeData} Z.~Huang, Source code and data for ``Exact replica-sector hierarchy of multi-resolvent spectral-weight fluctuations in random free fermions,'' Zenodo, \url{https://doi.org/10.5281/zenodo.22949584} (2026).

\bibitem{Muirhead} R.~J. Muirhead, \emph{Aspects of Multivariate
Statistical Theory} (Wiley, New York, 1982).

\bibitem{Forrester} P.~J. Forrester, \emph{Log-Gases and Random
Matrices} (Princeton University Press, Princeton, 2010).

\end{thebibliography}
\end{document}